\pdfoutput=1
\documentclass[aps,prd,preprint, onecolumn,superscriptaddress,preprintnumbers,floatfix,nofootinbib,longbibliography,eqsecnum, amsmath, amsfonts]{revtex4-2}

\usepackage[colorlinks,allcolors=blue]{hyperref}
\usepackage[german,english]{babel}
\usepackage{bm}
\usepackage{xcolor}
\usepackage{bbm}
\usepackage{bbold}
\usepackage{ulem}
\usepackage{braket}
\usepackage{tablefootnote}
\usepackage{slashed}
\usepackage{multirow}
\usepackage{enumerate}
\usepackage[mathlines]{lineno}
\usepackage{comment}
\usepackage{graphicx}
\usepackage{scalerel}

\newcommand{\cg}[6]{
\left\langle
\begin{matrix}
#1	& #3	\\[-5pt]
#2	& #4
\end{matrix}
\bigg\vert
\begin{matrix}
#5	\\[-5pt]
#6 
\end{matrix}
\right\rangle
}

\allowdisplaybreaks

\begin{document}
\count\footins = 1000
\preprint{TUM-EFT 198/25}
\title{Unravelling Pentaquarks with Born--Oppenheimer effective theory
}

\author{Nora Brambilla}
\email{nora.brambilla@tum.de}
\affiliation{Technical University of Munich,\\
TUM School of Natural Sciences, Physics Department,\\   
James-Franck-Str.~1, 85748 Garching, Germany.}
\affiliation{Technical University of Munich, Institute for Advanced Study, \\ 
Lichtenbergstrasse 2 a, 85748 Garching, Germany.}
\affiliation{Technical University of Munich, Munich Data Science Institute, \\ 
Walther-von-Dyck-Strasse 10, 85748 Garching, Germany.}

\author{Abhishek Mohapatra}
\email{abhishek.mohapatra@tum.de}
\affiliation{Technical University of Munich,\\
TUM School of Natural Sciences, Physics Department,\\   
James-Franck-Str.~1, 85748 Garching, Germany.}

\author{Antonio Vairo}
\email{antonio.vairo@tum.de}
\affiliation{Technical University of Munich,\\
TUM School of Natural Sciences, Physics Department,\\   
James-Franck-Str.~1, 85748 Garching, Germany.}

\date{\today}

\begin{abstract}
The hidden-charm pentaquark states $P_{c\bar{c}}\left(4312\right)^+$, $P_{c\bar{c}}\left(4380\right)^+$, $P_{c\bar{c}}\left(4440\right)^+$, and $P_{c\bar{c}}\left(4457\right)^+$, all with isospin $I = 1/2$, were discovered by the LHCb collaboration in the decay process $\Lambda_b^0 \to J/\psi p K^-$. 
Although their quantum numbers remain undetermined, these states have generated significant theoretical interest. 
We analyze their spectrum and decay patterns—including those of their spin partners—within the Born--Oppenheimer effective field theory (BOEFT), a framework grounded in QCD. 
At leading order in BOEFT, we identify these pentaquark states as bound states in BO potentials that exhibit at short-distance a repulsive octet behavior and a nonperturbative shift due to the  
adjoint baryons masses, while asymptotically approaching the $\Sigma_c\bar{D}$ threshold. 
We further incorporate ${\cal O}(1/m_Q)$ spin-dependent corrections to compute pentaquark multiplet spin splittings. 
Based on the spectrum, semi-inclusive decay widths to $J/\psi$ and $\eta_c$, and
the decay width ratios to $\Lambda_c\bar{D}$ and $\Lambda_c\bar{D}^*$, we
provide the first theoretical predictions for the adjoint baryon masses, which can
be confirmed by future lattice QCD studies. Moreover, our analysis supports the quantum number assignments:
$J^{P} = (1/2)^-$ for $P_{c\bar{c}}\left(4312\right)^+$,
$(3/2)^-$ for $P_{c\bar{c}}\left(4380\right)^+$,
$(1/2)^-$ for $P_{c\bar{c}}\left(4457\right)^+$, and
$(3/2)^-$ for $P_{c\bar{c}}\left(4440\right)^+$. 
We also present results for the lowest bottom pentaquarks.
\end{abstract}

\maketitle
\clearpage

\section{Introduction}

Quantum Chromodynamics (QCD), the established theory of the strong interaction, describes the dynamics of quarks and gluons. 
The property of color confinement confines quarks and gluons into color‑singlet hadrons and allows for a richer spectrum of bound states than conventional  mesons (quark-antiquark pairs) or baryons (three-quark states) \cite{GellMann:1964nj,Zweig:1964CERN,Jaffe:1975fd} including exotic hadrons such as tetraquarks (four-quark states), pentaquarks (five-quark states), hybrids (bound states of quarks with gluonic excitations), and glueballs (bound states composed purely of gluons) and so on, 
see Ref.~\cite{Brambilla:2019esw} for a review. 
The past two decades have seen a surge of discoveries of  $XYZ$ states that are exotic hadrons with  at least two heavy quarks and are often located near or above open-flavor thresholds. 
The Belle experiment in 2003 discovered the first XYZ state, $\chi_{c1}\left(3872\right)$, containing a charm quark-antiquark $\left(c\bar{c}\right)$ pair, in the $B\to K \pi^+\pi^-J/\psi$ decay channel \cite{Belle:2003nnu}. 
This breakthrough paved the way for the discovery of dozens of new $XYZ$ hadrons by various experiments at the B-factories (BaBar, Belle, Belle2), $\tau$-charm  facilities (BES, BESIII),  and also
proton-(anti)proton  colliders (CDF, D0, LHCb, ATLAS, CMS)  (see Refs.~\cite{Esposito:2016noz,Olsen:2017bmm, Guo:2017jvc, Ali:2017jda,Brambilla:2019esw,Chen:2022asf,Brambilla:2010cs} for reviews). 
To date, approximately $55$ $XYZ$  states with at least two heavy quarks have been observed \cite{Lebed:2023vnd}. 
Among them are six pentaquark states discovered by LHCb in the $c\bar{c}$ sector: $4$ isospin half-integer $\left(I=1/2\right)$ states $P_{c\bar{c}}\left(4312\right)^+$, $P_{c\bar{c}}(4380)^+$, $P_{c\bar{c}}(4440)^+$, and $P_{c\bar{c}}(4457)^+$ with quark content $c\bar{c}qqq$ and $2$ isospin integer $\left(I=0\right)$ states $P_{c\bar{c}}(4338)^0$ and $P_{c\bar{c}}(4459)^0$ with quark content $c\bar{c}sqq$, where $q\equiv\left(u, d\right)$ \cite{LHCb:2015yax, LHCb:2019kea, LHCb:2020jpq, LHCb:2022ogu}. 
Understanding the nature of these pentaquark states remains an elusive problem.

The LHCb collaboration first reported the discovery of two hidden-charm pentaquark states in the $J/\psi p$ invariant mass distribution of the $\Lambda_b^0 \to J/\psi p K^-$ decay \cite{LHCb:2015yax}. 
These included a broad state $P_{c\bar{c}}\left(4380\right)^+$ with mass $4380\pm 30$~MeV and width $210\pm 90$~MeV, and a narrower state $P_{c\bar{c}}\left(4450\right)^+$ with mass $4449.8\pm 3$~MeV and width $39\pm 20$~MeV. 
Four years later, in 2019, a subsequent updated analysis with larger data sample (Run 1 and Run 2) by the LHCb collaboration found a new narrow state $P_{c\bar{c}}\left(4312\right)^+$ and further resolved the originally reported $P_{c\bar{c}}\left(4450\right)^+$ into two distinct narrow resonances: $P_{c\bar{c}}\left(4440\right)^+$ and $P_{c\bar{c}}\left(4457\right)^+$ \cite{LHCb:2019kea}. 
The broad state $P_{c\bar{c}}\left(4380\right)^+$ was neither confirmed nor refuted in the updated analysis, which was only sensitive to narrow peaks.  
A clear signal for this state was not seen in the $J/\psi p$ mass spectrum in Ref.~\cite{LHCb:2019kea}, and  therefore the existence of $P_{c\bar{c}}\left(4380\right)^+$ needs further experimental confirmation. 
In our work, we refer to  a ``$P_{c\bar{c}}\left(4380\right)^+$'' state, whose mass aligns with the value reported in Ref.~\cite{LHCb:2015yax}, 
although our predicted decay width is significantly smaller and inconsistent with that analysis, which is now considered obsolete. 
Table~\ref{tab:exoticPenta} shows the four hidden-charm pentaquark states reported in the PDG \cite{ParticleDataGroup:2024cfk}. 
The quantum number $J^{P}$ assignment of these states is not yet known.

\begin{table}[h!]
\begin{center}
\small{\renewcommand{\arraystretch}{1.5}
\scriptsize
\begin{tabular}{|ccccc|}
\hline
   $\begin{array}{c} {\rm State}\\({\rm PDG})\end{array}$       &   $M$~(MeV)       & $\Gamma$~(MeV) &  $I\left(J^{P}\right)$    &  Decay modes      \\
   \hline
 $P_{c\bar{c}}\left(4312\right)^+$  & $4311.9^{+7.0}_{-0.9}$ & $10 \pm 5$ & $1/2\left(?^{?}\right)$ & $J/\psi\,p$\\
 $P_{c\bar{c}}\left(4380\right)^+$  & $4380 \pm 30$ & $210 \pm 90$\footnote{The reported width is based on the original analysis \cite{LHCb:2015yax}, which is nowadays considered obsolete.} & $1/2\left(?^{?}\right)$ & $J/\psi\,p$\\
$P_{c\bar{c}}\left(4440\right)^+$  & $4440^{+4.0}_{-5.0}$ & $21^{+10}_{-11}$ & $1/2\left(?^{?}\right)$ & $J/\psi\,p$\\  
$P_{c\bar{c}}\left(4457\right)^+$\footnote{Earlier this state was $P_{c\bar{c}}\left(4450\right)$} & $4457.3^{+4.0}_{-1.8}$ & $6.4^{+6.0}_{-2.8}$ & $1/2\left(?^{?}\right)$ & $J/\psi\,p$\\
    \hline
\end{tabular}
\caption{The list of non-strange pentaquark states with quark content $c\bar{c}qqq$, where $q=\left(u,d\right)$ that are listed on PDG \cite{ParticleDataGroup:2024cfk}. All the states have $I=1/2$ but $J^{P}$ quantum numbers are unknown. }
\label{tab:exoticPenta}}
\end{center}
\end{table}

The hidden-charm pentaquark states $P_{c\bar{c}}$ were theoretically anticipated well before their experimental observation \cite{Wu:2010jy,Wu:2010vk, Wang:2011rga,Yang:2011wz,Wu:2012md,Li:2014gra,Chen:2015loa,Karliner:2015ina}. 
After the LHCb collaboration discovery,  a wide range of theoretical frameworks has emerged to explain the nature of the pentaquark states such as hadronic molecules \cite{Chen:2016qju,Liu:2019tjn,Du:2019pij,Du:2021fmf, 
Chen:2019bip,Chen:2019asm,Guo:2019twa,He:2019ify,Guo:2019kdc,Shimizu:2019ptd,Xiao:2019pjg,Xiao:2019aya,Wang:2019nwt,Meng:2019ilv,Wu:2019adv,Voloshin:2019aut,Wang:2019hyc,Yamaguchi:2019seo,Liu:2019zvb,Lin:2019qiv,Wang:2019ato,Gutsche:2019mkg,Burns:2019iih,Wang:2019spc,Xu:2020gjl,Kuang:2020bnk,Peng:2020xrf,Xiao:2020frg,Dong:2021juy,Burns:2021jlu, Zhang:2023czx, Liu:2024ugt, Li:2025ejt}, compact pentaquark states \cite{Maiani:2015vwa,  Ali:2019npk,Ali:2019clg,Zhu:2019iwm,Wang:2019got,Giron:2019bcs,Cheng:2019obk,Stancu:2019qga, Kuang:2020bnk, Shi:2021wyt, Giron:2021fnl}, baryo-charmonia or hadro-charmonia \cite{Eides:2015dtr, Wu:2017weo, Eides:2019tgv,Ferretti:2018ojb, Garcilazo:2022kra, Germani:2024miu}, cusp effects \cite{Kuang:2020bnk}, triangle singularities \cite{Nakamura:2021qvy} and virtual states \cite{Fernandez-Ramirez:2019koa}. 
Given that the pentaquark states $P_{c\bar{c}}$ in Table~\ref{tab:exoticPenta} have masses within close proximity of the $\Sigma_c^{\left(*\right)}\bar{D}^{\left(*\right)}$ thresholds, the states have been widely interpreted as hadronic molecules with S-wave $\Sigma_c^{\left(*\right)}\bar{D}^{\left(*\right)}$ constituents. 
In this interpretation, $P_{c\bar{c}}\left(4312\right)^+$ is a $\left(1/2\right)^-$ $\Sigma_c\bar{D}$ state and $P_{c\bar{c}}\left(4380\right)$ is a $\left(3/2\right)^-$ $\Sigma_c^*\bar{D}$ state, albeit with a smaller width compared to the state observed in the original LHCb analysis \cite{Burns:2021jlu, Du:2019pij, Du:2021fmf}. 
$P_{c\bar{c}}\left(4440\right)^+$ and $P_{c\bar{c}}\left(4457\right)^+$ are both $\Sigma_c\bar{D}^*$ states, and can be assigned either $\left(1/2\right)^-$ and $\left(3/2\right)^-$, or $\left(3/2\right)^-$ and $\left(1/2\right)^-$, respectively. 
In contrast, the compact pentaquark model attributes positive parity to some of these states. 
Within  the molecular framework, the heavy quark spin symmetry implies that there is a  multiplet of seven pentaquark states (including three additional states near the $\Sigma_c^*\bar{D}^*$ threshold with respect to the four states listed in Table~\ref{tab:exoticPenta}). 
As remarked above, in the molecular framework, the mass spectrum alone is insufficient to determine the $J^P$ numbers of $P_c(4440)$ and $P_c(4457)$. 
Many studies have addressed this $J^P$ issue~\cite{PavonValderrama:2019nbk,Chen:2019asm,Yamaguchi:2019seo,Liu:2023wfo,Liu:2019zvb,Burns:2021jlu, Zhang:2023czx, Yang:2024nss, Xu:2025mhc}. 
We discuss the quantum number assignment in Sec.~\ref{sec:Pccspectrum}.

The  pentaquark states $P_{c\bar{c}}\left(4312\right)^+$, $P_{c\bar{c}}\left(4440\right)^+$, and $P_{c\bar{c}}\left(4457\right)^+$ have been investigated using lattice QCD by calculating the S-wave scattering of $\Sigma_c\bar{D}$ and $\Sigma_c\bar{D}^*$ via L\"uscher's method only in the $I\left(J^P\right)=\tfrac{1}{2}\left(1/2\right)^-$ channel at a pion mass of $294$~MeV and with a lattice spacing of $0.08$~fm. 
Two bound state poles were found with one corresponding to $P_{c\bar{c}}\left(4312\right)^+$ and the other to either $P_{c\bar{c}}\left(4440\right)^+$ or $P_{c\bar{c}}\left(4457\right)^+$ \cite{Xing:2022ijm}. 

Beyond spectroscopy, decay properties are also key to understanding the structure of the pentaquark states. 
Several approaches have been used to study the decay of charm pentaquark states $P_{c\bar{c}}$ into $J/\psi p$, $\eta_c p$, $\Lambda_c\bar{D}^{\left(*\right)}$ and $\Sigma_c\bar{D}^{\left(*\right)}$ thresholds \cite{Voloshin:2019aut,Sakai:2019qph, Xie:2022hhv, Yang:2024nss, Li:2025ejt, Lin:2019qiv, Liu:2024ugt, Du:2021fmf, Burns:2021jlu}.
Most recently, the LHCb collaboration performed a search for $P_{c\bar{c}}\left(4312\right)^+$,  $P_{c\bar{c}}\left(4440\right)^+$,  and  $P_{c\bar{c}}\left(4457\right)^+$ states in the prompt $\Lambda_c\bar{D}^{(*)}$, $\Lambda_c\bar{D}^{(*)}$, and $\Lambda_c\pi\bar{D}^{(*)}$ mass spectra. 
No significant signal was observed \cite{LHCb:2024pnt}. 
Thus far, the only observed decay channel for the  $P_{c\bar{c}}$ pentaquark states remain $J/\psi p$.

In this paper, we use the QCD effective field theory called \textit{Born--Oppenheimer EFT} (BOEFT) to address both the spectra and decays of  quarkonium pentaquark states $Q\bar{Q}qqq$, without making any a priori assumption on their quark configurations. 
The BOEFT is derived from QCD on the basis of symmetries and scale separations and results in coupled channel Schr\"odinger equations governing 
the dynamics of the states \cite{Berwein:2024ztx}. 
A crucial input to the Schr\"odinger equations are the static energies or potentials between the heavy quarks due to light quarks or gluons. 
These are nonperturbative functions of the heavy quark separation to be computed in lattice QCD.
Currently, there are no lattice QCD results for the pentaquark static energies. However, BOEFT constraints the behavior of the static energies based on the symmetries both at short and long distances. 
While BOEFT requires lattice input for the static energies computed from some generalized Wilson loops, the factorization inherent the framework ensures
that only a small set of universal, flavor-independent nonperturbative correlators are needed, greatly simplifying the problem. 

The \textit{Born--Oppenheimer picture} has been used for a long time to  study QCD bound states such as heavy quarkonium hybrids \cite{Griffiths:1983ah, Juge:1999ie,Brambilla:1999xf, Juge:2002br, Braaten:2013boa, Braaten:2014qka, Braaten:2014ita, Meyer:2015eta, Alasiri:2024nue}.  
It has been recast into an EFT framework known as BOEFT \cite{Berwein:2015vca, Brambilla:2017uyf, TarrusCastella:2019rit, Soto:2020xpm,  Berwein:2024ztx, Mohapatra:2025iar}.
For hybrids, the BOEFT has been used to study the spectrum, incorporating spin corrections in the hybrid multiplets through spin-dependent potentials~\cite{Berwein:2015vca, Oncala:2017hop, Brambilla:2018pyn, Brambilla:2019jfi, Soto:2023lbh, Schlosser:2025tca}, and computing semi-inclusive decays to low-lying quarkonium states~\cite{Oncala:2017hop, Brambilla:2022hhi, TarrusCastella:2021pld}. 
In addition, decays of hybrids into open-flavor threshold states have been investigated within this framework~\cite{Bruschini:2023tmm, TarrusCastella:2024zps, Braaten:2024stn}. 
The BOEFT formalism has been also applied to study spectra and decays of doubly heavy baryons~\cite{Brambilla:2005yk, Soto:2020pfa, Soto:2021cgk,  Bruschini:2024fyj} and recently the lowest-lying tetraquark multiplets, including their spin splittings~\cite{Brambilla:2024thx, Braaten:2024tbm}.

The paper is organized as follows. 
In  Sec.~\ref{sec:Born--Oppenheimer}, we give a brief description of the BOEFT, information on the quantum numbers, expressions of the pentaquark static energies relevant for lattice QCD computation, and the expected short and long distance behavior of the static energies.
In  Sec.~\ref{sec:Pccspectrum}, we first write the coupled Schr\"odinger equations that follow
from BOEFT for pentaquark states ignoring spin corrections and then we include spin corrections using first-order perturbation theory to identify the experimental states. 
We discuss the different scenarios for the quantum number $J^{P}$ assignments to the pentaquark states in the charm sector. 
In Sec.~\ref{sec:decays}, we examine the decays of the pentaquark states  into $J/\psi$, $\eta_c$, $\Lambda_c\bar{D}$ and $\Lambda_c\bar{D}^*$ based on the scenarios in Sec.~\ref{sec:Pccspectrum}. 
In Sec.~\ref{sec:Pbb}, we give the results for the bottom pentaquark states $P_{b\bar{b}}$ based on the preferred scenario fixed by the spectrum and decays of the charm pentaquark $P_{c\bar{c}}$ states. 
Finally, Sec.~\ref{sec:conclusion} contains discussion and conclusion.

\section{Born--Oppenheimer effective theory}
\label{sec:Born--Oppenheimer}

Heavy quarkonium systems and hidden-heavy systems are hadrons made of a heavy quark-antiquark pair ($c\bar{c}$ or $b\bar{b}$) bound to light degrees of freedom (LDF) such as gluons $g$ or light quarks $q$ or antiquarks $\bar{q}$. 
The large heavy quark mass scale, $m_Q$, introduces significant simplifications: the scale is perturbative ($m_Q \gg \Lambda_{\mathrm{QCD}}$, 
with $\Lambda_{\mathrm{QCD}}$ the nonperturbative hadronic scale), and the quarks move nonrelativistically, 
i.e. $v\ll 1$, where $v$ is the relative velocity of the heavy quark-antiquark pair in the bound system.

The simplest hidden-heavy systems are quarkonium states $(Q\bar{Q})$, which are color singlet heavy quark-antiquark bound states with no valence LDF in the bound state. 
In contrast, pentaquark states $Q\bar{Q}qqq$ are examples of hidden-heavy systems with three light quarks as LDF bound  to the $Q\bar{Q}$ pair to form a color-singlet state. Other examples include  hybrids ($Q\bar{Q}g$), where a valence gluon contributes to the dynamics and tetraquarks ($Q\bar{Q} q \bar{q}$), which contain light quark-antiquark pair as valence LDF. 
The relevant energy scales to describe   such systems made of  two nonrelativistic heavy quarks ($Q\bar{Q}$ or even $QQ$) are the mass scale $m_Q$, the   relative momentum  scale $m_Q v\sim 1/r$, where $r$ is the relative distance between the  heavy quarks, and 
the heavy-quark-antiquark binding energy scale $m_Q v^2$. 
Such scales are hierarchically ordered as $m_Q \gg m_Qv \gg m_Qv^2$,  where $v\ll 1$ is the relative velocity of the heavy (anti)quark in the bound system.
In addition, in QCD, there is the nonperturbative energy scale of the LDF, $\Lambda_{\mathrm{QCD}}$. 
Apart from  the heavy quark mass, the treatment of the other energy scales depends on their proximity to $\Lambda_{\mathrm{QCD}}$, as nonperturbative methods have to be used if the scales are close to it. 
The challenge of dealing with multiple intertwined energy scales in nonrelativistic bound states of QCD has been systematically addressed by substituting
QCD with simpler yet equivalent nonrelativistic effective field theories (NREFTs)~\cite{Caswell:1985ui, Bodwin:1994jh,Pineda:1997bj,Brambilla:1999xf,Brambilla:2000gk,Pineda:2000sz,Brambilla:2004jw}.

The suitable EFT  for describing systems with two heavy quarks after integrating out the heavy quark mass scale $m_Q$ is  nonrelativistic QCD (NRQCD)~\mbox{\cite{Caswell:1985ui,Bodwin:1994jh}}. 
NQRCD is formulated as a systematic expansion in $1/m_Q$. 
For exotic states such as pentaquarks, which are typically extended objects, we assume that the LDF responsible for their binding satisfy the hierarchy condition $m_Q v \gtrsim \Lambda_{\mathrm {QCD}}\gg m_Qv^2$. 
This hierarchy implies that the nonrelativistic motion of the $Q\bar{Q}$ pair evolves with a much larger time scale compared to the LDF time scale, $1/\Lambda_{\rm QCD}$.  
It also implies that mixing between states separated by energy gaps of order $\Lambda_{\mathrm {QCD}}$ is parametrically suppressed.\footnote{Note that states like quarkonium and pentaquarks cannot mix due to differences in quantum numbers such as isospin.}
This separation of scales naturally motivates the use of the Born--Oppenheimer (BO) approximation, well known from molecular physics~\cite{Born-Oppenheimer, Landau:1991wop}. 
In the context of QCD bound states, focusing on the lower energy dynamics at the scale $m_Qv^2$ by integrating out all the higher energy scales above $m_Qv^2$  leads to the Born--Oppenheimer effective field theory (BOEFT) \cite{Berwein:2015vca, Brambilla:2017uyf, TarrusCastella:2019rit, Soto:2020xpm,  Berwein:2024ztx, Mohapatra:2025iar}. 
In BOEFT, the LDF dynamics is encoded in the static potentials between the heavy quarks. 
In this work, we apply the formalism developed in \cite{Berwein:2024ztx, Brambilla:2022hhi, Braaten:2024stn} to the study of spectrum and decays of $Q\bar{Q}qqq$  pentaquark states. 
We consider both semi-inclusive decays to quarkonium states like $J/\psi$, $\eta_c$, and decays to baryon-antimeson $\Lambda_c\bar{D}$, and $\Lambda_c\bar{D}^*$ thresholds.

\subsection{BO quantum numbers}

In the \textit{static limit}, eigenstates and eigenvalues of the QCD Hamiltonian in the $Q\bar{Q}$ sector are referred to as \textit{static states} and \textit{static energies}, respectively. 
In this limit, the heavy quarks act as static color sources,  while the different LDF configurations identify different static energies that are labeled by  the \textit{BO quantum numbers} $\Lambda^{\sigma}_\eta$, representations of the cylindrical symmetry group $D_{\infty h}$.
The quantity $\Lambda$ is defined as $\Lambda \equiv |\lambda| = |\bm{K}\cdot\hat{\bm{r}}|$, where $\lambda$ is the eigenvalue of the projection of the LDF total angular momentum (spin) $\bm K$ along the heavy quark pair axis $\hat{\bm r}$. 
For integer values of $\Lambda$, we use the standard notation of  capital Greek letters: $\Sigma$, $\Pi$, $\Delta,...$ for $\Lambda=0,1,2,...$~. 
The index $\eta$ is the $CP$ eigenvalue, if $CP$ is a good quantum number, elsewhere it is the parity $P$, and is denoted by $g = + 1$ and $u = - 1$. 
The index $\sigma=\pm 1$ is the eigenvalue of the reflection operator  with respect to a plane passing through the $\hat{\bm{r}}$ axis. 
The index $\sigma$ is explicitly written only for $\Sigma$ states, which are not degenerate under reflection. 
The static energies are also characterized by the flavor quantum numbers of the LDFs, such as isospin $I$ and projection $m_I$, baryon number $b$, and so on. 

Assuming that the angular momentum operator $\bm{K}^2$ has eigenvalues $k(k+1)$, which restricts the projection to $\Lambda\leq k$, we introduce the shorthand notation
\begin{equation}
    \kappa\equiv\{k^{PC}, f\}\,,
    \label{label-n}
\end{equation}
where  $f$ denotes the LDF flavor indices.\footnote{For notational ease, we explicitly mention whenever we suppress the flavor index $f$.}
The quarkonium states are isoscalar states with $\kappa=\{0^{++}, I=0\}$, while pentaquark states have non-zero isospin and non-zero LDF spin $k$.
It is important to note that $k$ is a good quantum number only at short distances ($r \to 0$), where the symmetry group of the system becomes the spherical symmetry $O(3)$, 
but not at large distances, where the symmetry of the system is $D_{\infty h}$.

For $Q\bar{Q}qqq$ pentaquarks, the total spin ${\bm K}$ of the three light quarks can only assume half-integer values. 
As a result, the standard notation for the BO-quantum numbers $\Lambda_\eta^\sigma$ is not applicable: 
there are no established labels for $\Lambda$ in the half-integer case, the reflection quantum number $\sigma$ does not appear without a $\lambda=0$ state, 
and, because $CP$ transforms a light $qqq$ state into a $\bar{q}\bar{q}\bar{q}$ state, 
the $CP$ eigenstates are even and odd linear combinations of the two.\footnote{See Sec.~IV C in Ref.~\cite{Berwein:2024ztx} for more details.} 
For notation, we choose to label the LDF states by $k^P$ and represent the BO quantum numbers as $\left(\Lambda\right)_\eta$, where  $\eta$ refers only to parity $P$, 
and $\Lambda = |\bm{K}\cdot\hat{\bm{r}}|$. 
The ground state with given BO quantum number is labeled $\left(\Lambda\right)_\eta$, whereas excited states with the same quantum numbers are labeled 
$\left(\Lambda\right)^{\prime}_\eta$, $\left(\Lambda\right)^{\prime\prime}_\eta$, \ldots~.

\subsection{Pentaquark static energies: operator and behavior at short-distance}
\label{subsec:Pstatic}
While quarkonium $Q\bar{Q}$ states are bound states of a heavy quark-antiquark pair with no LDF, pentaquark $Q\bar{Q}qqq$ states are bound states with three light quarks as LDF. 
In the $r\to 0$ (short-distance) limit, where $r$ is the separation between the heavy quark and antiquark pair, the color quantum number combination of the $Q\bar{Q}$ pair results in both color singlet and color octet configurations.  
In the color octet case, the $\left(Q\bar{Q}\right)_8$ pair binds with the LDF that are in the adjoint representation of $SU(3)$ to form color singlet hadrons. 
We refer to those LDF states with quantum numbers $k^{P}$ in the adjoint representation as \textit{adjoint hadrons}.  
In the context of the current work, the color octet case describes pentaquarks $\left[\left(Q\bar{Q}\right)_8+\left(qqq\right)_8\right]$, while the color singlet case corresponds to quarkonia $\left(Q\bar{Q}\right)_1$ or quarkonia with light hadrons  $\left[\left(Q\bar{Q}\right)_1+\left(qqq\right)_1\right]$~\cite{Berwein:2024ztx, Mohapatra:2025iar}.\footnote{Lattice QCD calculations show that these states are not sufficiently bound to form a multiquark state~\cite{Alberti:2016dru, Prelovsek:2019ywc}.}

The static energy -- also known as the \textit{BO potential} -- for the pentaquark state with BO-quantum number $\left(\Lambda\right)_{\eta}$ can be computed from the large time behavior of the logarithm of an appropriate gauge-invariant correlator:
\begin{align}
E_{\left(\Lambda\right)_{\eta}}(r)\equiv E_{\kappa, |\lambda|}(r)=\lim_{T\to\infty}\frac{i}{T}\,\log\left[ \langle \mathrm{vac}| \mathcal{O}_{\kappa, \lambda}(T/2,\,\bm{r},\,\bm{R})\,\mathcal{O}^{\dagger}_{\kappa, \lambda}(-T/2,\,\bm{r},\,\bm{R})|\mathrm{vac}\rangle\right],
\label{eq:En}
\end{align}
where $\kappa$ is given by Eq.~\eqref{label-n} and $|\mathrm{vac}\rangle$ denotes the NRQCD vacuum. 
The gauge-invariant interpolating operator $\mathcal{O}_{\kappa, \lambda}$ is a function of  the relative coordinate $\bm{r}\equiv {\bm x}_1-{\bm x}_2$ and the center of mass coordinate $\bm{R}\equiv \left({\bm x}_1+{\bm x}_2\right)/2$ of the $Q\bar{Q}$ pair, ${\bm x}_1$ and ${\bm x}_2$ being the space locations of the quark and antiquark. 
The LDF isospin or flavour quantum numbers are not explicitly written. 
For a heavy quark-antiquark octet, $\mathcal{O}_{\kappa, \lambda}$ can be given in terms of NRQCD fields by
\begin{align}
\mathcal{O}_{\kappa, \lambda}\left(t, \bm{r}, \bm{R}\right)&=\nonumber\\
&\hspace{-1.7 cm}\chi^{\dagger}\left(t, \bm{R}+\bm{r}/2\right)\phi\left(t; \bm{R}+\bm{r}/2,\bm{R}\right)P^{\alpha \dag}_{\kappa, \lambda}H_{8,\,\kappa}^{\alpha,\,a}(t, \bm{R})\,T^a\,\phi\left(t; \bm{R},\bm{R}-\bm{r}/2\right)\psi\left(t, \bm{R}-\bm{r}/2\right),
\label{eq:intop}
\end{align}
where the field $\psi$ is the Pauli spinor that annihilates the heavy quark and $\chi$ is the Pauli spinor that creates the heavy antiquark; they satisfy canonical equal time anticommutation relations. 
The matrices $T^a\,\left(a=1,\dots,8\right)$ are the $SU(3)$ generators and $\phi\left(t;\bm{x}, \bm{y}\right)$ is the Wilson line 
\begin{align}
    \phi\left(t; \bm{x}, \bm{y}\right)=\mathcal{P}\exp\left[-ig\int_{\bm{y}}^{\bm{x}}\,d\bm{z}\cdot \bm{A}\left(t, \bm{z}\right)\right],
\end{align}
with $\mathcal{P}$ the path ordering operator. 
The operators $H_{8,\,\kappa}^{\alpha,\,a}(\bm{R})$  are related to adjoint baryons when the LDF are three light quarks. 
The projection vectors $P^{\alpha}_{\kappa, \lambda}$ with $\alpha$ the vector or spin index project onto eigenstates of $\bm{K}\cdot\hat{\bm{r}}$
with eigenvalue $\lambda$, thereby fixing the cylindrical symmetry group $D_{\infty h}$ quantum numbers. 
They are are given in terms of Wigner D-matrices and read \cite{Berwein:2024ztx}:
\begin{equation}
 P^{\alpha}_{\kappa\lambda}\left(\theta, \varphi\right)=\sqrt{\frac{8\pi^2}{2k+1}} D^{\lambda*}_{k\,\alpha}\left(0, \theta, \varphi\right),
 \label{eq:Pdef2}
\end{equation}
where $\alpha = k, \dots, 0, \dots, -k$, as we are using a spherical basis. Specific expressions of  different projection vectors  are in Appendix~F of Ref.~\cite{Berwein:2024ztx}.  
For quarkonium, the quantum numbers of the LDF are $\kappa=0^{++}$, which implies a trivial form of the projection vector: $P_{00}=\mathbb{1}$. 
Indeed, setting $P^\alpha_{\kappa, \lambda}\,H_{8,\,\kappa}^{\alpha,\,a}(t, \bm{R})\,T^a = \mathbbm{1}$ in Eq.~\eqref{eq:intop} identifies the color singlet $\left(Q\bar{Q}\right)_1$ state corresponding to quarkonium.

In the current work, we consider the lowest pentaquark states $Q\bar{Q}qqq$, where $q=\left(u, d\right)$, listed in Table~\ref{tab:exoticPenta}, all of which have isospin $I=1/2$. 
Without orbital excitations, the three light quarks  have positive parity and can have total spin either as a spin doublet [$k^P = (1/2)^+$] or a spin quartet [$k^P = (3/2)^+$]. 
Their isospin can similarly be either a  doublet ($I = 1/2$) or  quartet ($I = 3/2$). 
However, the light quarks being indistinguishable, the Pauli exclusion principle imposes constraints on the whole color-spin-isospin combinations, which must be totally antisymmetric under quark exchange. 
As a result, in the color-octet sector, all combinations of spin and isospin are allowed except for the case where both are quartets ($I = 3/2$, $k = 3/2$).\footnote{
There is no fully antisymmetric combination in color for three identical quarks in a color octet configuration. 
For details, see Appendix~H in \cite{Berwein:2024ztx}.}
For light quark spin $k^P = (1/2)^+$, we denote the BO potential $E_{\left(1/2\right)_g}\left(r\right)$ and for light quark spin $k^P = (3/2)^+$, 
we denote the two BO potentials corresponding to the two projections along the $Q\bar{Q}$ axis as $E_{\left(1/2\right)_g^{\prime}}\left(r\right)$ and  $E_{\left(3/2\right)_g}\left(r\right)$.

The light-quark interpolating operator $H_{8,\,\kappa}^{\alpha,\,a}(t, \bm{x})$ for $I=1/2$, $I_3=\pm 1/2$, and $k=1/2$ is given by \cite{Berwein:2024ztx}
\begin{align}
&H^{\alpha,\,a}_{8,I_3=\pm 1/2, (1/2)^{+}}(t,\bm{x})=\nonumber\\
&\hspace{1.5 cm}\Bigg[\left(\delta_{\alpha\beta_1}\sigma^2_{\beta_2\beta_3}+\delta_{\alpha\beta_2}\sigma^2_{\beta_1\beta_3}+\delta_{\alpha\beta_3}\sigma^2_{\beta_1\beta_2}\right)\left(\delta_{I_3f_1}\tau^2_{f_2f_3}+\delta_{I_3f_2}\tau^2_{f_1f_3}+\delta_{I_3f_3}\tau^2_{f_1f_2}\right)\left(T_2\right)^a_{l_1,l_2,l_3}\nonumber\\
&\hspace{1.5 cm}+\left(\delta_{\alpha\beta_1}\sigma^2_{\beta_2\beta_3}+\delta_{\alpha\beta_2}\sigma^2_{\beta_3\beta_1}+\delta_{\alpha\beta_3}\sigma^2_{\beta_2\beta_1}\right)\left(\delta_{I_3f_1}\tau^2_{f_2f_3}+\delta_{I_3f_2}\tau^2_{f_3f_1}+\delta_{I_3f_3}\tau^2_{f_2f_1}\right)\left(T_3\right)^a_{l_1,l_2,l_3}\nonumber\\
&\hspace{1.5 cm}+\left(\delta_{\alpha\beta_1}\sigma^2_{\beta_3\beta_2}+\delta_{\alpha\beta_2}\sigma^2_{\beta_3\beta_1}+\delta_{\alpha\beta_3}\sigma^2_{\beta_1\beta_2}\right)\left(\delta_{I_3f_1}\tau^2_{f_3f_2}+\delta_{I_3f_2}\tau^2_{f_3f_1}+\delta_{I_3f_3}\tau^2_{f_1f_2}\right)\left(T_1\right)^a_{l_1,l_2,l_3}\Bigg]\nonumber\\
&\hspace{2.5 cm}\left(P_+q_{l_1f_1}(t,\bm{x})\right)^{\beta_1} \left(P_+q_{l_2f_2}(t,\bm{x})\right)^{\beta_2} \left(P_+q_{l_3f_3}(t,\bm{x})\right)^{\beta_3}\,,
\label{eq:OpQQbarqqq}
\end{align}
and for $I=1/2$, $I_3=\pm 1/2$, and $k=3/2$ it is given by
\begin{align}
&H^{\alpha,\,a}_{8,I_3=\pm 1/2, (3/2)^{+}}(t,\bm{x})=\Bigg[\nonumber\\
&\left({\cal C}^{3/2\,\alpha}_{1\,m\,1/2\,\beta_1}\left({\bm e_m}\cdot{\bm \sigma}i\sigma^2\right)_{\beta_2\beta_3}+ {\cal C}^{3/2\,\alpha}_{1\,m\,1/2\,\beta_2}\left({\bm e_m}\cdot{\bm \sigma}i\sigma^2\right)_{\beta_1\beta_3}+{\cal C}^{3/2\,\alpha}_{1\,m\,1/2\,\beta_3}\left({\bm e_m}\cdot{\bm \sigma}i\sigma^2\right)_{\beta_1\beta_2}\right)\times\nonumber\\
&\left({\cal C}^{1/2\,I_3}_{1\,m\,1/2\,f_1}\left({\bm e_m}\cdot{\bm \tau}i\tau^2\right)_{f_2f_3}+ {\cal C}^{1/2\,I_3}_{1\,m\,1/2\,f_2}\left({\bm e_m}\cdot{\bm \tau}i\tau^2\right)_{f_1f_3}+{\cal C}^{1/2\,I_3}_{1\,m\,1/2\,f_3}\left({\bm e_m}\cdot{\bm \tau}i\tau^2\right)_{f_1f_2}\right)
\left(T_2\right)^a_{l_1,l_2,l_3}\nonumber\\
&\hspace{8.0 cm}+\nonumber\\
&\left({\cal C}^{3/2\,\alpha}_{1\,m\,1/2\,\beta_1}\left({\bm e_m}\cdot{\bm \sigma}i\sigma^2\right)_{\beta_2\beta_3}+ {\cal C}^{3/2\,\alpha}_{1\,m\,1/2\,\beta_2}\left({\bm e_m}\cdot{\bm \sigma}i\sigma^2\right)_{\beta_3\beta_1}+{\cal C}^{3/2\,\alpha}_{1\,m\,1/2\,\beta_3}\left({\bm e_m}\cdot{\bm \sigma}i\sigma^2\right)_{\beta_2\beta_1}\right)\times\nonumber\\
&\left({\cal C}^{1/2\,I_3}_{1\,m\,1/2\,f_1}\left({\bm e_m}\cdot{\bm \tau}i\tau^2\right)_{f_2f_3}+ {\cal C}^{1/2\,I_3}_{1\,m\,1/2\,f_2}\left({\bm e_m}\cdot{\bm \tau}i\tau^2\right)_{f_3f_1}+{\cal C}^{1/2\,I_3}_{1\,m\,1/2\,f_3}\left({\bm e_m}\cdot{\bm \tau}i\tau^2\right)_{f_2f_1}\right)
\left(T_3\right)^a_{l_1,l_2,l_3}\nonumber\\
&\hspace{8.0 cm}+\nonumber\\
&\left({\cal C}^{3/2\,\alpha}_{1\,m\,1/2\,\beta_1}\left({\bm e_m}\cdot{\bm \sigma}i\sigma^2\right)_{\beta_3\beta_2}+ {\cal C}^{3/2\,\alpha}_{1\,m\,1/2\,\beta_2}\left({\bm e_m}\cdot{\bm \sigma}i\sigma^2\right)_{\beta_3\beta_1}+{\cal C}^{3/2\,\alpha}_{1\,m\,1/2\,\beta_3}\left({\bm e_m}\cdot{\bm \sigma}i\sigma^2\right)_{\beta_1\beta_2}\right)\times\nonumber\\
&\left({\cal C}^{1/2\,I_3}_{1\,m\,1/2\,f_1}\left({\bm e_m}\cdot{\bm \tau}i\tau^2\right)_{f_3f_2}+ {\cal C}^{1/2\,I_3}_{1\,m\,1/2\,f_2}\left({\bm e_m}\cdot{\bm \tau}i\tau^2\right)_{f_3f_1}+{\cal C}^{1/2\,I_3}_{1\,m\,1/2\,f_3}\left({\bm e_m}\cdot{\bm \tau}i\tau^2\right)_{f_1f_2}\right)
\left(T_1\right)^a_{l_1,l_2,l_3}\nonumber\\
&\hspace{4.5 cm}\Bigg]\left(P_+q_{l_1f_1}(t,\bm{x})\right)^{\beta_1} \left(P_+q_{l_2f_2}(t,\bm{x})\right)^{\beta_2} \left(P_+q_{l_3f_3}(t,\bm{x})\right)^{\beta_3}\,,
\label{eq:OpQQbarqqq-1}
\end{align}
where repeated indices are summed over, $\alpha, \beta_i\,\left(i=1, 2, 3\right)$ are the spin or vector indices, $f_i\,\left(i=1, 2, 3\right)$ are the isospin or flavor index, $l_i\,\left(i=1, 2, 3\right)$ are the color index, ${\cal C}^{3/2\,\alpha}_{1\,m\,1/2\,\beta_i}$ and ${\cal C}^{1/2\, I_3}_{1\,m\,1/2\,f_i}$ are Clebsch--Gordan coefficients and the color matrices $\left(T_i\right)^a\,\left(i=1, 2, 3\right)$ are defined as
\begin{align}
&\left(T_1\right)^a_{ijk}=\epsilon_{ljk}T_{li}^a,\qquad 
\left(T_2\right)^a_{ijk}=\epsilon_{ilk}T_{lj}^a,\qquad \left(T_3\right)^a_{ijk}=\epsilon_{ijl}T_{lk}^a.
\label{eq:T1T2T3}
\end{align}
The projector $P_{+}=\left(1+\gamma^0\right)/2$ in Eqs.~\eqref{eq:OpQQbarqqq} and \eqref{eq:OpQQbarqqq-1} is required due to positive parity \cite{Soto:2020xpm, Sadl:2021bme}, 
it selects a two component Pauli spinor, 
$\sigma^2$ and $\tau^2$ are the antisymmetric spin and isospin Pauli matrices, $\sigma^i$ and $\tau^i$ are the spin and isospin Pauli matrices and $\bm{e}_{m}$ are the spherical basis vectors:
\begin{equation}
    {\bm e}_0=\left(0, 0, 1\right),\qquad {\bm e}_{-1}=-\left(1, i, 0\right)/\sqrt{2},\qquad {\bm e}_{+1}=\left(1, -i, 0\right)/\sqrt{2}.
    \label{eq:em}
\end{equation}
By replacing the Clebsch–-Gordan coefficient ${\cal C}^{3/2,\alpha}_{1,m,1/2,\beta_i}$ in Eq.~\eqref{eq:OpQQbarqqq-1} with ${\cal C}^{1/2,\alpha}_{1,m,1/2,\beta_i}$ yields another light-quark interpolating operator, $H^{\alpha,\,a}_{8,I_3=\pm 1/2, (3/2)^{+}}(t,\bm{x})$, for $I=1/2$, $I_3=\pm 1/2$, and $k=1/2$. 
As discussed in Secs.~\ref{subsec:multiplets} and \ref{subsubsec:mixing}, and summarized in Tables~\ref{tab:QQbarpenta} and \ref{tab:qqq}, there are in fact two distinct $k^P = (1/2)^+$ configurations. 
The interpolating operator in Eq.~\eqref{eq:OpQQbarqqq} corresponds to the light quark pair-quark combination $k_{qq}^P\otimes k_q^P = 0^+ \otimes \left(1/2\right)^+$, while the operator obtained via the above substitution corresponds to $k_{qq}^P \otimes k_q^P = 1^+ \otimes (1/2)^+$.

Following Ref.~\cite{Berwein:2024ztx}, the short-distance behavior of the pentaquark BO potential $E^{(0)}_{\left(\Lambda\right)_{\eta}}\left(r\right)$ can be written as
\begin{align}
	E_{\left(\Lambda\right)_{\eta}}\left(r\right)= V_o(r) + \Lambda_{\kappa} + {\cal O}(r^2).
	\label{eq:QQbarpot-short}
\end{align}
where $V_o(r) = \alpha_s/\left(6r\right)$ is the color octet potential and $\Lambda_{\kappa}$ is the adjoint baryon mass with $\kappa=\{\left(1/2\right)^+,\, I=1/2\}$ and $\{\left(3/2\right)^+,\, I=1/2\}$. 
Equation~\eqref{eq:QQbarpot-short} incorporates the information that, at leading order in the multipole expansion, several BO potentials become degenerate as they depend solely on the  adjoint baryon quantum number $\kappa\equiv k^{P}$ (see second column of Table~\ref{tab:QQbarpenta}). 
The degeneracy is broken by the ${\cal O}\left(r^2\right)$ terms that arise from the multipole expansion \cite{Brambilla:1999xf}.

\begin{table}[ht]
	\begin{tabular}{||c|c|c||c|c||}
		\hline\hline
	\multirow{2}{*}{\hspace{2pt}$\begin{array}{c} Q\bar{Q}\\\text{color state}\end{array}$\hspace{2pt}} & \multirow{2}{*}{\hspace{2pt}$\begin{array}{c}\text{Light spin}\\k^{P}\end{array}$\hspace{2pt}} & \multirow{2}{*}{\hspace{2pt} $\begin{array}{c}\text{BO quantum \#}\\D_{\infty h}\end{array}$\hspace{2pt}} &\multirow{2}{*}{\hspace{2pt} $l$\hspace{2pt}}& \multirow{2}{*}{\hspace{2pt} $\begin{array}{c}J^{P}\\\{s=0, s=1\}\end{array}$\hspace{2pt}}\\
		& &  &  &  \\
		\hline\hline
		\multirow{2}{*} {\hspace{2pt}$\begin{array}{c} \text{Octet}\\\mathbf{8}\end{array}$\hspace{2pt}} &\multirow{1}{*}{$(1/2)^{+}$}& \multirow{1}{*}{$\left(1/2\right)_g$}\hspace{2pt}  & \hspace{2pt}$1/2$\hspace{2pt} & \hspace{2pt}$\{1/2^{-}, \left(1/2, 3/2\right)^-\}$\hspace{2pt}
		\\
		\cline{2-5}
		&\multirow{1}{*}{$(3/2)^{+}$}  & \multirow{1}{*}\{$\left(1/2\right)_g^{\prime}$, $\left(3/2\right)_g$\}\hspace{2pt} & \hspace{2pt}$3/2$\hspace{2pt} & \hspace{2pt}$\{3/2^{-}, (1/2, 3/2, 5/2)^{-}\}$\hspace{2pt}\\
		\hline\hline
	\end{tabular}
	\caption{$J^{P}$ multiplets for the lowest  pentaquark states $Q\bar{Q}qqq$ where $q=\left(u, d\right)$ and ${P}$ denotes the parity of the state. 
    We represent the BO quantum numbers ($D_{\infty h}$ representation) in the third column as $\left(\Lambda\right)_\eta$, where  $\eta = g$ denotes positive parity and $\Lambda=|{\bm K}\cdot{\bm r}|$. 
    There are two distinct multiplets corresponding to two distinct adjoint baryons with light-spin-parity $\left(1/2\right)^+$: 
    one multiplet with $k^P=\left(1/2\right)^+$ corresponds to states near the $\Lambda_c/\Lambda_b-\bar{D}/\bar{B}$ thresholds, while the other multiplet with $k^P=\left(1/2\right)^+$ together with the multiplet with $k^P=\left(3/2\right)^+$ corresponds to states near the $\Sigma_c/\Sigma_b-\bar{D}/\bar{B}$ thresholds. 
    In the table, we only show the $J^{P}$ multiplets associated with the lowest pentaquark states near the $\Sigma_c/\Sigma_b-\bar{D}/\bar{B}$ thresholds, 
    since no experimental states have been seen observed near the $\Lambda_c/\Lambda_b-\bar{D}/\bar{B}$ thresholds. }
	\label{tab:QQbarpenta}
\end{table}

\subsection{Pentaquark multiplets}
\label{subsec:multiplets} 
The lowest quarkonium pentaquarks are bound states in the BO potentials (static energies) with BO-quantum number $(1/2)_g$ corresponding to the adjoint baryon  $k^{P} = (1/2)^{+}$, 
or BO potentials with BO-quantum numbers $(1/2)_g^{\prime}$ and $(3/2)_g$ corresponding to the adjoint baryon  $k^{P} = (3/2)^{+}$ (see Table~\ref{tab:QQbarpenta}). 
The ordering of the adjoint baryon masses for $(1/2)^{+}$ and $(3/2)^{+}$ is not known from lattice QCD. 
We consider both adjoint baryons for the ground state. For $k^P=(3/2)^{+}$, we need to consider both static energies $(1/2)_g^{\prime}$ and $(3/2)_g$ because they are degenerate in the short distance limit.\footnote{
An analogous degeneracy at short distances occurs in the case of the hybrid static energies $\Sigma_u^-$ and $\Pi_u$ \cite{Foster:1998wu,Juge:2002br,Brambilla:1999xf} and the tetraquark static energies $\Sigma_g^{+\prime}$ and $\Pi_g$ \cite{Berwein:2024ztx, Brambilla:2024thx}.}
The states associated with the static energies $(1/2)_g^{\prime}$ and $(3/2)_g$ mix and their dynamics is governed by a set of coupled Schr\"odinger equations~\cite{Berwein:2024ztx}.
Higher static energies are assumed to be separated from the static energies $(1/2)_g$, $(1/2)_{g^{\prime}}$, and $(3/2)_g$ by a gap of order $\Lambda_{\mathrm{QCD}}$ (at least true for the hybrid static energies), and their modes are integrated out when integrating out LDF of energy or momentum of order $\Lambda_{\mathrm {QCD}}$.

We define the total angular momentum of the state as $\bm{J}=\bm{L}+\bm{S}$, where the angular momentum $\bm{L}=\bm{L}_{Q\bar{Q}}+\bm{K}$ is the 
sum of the orbital angular momentum ${\bm L}_{Q\bar{Q}}$ of the $Q\bar{Q}$  pair and the spin ${\bm K}$ of the LDF, and 
${\bm S}={\bm S_1}+{\bm S_2}$ is the total spin of the $Q\bar{Q}$ pair. 
The associated quantum numbers are as follows: $J(J+1)$ and $m_J$ are the eigenvalue of $\bm{J}^2$ and $J_3$ respectively, $l(l+1)$ is the eigenvalue of ${\bm L}^2$, and $s(s+1)$ is the eigenvalue of ${\bm S}^2$. 
In addition, we denote the quantum number of the angular momentum ${\bm L}_{Q\bar{Q}}$ by $L_Q$. 
In Table.~\ref{tab:QQbarpenta}, we show the results for the $J^{P}$ multiplets of the lowest $Q\bar{Q}qqq$ pentaquark states, where ${P}$ denotes the parity of the state. 
Notably, there are two distinct multiplets corresponding to two different adjoint baryon configurations with light spin-parity $k^P=\left(1/2\right)^+$: one multiplet is associated with states near the $\Lambda_c/\Lambda_b-\bar{D}/\bar{B}$ thresholds, while the other—along with the  $k^P=\left(3/2\right)^+$ multiplet—is associated with states near the $\Sigma_c/\Sigma_b-\bar{D}/\bar{B}$ thresholds. 
Hence, in total, there are $10$ possible $J^{P}$ states. 
However, three of these states are excluded as no experimental states have been seen observed near the $\Lambda_c/\Lambda_b-\bar{D}/\bar{B}$ thresholds. 
As we will see in Sec.~\ref{BO-parametrization}, this has implications on the behavior of the BO potentials.

As seen in Table~\ref {tab:QQbarpenta}, all the ground state pentaquark states have negative parity given by ${P}=\left(-1\right)^{l+k}$ \cite{Berwein:2024ztx}. 
The $Q\bar{Q}$ pair has negative intrinsic parity, while the three light quarks have positive intrinsic parity.  
For $k^P=\left(1/2\right)^+$ with $l=1/2$, the negative parity ${P}$ implies $L_{Q}=0$ and for $k^P=\left(3/2\right)^+$ with $l=3/2$, the negative parity implies $L_{Q}=0, 2$.

\subsection{Behavior of pentaquark BO potentials and  parametrization}
\label{BO-parametrization}

\subsubsection{Mixing with baryon-antimeson threshold}
\label{subsubsec:mixing}

The BOEFT constrains the behavior of the BO (static) potentials  at short and large distances based on the BO quantum number \cite{Berwein:2024ztx}.  
At short distances $\left(r\rightarrow 0\right)$, the form of the potential is fixed by the pNRQCD multipole expansion: for color octet $\left(Q\bar{Q}\right)_8$, the potential is the sum of the repulsive color octet potential, adjoint hadron mass 
and higher-order corrections  of ${\cal O}\left(\Lambda_{\rm QCD}^3\,r^2\right)$ \cite{Brambilla:1999xf, Berwein:2024ztx}. 
At large distances $\left(r\rightarrow \infty\right)$, the form  depends on the LDF. 
In the quenched approximation, where the LDF consist solely of gluons (quarkonium and hybrid configurations), lattice QCD calculations show that the static potentials grow linearly with the interquark distance $r$ (see recent lattice results in \cite{Brambilla:2022het, Schlosser:2021wnr, Sharifian:2023idc}). 
In contrast, in the unquenched case, the LDF include light quarks, and the relevant long-distance behavior is governed by the heavy-light hadron pair thresholds, such as the heavy baryon–heavy antimeson threshold (denoted by $\rm{B\overline{M}}$ for notational ease) in the $Q\bar{Q}qqq$ pentaquark system.

\begin{table}[h!]
\begin{center}
\begin{tabular}{|c|c||c|}  \hline\hline
\multirow{2}{*}{\hspace{2pt} $\begin{array}{c} k_{qq}^{P}\otimes k_q^{P}\end{array} $ \hspace{2pt}} & \multirow{2}{*}{\hspace{2pt} $\begin{array}{c}k^{P}\end{array} $ \hspace{2pt}}&  \multirow{2}{*}{\hspace{2pt} $\begin{array}{c} \text{BO quantum \#}\\D_{\infty h}\end{array} $ \hspace{2pt}}\\& & \\
\hline\hline
$0^+\otimes(1/2)^+ $         & $\left(1/2\right)^{+}$ & $\left(1/2\right)_g$ \\\hline
$1^+\otimes(1/2)^+$         & $\left(1/2\right)^{+}$ & $\left(1/2\right)_{g}$\\                                 
                                & $\left(3/2\right)^{+}$ & $\{\left(1/2\right)_g^\prime, \left(3/2\right)_g\}$
                                \\\hline\hline
\end{tabular}
\caption{The total LDF spin-parity $k^{P}$ quantum numbers of light quark pair-quark ($qq\otimes q$) combinations for the lowest light quark states forming  $\rm{B\overline{M}}$ pair thresholds. 
The light quark states have quantum numbers $k_{qq}^P$ and $k_{q}^P$.  The  BO quantum numbers corresponding to $k^P$ are listed in the third column. 
The prime in the third row indicates the excited state of the same BO quantum number. 
The first row, which accounts for the spin singlet $qq$ pair, is relevant for the $\Lambda_c\bar{D}$ or $\Lambda_b\bar{B}$ thresholds with isospin $I=1/2$, 
while the second and third rows, which account for the spin triplet $qq$ pair, are relevant for the $\Sigma_c\bar{D}$ or $\Sigma_b\bar{B}$ thresholds, and include both isospin $I=1/2$ and $I=3/2$.}
\label{tab:qqq}
\end{center}
\end{table}

The LDF quantum numbers of the $\rm{B\overline{M}}$ thresholds are inferred from the light quark spin, $k^P_q$, in the heavy antimeson and the light quark pair spin, $k^P_{qq}$, in the heavy baryon: $k^P_q=(1/2)^+$ for the ground state heavy antimeson  and $k^P_{qq}=0^+$ and $1^+$ for the ground state heavy baryon, where we assume that the $k^P=0^+$ state is lower in energy.\footnote{
This is supported by the observation that $\Lambda$-baryons have lower mass than $\Sigma$-baryons.}
The total LDF spin-parity $k^P$ and the corresponding BO quantum numbers in the $\Lambda^{\sigma}_{\eta}$ representation for the lowest $\rm{B\overline{M}}$ thresholds are listed in Table~\ref{tab:qqq}. 
The BO quantum number conservation between adjoint baryons (Table~\ref{tab:QQbarpenta}) and $\rm{B\overline{M}}$ thresholds (Table~\ref{tab:qqq}) implies that the $Q\bar{Q}$ pentaquark BO static potentials with repulsive color octet behavior at $r\rightarrow 0$ decrease as $r$ increases and eventually evolve smoothly into $\rm{B\overline{M}}$ thresholds at $r\rightarrow \infty$ \cite{Berwein:2024ztx, Braaten:2024tbm}. 
This constraint severely restricts the potentials capable of supporting bound states. 
These are only the pentaquark potentials that cross the $\rm{B\overline{M}}$ threshold and then approaches it from below. 
If the pentaquark potential decreases monotonically as $r$ increases and connects to the $\rm{B\overline{M}}$ threshold from above, it cannot support bound states \cite{Berwein:2024ztx, Braaten:2024tbm}. 

\begin{figure}[h!]
\begin{center}
\includegraphics[width=0.55\textwidth,angle=0,clip]{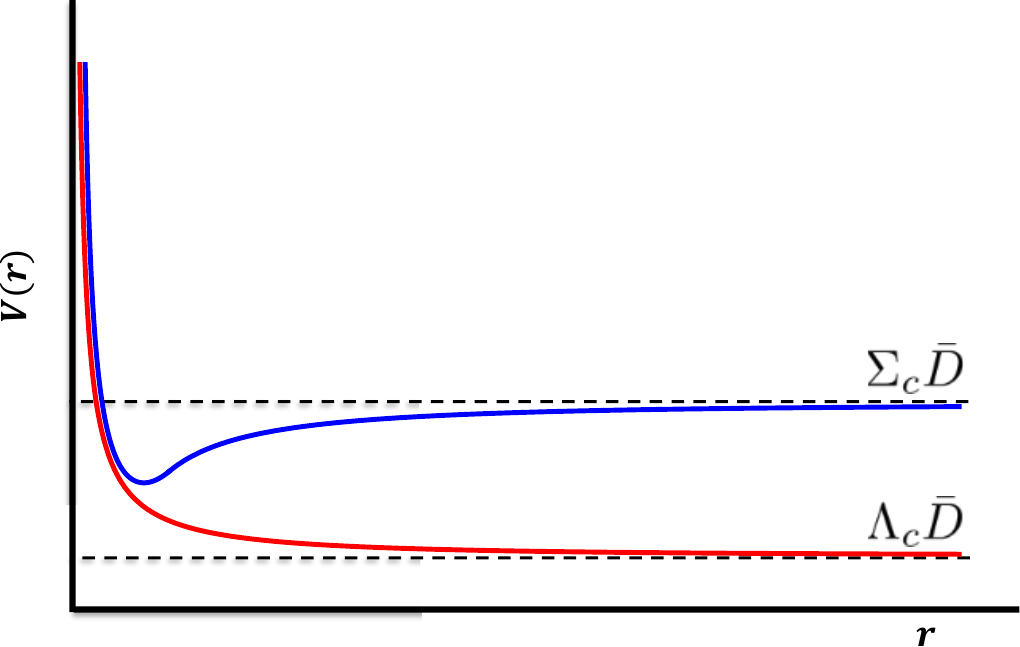}
\caption{Illustrative figure showing BO potentials that decrease monotonically with $r$ and connect to the baryon-antimeson thresholds at large distances. 
The blue curve represents a BO potential that approaches the $\Sigma_c\bar{D}$ threshold from below, potentially supporting a bound state. 
In contrast, the red curve corresponds to a BO potential that approaches the  $\Lambda_c\bar{D}$ threshold from above, in which case, making it unlikely to support any bound state. 
Furthermore, though not shown here, a BO potential may dip slightly below a threshold, such as the $\Lambda_c\bar{D}$ one, cross it again, and then approach it from above, in which case it could support a resonant state \cite{Braaten:2024tbm}.}
\label{fig:illustration}
\end{center}
\end{figure}

The LHCb experiment has observed $4$ half-integer isospin  $\left(I=1/2\right)$ pentaquark states: $P_{c\bar{c}}\left(4312\right)^+$, $P_{c\bar{c}}(4380)^+$, $P_{c\bar{c}}(4440)^+$, and $P_{c\bar{c}}(4457)^+$, all with quark content $c\bar{c}qqq$, where $q\equiv\left(u, d\right)$ \cite{LHCb:2015yax, LHCb:2019kea}. 
The $J^P$ quantum numbers are unknown. 
All of these states are below  the spin-isospin averaged $\Sigma_c\bar{D}$ threshold (which includes $\Sigma_c\bar{D}$, $\Sigma_c^*\bar{D}$, $\Sigma_c\bar{D}^*$, and $\Sigma_c^*\bar{D}^*$) 
and are above the spin-isospin averaged $\Lambda_c\bar{D}$ threshold (which includes $\Lambda_c\bar{D}$ and $\Lambda_c\bar{D}^*$). 
Notably, no states have been observed near the spin-isospin averaged $\Lambda_c\bar{D}$ or $\Lambda_b\bar{B}$ thresholds. 
This suggests that the static potential with BO quantum number $\left(1/2\right)_g$ corresponding to the adjoint baryon $\left(1/2\right)^+$ for $k^P_{qq}=0^+$ (first row in Table~\ref{tab:qqq}) decreases monotonically with $r$ and joins the $\Lambda_c/\Lambda_b$–$\bar{D}/\bar{B}$ thresholds from above--thus failing to support bound states, as sketched  in Fig.~\ref{fig:illustration}. 
In contrast, for the BO potentials with quantum numbers $(1/2)_g$ and ${(1/2)_g', (3/2)_g}$ corresponding to the two adjoint baryons $(1/2)^+$ and $(3/2)^+$ with $k^P_{qq} = 1^+$ (second and third rows of Table~\ref{tab:qqq}) and asymptotically connecting to the $\Sigma_c/\Sigma_b$–$\bar{D}/\bar{B}$ thresholds, two scenarios appear possible. 
(1) Only the potentials with BO quantum numbers $\{\left(1/2\right)_g^\prime, \left(3/2\right)_g\}$ cross the $\Sigma_c/\Sigma_b$–$\bar{D}/\bar{B}$ thresholds and then approach them from below, supporting bound states (see Fig.~\ref{fig:illustration}), while the potential with BO quantum number $\left(1/2\right)_g$ joins the $\Sigma_c/\Sigma_b$–$\bar{D}/\bar{B}$ thresholds from above, thus not supporting any bound state. 
This implies that the lowest pentaquark states are only four, those associated with the multiplet of the adjoint baryon $\left(3/2\right)^+$ (see Table~\ref{tab:QQbarpenta}). 
This number matches the number of observed pentaquark states reported in Table \ref{tab:exoticPenta}.
(2) All the static potentials with BO quantum numbers $\left(1/2\right)_g$ and $\{\left(1/2\right)_g^\prime, \left(3/2\right)_g\}$ cross the $\Sigma_c/\Sigma_b$–$\bar{D}/\bar{B}$ thresholds and then approach them from below, supporting bound states near the potential minima (see Fig.~\ref{fig:illustration}). 
This implies that the lowest pentaquark states are the seven pentaquark states associated with the multiplet of the adjoint baryons  $\left(1/2\right)^+$ and 
$\left(3/2\right)^+$ (see Table~\ref{tab:QQbarpenta}). 
Out of these seven states, four can be associated with the observed pentaquark states reported in Table~\ref{tab:exoticPenta}, the remaining three must be understood as yet undetected.
The first scenario has been analyzed in Ref.~\cite{Alasiri:2025roh}. 
In the current work, we analyze the second one. 
We only consider $I=1/2$ adjoint baryons and exclusively focus on the static potentials that asymptotically connect to the $\Sigma_c/\Sigma_b$–$\bar{D}/\bar{B}$ thresholds.

\subsubsection{Quark masses and parameterization}
\label{subsubsec:parameterization}
The charm and bottom quark masses that we use in the current work are the renormalon-subtracted (RS) charm and bottom masses~\cite{Pineda:2001zq,Bali:2003jq}
defined at the renormalon subtraction scale $\nu_f=1\,\mathrm{GeV}$ and computed in \cite{Berwein:2015vca}: $m_c^{\mathrm{RS}}= 1.477\,\mathrm{GeV}$ and $m_b^{\mathrm{RS}}=4.863\,\mathrm{GeV}$. 
The physical spin-isospin averaged  $\rm{B\overline{M}}$ thresholds are $E_{\Sigma_c\bar{D}}=4.470$~GeV, and $E_{\Sigma_b\bar{B}}=11.140$~GeV~\cite{ParticleDataGroup:2024cfk}. 
We choose the zero of energy to be the spin-isospin-averaged $\Sigma_c\bar{D}$ threshold for charm and $\Sigma_b\bar{B}$ threshold for bottom. For completeness, the spin-isospin averaged $\Lambda_c\bar{D}$ and $\Lambda_b\bar{B}$ thresholds are $E_{\Lambda_c\bar{D}}=4.260$~GeV and 
$E_{\Lambda_b\bar{B}}=10.933$~GeV \cite{ParticleDataGroup:2024cfk}.

There are no lattice QCD computations of the pentaquark static potentials. 
Therefore, we model the potentials based  on the  short-distance and long-distance behavior dictated by BOEFT \cite{Berwein:2024ztx}:
\begin{align}
E_{\left(\Lambda\right)_\eta}(r) =
   \begin{cases}
V^{\mathrm{RS}}_o(r, \nu_f) +  \Lambda_{\kappa} + A_{\Lambda_\eta}\, r^2 \, & r < R_{\Lambda_\eta}
\\
F_{\Lambda_\eta}\,e^{-r/d}/r \, &r > R_{\Lambda_\eta}
\end{cases}\,,
\label{eq:QQbar_V}
\end{align}
where $\Lambda_\eta\in\left[\left(1/2\right)_g, \{\left(1/2\right)_g^{+\prime}, \left(3/2\right)_g\}\right]$. 
For the short-distance part $(r < R_{\Lambda_\eta})$,  we use the RS octet potential $V^{\mathrm{RS}}_o(r, \nu_f)$ up to order $\alpha^3_s$ in perturbation theory\footnote{
An expression of the RS octet potential can be found in Appendix B of Ref.~\cite{Berwein:2015vca}.} 
and $\nu_f=1$~GeV is the renormalon subtraction scale. 
The constant $\Lambda_{\kappa}$ is the adjoint baryon mass with $\kappa=\{\left(1/2\right)^+,\, I=1/2\}$ and $\{\left(3/2\right)^+,\, I=1/2\}$. 
Due to the lack of lattice data, we choose the parameters $A_{\Lambda_\eta}$ in Eq.~\eqref{eq:QQbar_V} to be the same as the parameters for the lowest hybrid potentials in Ref.~\cite{Alasiri:2024nue}: 
\begin{equation}
A_{\left(1/2\right)_g}=0.042\,\mathrm{GeV}^3,\qquad 
A_{\left(1/2\right)_g^\prime}= 0.0065 \,\mathrm{GeV}^3,\qquad
A_{\left(3/2\right)_g}= 0.0726\,\mathrm{GeV}^3.
\label{eq:ALambda}
\end{equation}
For the long-distance part $(r > R_{\Lambda_\eta})$ in Eq.~\eqref{eq:QQbar_V}, we use the one-pion exchange potential motivated from molecular models \cite{Du:2019pij, Liu:2019zvb, Peng:2020xrf, Du:2021fmf, Yalikun:2021bfm,Xu:2025mhc}. 
The parameters $F_{\Lambda_\eta}$ and $R_{\Lambda_\eta}$ are determined by imposing continuity up to the first derivatives. 
We treat the adjoint baryon masses $\Lambda_{\left(1/2\right)^+}$ and $\Lambda_{\left(3/2\right)^+}$ as adjustable parameters to reproduce the $P_{c\bar{c}}\left(4312\right)^+$, $P_{c\bar{c}}(4380)^+$, $P_{c\bar{c}}(4440)^+$, and $P_{c\bar{c}}(4457)^+$ pentaquark masses within the experimental uncertainties.

\section{Pentaquark spectrum and identification with experimental states}
\label{sec:Pccspectrum}

\subsection{Schr\"odinger equations without spin splittings}
\label{subsec:Scheqs}
Pentaquarks $(Q\bar{Q}qqq)$ are exotic hadrons that are color-singlet bound states of a color octet $Q\bar{Q}$ pair coupled to three light quarks $qqq$. 
We focus here on the lowest-lying pentaquark states that can be built from the $\left(1/2\right)_g$ and $\{\left(1/2\right)_g^\prime, \left(3/2\right)_g\}$ static energies corresponding to adjoint baryons  with quantum numbers $\kappa=\{\left(1/2\right)^+,\, I=1/2\}$ and $\{\left(3/2\right)^+,\, I=1/2\}$ (see multiplets in Table~\ref{tab:QQbarpenta}).
For $k^P=\left(1/2\right)^{+}$, two values of $\lambda$ ($\pm 1/2$) are possible and for $k^P=\left(3/2\right)^{+}$, four values of $\lambda$ ($\pm 1/2$ and  $\pm 3/2$) are possible.
The pentaquark state $|P_N\rangle$ with quantum numbers $ N \equiv \{n, J, m_J, l, s\}$, where $n$ is the principal quantum number, can be written in the rest frame as~\cite{Berwein:2024ztx}
\begin{equation}
|P_N\rangle = \sum_\lambda\int d^3r\ket{\bm{r}}\ket{
\kappa,
\lambda}\Psi^{(N)}_{\kappa\lambda}(\bm{r}).
\label{eq:Psiwf2}
\end{equation}
The state $\ket{\bm{r}}$ stands for the static heavy quark-antiquark pair at the center of mass $\left({\bm R}=0\right)$ with relative separation $\bm{r}$, $\ket{\kappa,\lambda}$ for the LDF identified by the quantum numbers $\kappa$  and $\lambda$, and the integration over the coordinates ${\bm r}$ implies that we are beyond the static limit.
In the short-distance limit $r\to0$, the static states are approximate eigenstates of $\bm{K}^2$, whose eigenvalues are labeled by $k$.

The wavefunctions $\Psi^{(N)}_{\kappa\lambda}$ in Eq.~\eqref{eq:Psiwf2} are eigenfunctions of $\bm{K}\cdot\hat{\bm{r}}$ but not of parity.
Under parity, the index $\lambda$ goes into $-\lambda$, so, the parity eigenfunctions are constructed as a linear combinations of $\Psi^{(N)}_{\kappa\lambda}(\bm{r})$ and $\Psi^{(N)}_{\kappa-\lambda}(\bm{r})$.\footnote{
The projection vectors $P^{\alpha}_{\kappa\lambda}$ in Eq.~\eqref{eq:Pdef2} project on states with definite  $\bm{K}\cdot\hat{\bm{r}}$.
States labeled by the index $\lambda=0$ are eigenstates of $\bm{K}\cdot\hat{\bm{r}}$ and parity as well~\cite{Berwein:2015vca, Berwein:2024ztx}.} 
Following Ref.~\cite{Berwein:2024ztx}, the parity eigenfunctions for the pentaquark mulitiplet in Table~\ref{tab:QQbarpenta} corresponding to $k^P=\left(1/2\right)^+$ can be written as 
\begin{align}
\Psi_{\left(1/2\right)^+, {P}=-}^{(N)}\left({\bm r}\right) = \frac{1}{\sqrt{2}}
  \sum_{m_l,\,m_s}\,{\cal C}_{J m_J l s}^{m_l m_s} \left({\bm P}_{\frac{1}{2}\frac{1}{2}}v_{l,\,m_l}^{1/2}\left(\theta,\phi\right)+{\bm P}_{\frac{1}{2}-\frac{1}{2}}v_{l,\,m_l}^{-1/2}\left(\theta,\phi\right)\right)\,\psi^{(N)}_{\left(1/2\right)^+}(r)\,\chi_{s m_s}.
\label{eq:Pwf-1/2}
\end{align}
Similarly, the parity eigenfunctions for the pentaquark multiplet in Table~\ref{tab:QQbarpenta} corresponding to $k^P=\left(3/2\right)^+$ are
\begin{align}
 & \Psi_{\left(3/2\right)^+, {P}=-}^{(N)}\left({\bm r}\right) = \frac{1}{\sqrt{2}}
  \sum_{m_l,\,m_s}\,{\cal C}_{J m_J l s}^{m_l m_s} \Bigg[ \left({\bm P}_{\frac{3}{2}\frac{3}{2}}v_{l,\,m_l}^{3/2}\left(\theta,\phi\right)+{\bm P}_{\frac{3}{2}-\frac{3}{2}}v_{l,\,m_l}^{-3/2}\left(\theta,\phi\right)\right)\,\psi^{(N)}_{\left(3/2\right)^+}(r)\nonumber\\
  &\hspace{5.0 cm}+\left({\bm P}_{\frac{3}{2}\frac{1}{2}}v_{l,\,m_l}^{1/2}\left(\theta,\phi\right)+{\bm P}_{\frac{3}{2}-\frac{1}{2}}v_{l,\,m_l}^{-1/2}\left(\theta,\phi\right)\right)\,\psi^{(N)}_{\left(1/2\right)^{+\prime}}(r)\Bigg]\,\chi_{s m_s},
 \label{eq:Pwf-3/2}
\end{align}
where $\psi^{(N)}_{\left(1/2\right)^+}\left(r\right)$, $\psi^{(N)}_{\left(3/2\right)^+}\left(r\right)$, and $\psi^{(N)}_{\left(1/2\right)^{+\prime}}\left(r\right)$ are the radial wavefucntions,   $\chi_{s m_s}$ denotes the  heavy quark-antiquark $Q\bar{Q}$ pair spin wavefunction, $\mathcal{C}_{J m_J l s}^{m_l m_s}$ are suitable Clebsch--Gordan coefficients, 
and the index ${P}$ in the subscript on the left-hand side of Eqs.~\eqref{eq:Pwf-1/2} and \eqref{eq:Pwf-3/2} denotes the parity of the states in Table~\ref{tab:QQbarpenta}. 
The angular eigenfunctions $v_{l,\,m_l}^{\lambda}\left(\theta, \phi\right)$ are generalizations of the spherical harmonics for systems with cylindrical symmetry~\cite{Landau:1991wop} and are given by
\begin{align}
v_{lm}^\lambda(\theta,\varphi)&=\frac{(-1)^{l+m}}{2^l}\sqrt{\frac{2l+1}{4\pi}\frac{(l-m)!}{(l+m)!(l-\lambda)!(l+\lambda)!}}P_{lm}^\lambda(\cos\theta)e^{im\varphi}\,,
\label{CoeffExplicit}\\
P_{lm}^\lambda(x)&=(1-x)^{(m-\lambda)/2}(1+x)^{(m+\lambda)/2}\partial_x^{l+m}(1-x)^{l+\lambda}(1+x)^{l-\lambda}\,.
\end{align}

The pentaquark wavefunctions in Eqs.~\eqref{eq:Pwf-1/2} and \eqref{eq:Pwf-3/2} are tensor wavefunctions. 
Their Schr\"odinger equations can be derived from the equations of motion of the BOEFT Lagrangian~\cite{Berwein:2024ztx}. 
The radial Schr\"odinger equation for the pentaquark multiplet in Table~\ref{tab:QQbarpenta} with parity ${P}=-$ and associated with the adjoint baryon $k^P=\left(1/2\right)^+$ is
\begin{equation}
\left[-\frac{1}{m_Qr^2}\,\partial_r\,r^2\,\partial_r+\frac{(l-1/2)(l+1/2)}{m_Qr^2}+E_{\left(1/2\right)_g}\right]\psi_{ \left(1/2\right)^+}^{(N)}=\mathcal{E}_{1/2}\,\psi_{ \left(1/2\right)^+}^{(N)}\,.
\label{eq:Sch12}
\end{equation}
For the pentaquark multiplet in Table~\ref{tab:QQbarpenta} with parity ${P}=-$ and associated with the adjoint baryon $k^P=\left(3/2\right)^+$, the two static energies $E_{\left(1/2\right)_g^\prime}$ and $E_{\left(3/2\right)_g}$ mix at short distances, so, the radial Schr\"odinger equations are a set of coupled equations:
\begin{align}
 &\hspace{-2.0 cm}\Bigg[-\frac{1}{m_Q r^2}\,\partial_rr^2\partial_r+\frac{1}{m_Qr^2}\begin{pmatrix} l(l-1)+\frac{9}{4} & -\sqrt{3l(l+1)-\frac{9}{4}} \\ -\sqrt{3l(l+1)-\frac{9}{4}} & l(l+1)-\frac{3}{4} \end{pmatrix} \nonumber\\
	&\hspace{3.0 cm}+\begin{pmatrix} E_{(1/2)_g^{\prime}} & 0 \\ 0 & E_{(3/2)_g} \end{pmatrix}\Bigg]\hspace{-2pt}\begin{pmatrix} \psi_{\left(1/2\right)^{+\prime}}^{(N)} \\ \psi_{\left(3/2\right)^+}^{(N)}\end{pmatrix}
	=\mathcal{E}_{3/2}\begin{pmatrix} \psi_{\left(1/2\right)^{+\prime}}^{(N)} \\ \psi_{\left(3/2\right)^+}^{(N)}\end{pmatrix}\,,
	\label{eq:Sch32}
\end{align}
where ${\cal E}_{1/2}$ and ${\cal E}_{3/2}$ are the eigenenergies for which we have suppressed the label $\left(N\right)$ for simplicity. 
These eigenenergies provide the binding energies with respect to the spin-isospin averaged $\rm{B\overline{M}}$ threshold -- specifically, the $\Sigma_c\bar{D}$ and $\Sigma_b\bar{B}$ thresholds -- that are set to zero. 
The static potentials $E_{\left(1/2\right)_g}$, $E_{\left(1/2\right)_g^{\prime}}$ and $E_{\left(3/2\right)_g}$ are given by Eqs.~\eqref{eq:QQbar_V} and \eqref{eq:ALambda} and are independent of the heavy quark spin. 
Since we treat $\Lambda_{\kappa}$ in Eq.~\eqref{eq:QQbar_V} as a free parameter, the eigenenergies ${\cal E}_{1/2}$ and ${\cal E}_{3/2}$ are functions of the respective adjoint baryon masses.

Without spin-dependent corrections, we obtain the spin-averaged pentaquark spectrum with masses given by $E_{\Sigma_c\bar{D}}+{\cal E}_{1/2}$ and $E_{\Sigma_c\bar{D}}+{\cal E}_{3/2}$ in the charm sector and  $E_{\Sigma_b\bar{B}}+{\cal E}_{1/2}$ and $E_{\Sigma_b\bar{B}}+{\cal E}_{3/2}$ in the bottom sector. 
For ${\cal E}_{1/2}={\cal E}_{3/2}=0$, the seven pentaquark states in Table~\ref{tab:QQbarpenta} lie exactly at the spin-isospin averaged $\Sigma_c\bar{D}$ or $\Sigma_b\bar{B}$ thresholds.

\subsection{Spin-splittings}
\label{subsec:spin}
Quantitative results of the pentaquark spectrum that reproduce the experimental masses of the states listed in Table~\ref{tab:exoticPenta} require incorporating spin-dependent corrections, which depend on the heavy quark spins ${\bm S}_1$, ${\bm S}_2$, or their total spin ${\bm S}$.  
The BOEFT-potential at ${\cal O}\left(1/m_Q^0\right)$ includes only static contributions, and solving the Schr\"odinger equation in this potential yields degenerate spin multiplets. 
However, BOEFT allows for the systematic inclusion of the spin-dependent corrections to the BO potential, which for hybrids and tetraquarks appear at order $1/m_Q$  \cite{Oncala:2017hop, Brambilla:2018pyn,Brambilla:2019jfi, Soto:2023lbh,Schlosser:2025tca,Soto:2020xpm, Braaten:2024tbm, Brambilla:2024thx}.
The general form of the spin-corrections to the BO potentials in terms of generalized static Wilson loops is known \cite{Soto:2020xpm},
but preliminary lattice calculations of them exist at the moment only for hybrids~\cite{Schlosser:2025tca}.  

In the absence of lattice data, we evaluate the spin splittings within the ground-state $Q\bar{Q}$ pentaquark multiplets in Table~\ref{tab:QQbarpenta} using a method analogous to the one developed for the $Q\bar{Q}$ tetraquarks in Refs.~\cite{Braaten:2024tbm, Brambilla:2024thx}.  The spin splittings in the multiplet are estimated from the spin splittings in the heavy-light baryon-antimeson  pair states $\rm{B\overline{M}}$. 
As discussed, the BO static potentials for the $Q\bar{Q}$ pentaquark asymptotically approach (smoothly connect to) the $\rm{B\overline{M}}$ energy at large separations ($r \to \infty$), owing to the conservation of BO quantum numbers~\cite{Berwein:2024ztx}. 
Since the spin-dependent interactions responsible for the $\rm{B\overline{M}}$ threshold splittings arise at $\mathcal{O}(1/m_Q)$, the pentaquark spin-dependent BO potential at $\mathcal{O}(1/m_Q)$ reduces at large $r$ to a constant potential $V_{SS}$  
that depends on the heavy quark and antiquark spins ${\bm S}_1$ and ${\bm S}_2$,
\begin{equation}
 V_{SS} = \frac{2\Delta_1^Q}{3}\,{\bm S}_1\cdot{\bm K}_1+ \Delta_2^Q\,{\bm S}_2\cdot{\bm K}_2,
 \label{eq:VSS}
\end{equation}
where ${\bm K}_1$ and ${\bm K}_2$ denote the total light quark spin in the heavy baryon ${\rm B}$ and heavy antimeson ${\rm \overline{M}}$, respectively. 
The prefactor $2/3$ in the first term on the right-hand side of Eq.~\eqref{eq:VSS} is such that the parameter $\Delta_1^Q$, which is proportional to $1/m_Q$, is equal to the spin splitting between the heavy baryons ${\rm B}^*$ and ${\rm B}$: 
$\Delta_{1}^c=64.6$~MeV for $\Sigma_c$ baryons  and $\Delta^b_1=19.4$~MeV for $\Sigma_b$ baryons, and $\Delta_1^{c/b}=0$~MeV for $\Lambda_c$ or $\Lambda_b$ baryons~\cite{ParticleDataGroup:2024cfk}. 
The parameter $\Delta_2^Q$ is proportional to $1/m_Q$ and is equal to the spin splitting between the heavy mesons ${\rm M}^*$ and ${\rm M}$:  $\Delta_{2}^c=141.3$~MeV for $D$ mesons  and $\Delta^b_2=45.2$~MeV for $B$ mesons~\cite{ParticleDataGroup:2024cfk}. 
The energies of the $\Sigma^*_{c/b}$ and $\Sigma_{c/b}$ baryons relative to spin-isospin average are $\frac{\Delta_1^Q}{3}$ and $-\frac{2\Delta_1^Q}{3}$. 
The energies of the $D^*/B^*$ and $D/B$ mesons relative to spin-isospin average are $\frac{\Delta_2^Q}{4}$ and $-\frac{3\Delta_2^Q}{4}$. 
The different heavy-light  baryon  antimeson thresholds relative to spin-isospin average threshold are $\left(-\frac{2\Delta_1^Q}{3}-\frac{3\Delta_2^Q}{4}\right)$ for ${\rm B\overline{M}}$, $\left(\frac{\Delta_1^Q}{3}-\frac{3\Delta_2^Q}{4}\right)$ for ${\rm B^*\overline{M}}$, $\left(-\frac{2\Delta_1^Q}{3}+\frac{\Delta_2^Q}{4}\right)$ for ${\rm B\overline{M}^*}$, $\left(\frac{\Delta_1^Q}{3}+\frac{\Delta_2^Q}{4}\right)$ for ${\rm B^*\overline{M}^*}$, where ${\rm B}=\{\Sigma_{c}, \Sigma_b\}$ and ${\rm\overline{M}}=\{\bar{D}, \bar{B}\}$.

Because of the form of the assumed spin interaction in Eq. ~\eqref{eq:VSS}, it is convenient to move from a basis of the total heavy quark spin, labeled by $s$, and light quark spin, labeled by $k$, to a basis of the $\bar{Q}q$ and $Qqq$ total angular momenta labeled by $j_{\bar{Q}q}$ and $j_{Qqq}$, respectively. 
These quantum numbers match the ones of the  heavy baryon-heavy antimeson pair states $\rm {B\overline{M}}$, $\rm {B^*\overline{M}}$, $\rm {B\overline{M}^*}$, and $\rm {B^*\overline{M}^*}$.
The change of basis may be expressed in terms of Wigner-9j symbols and found, e.g., in \cite{Sakai:2019qph,Du:2021fmf,Braaten:2024stn}.
It reads for the pentaquark states listed in Table \ref{tab:QQbarpenta} with $J^P = (1/2)^{-}$ 
\begin{align}
|s=0, k=1/2 \rangle_{J^{P}=\left(1/2\right)^-}&= \frac{1}{2}\,|j_{Qqq}=1/2, j_{\bar{Q}q}=0 \rangle +\frac{1}{2\sqrt{3}}\,|j_{Qqq}=1/2, j_{\bar{Q}q}=1 \rangle\nonumber\\
&+\sqrt{\frac{2}{3}}\,|j_{Qqq}=3/2, j_{\bar{Q}q}=1 \rangle,\nonumber\\
|s=1, k=1/2 \rangle_{J^{P}=\left(1/2\right)^-}&= \frac{1}{2\sqrt{3}}\,|j_{Qqq}=1/2, j_{\bar{Q}q}=0 \rangle +\frac{5}{6}\,|j_{Qqq}=1/2, j_{\bar{Q}q}=1 \rangle\nonumber\\
&-\frac{\sqrt{2}}{3}\,|j_{Qqq}=3/2, j_{\bar{Q}q}=1 \rangle,\nonumber\\
|s=1, k=3/2 \rangle_{J^{P}=\left(1/2\right)^-}&= \sqrt{\frac{2}{3}}\,|j_{Qqq}=1/2, j_{\bar{Q}q}=0 \rangle -\frac{\sqrt{2}}{3}\,|j_{Qqq}=1/2, j_{\bar{Q}q}=1 \rangle\nonumber\\
&-\frac{1}{3}\,|j_{Qqq}=3/2, j_{\bar{Q}q}=1 \rangle.
\label{eq:Fierz1/2}
\end{align}
Similarly, for $J^{P}=\left(3/2\right)^-$ states, we have
\begin{align}
|s=1, k=1/2 \rangle_{J^{P}=\left(3/2\right)^-}&= \frac{1}{\sqrt{3}}\,|j_{Qqq}=3/2, j_{\bar{Q}q}=0 \rangle -\frac{1}{3}\,|j_{Qqq}=1/2, j_{\bar{Q}q}=1 \rangle\nonumber\\
&+\frac{\sqrt{5}}{3}\,|j_{Qqq}=3/2, j_{\bar{Q}q}=1 \rangle,\nonumber\\
|s=0, k=3/2 \rangle_{J^{P}=\left(3/2\right)^-}&= -\frac{1}{2}\,|j_{Qqq}=3/2, j_{\bar{Q}q}=0 \rangle +\frac{1}{\sqrt{3}}\,|j_{Qqq}=1/2, j_{\bar{Q}q}=1 \rangle\nonumber\\
&+\sqrt{\frac{5}{12}}\,|j_{Qqq}=3/2, j_{\bar{Q}q}=1 \rangle,\nonumber\\
|s=1, k=3/2 \rangle_{J^{P}=\left(3/2\right)^-}&= \sqrt{\frac{5}{12}}\,|j_{Qqq}=3/2, j_{\bar{Q}q}=0 \rangle +\frac{\sqrt{5}}{3}\,|j_{Qqq}=1/2, j_{\bar{Q}q}=1 \rangle\nonumber\\
&-\frac{1}{6}\,|j_{Qqq}=3/2, j_{\bar{Q}q}=1 \rangle,
\label{eq:Fierz3/2}
\end{align}
and for $J^{P}=\left(5/2\right)^-$:
\begin{equation}
 |s=1, k=3/2 \rangle_{J^{P}=\left(5/2\right)^-}=|j_{Qqq}=3/2, j_{\bar{Q}q}=1 \rangle.
\end{equation}

To compute the effects of the spin-dependent potential $V_{SS}$, we use first-order perturbation theory. 
We first consider the pentaquark multiplet in Table~\ref{tab:QQbarpenta} associated with the adjoint baryon $k^P=\left(1/2\right)^+$. 
To first order in $\Delta_1^Q$ and $\Delta_2^Q$, the energies $E_{kJ}^s$ of the three states in the spin multiplet are 
\begin{align}
&E^{s=0}_{\frac{1}{2}\frac{1}{2}}={\cal E}_{1/2},\qquad\qquad E^{s=1}_{\frac{1}{2}\frac{1}{2}}={\cal E}_{1/2}-\frac{4\Delta_1^Q}{9}+\frac{\Delta_2^Q}{6},\nonumber\\
&\hspace{1.0 cm}E^{s=1}_{\frac{1}{2}\frac{3}{2}}={\cal E}_{1/2}+\frac{2\Delta_1^Q}{9}-\frac{\Delta_2^Q}{12},
\label{eq:VSS12}
\end{align}
where ${\cal E}_{1/2}$ is the eigenenergy obtained from the Schr\"odinger equation \eqref{eq:Sch12}, the first subscript corresponds to $k$ and the second subscript corresponds to $J$.
We next consider the pentaquark multiplet in Table~\ref{tab:QQbarpenta} associated with the adjoint baryon $k^P=\left(3/2\right)^+$. To first order in $\Delta_1^Q$ and $\Delta_2^Q$, the energies of the four states in the spin multiplet are 
\begin{align}
&E^{s=0}_{\frac{3}{2}\frac{3}{2}}={\cal E}_{3/2},\qquad\qquad E^{s=1}_{\frac{3}{2}\frac{1}{2}}={\cal E}_{3/2}-\frac{5\Delta_1^Q}{9}-\frac{5\Delta_2^Q}{12},\nonumber\\
&E^{s=1}_{\frac{3}{2}\frac{3}{2}}={\cal E}_{3/2}-\frac{2\Delta_1^Q}{9}-\frac{\Delta_2^Q}{6},\qquad\qquad E^{s=1}_{\frac{3}{2}\frac{5}{2}}={\cal E}_{3/2}+\frac{\Delta_1^Q}{3}+\frac{\Delta_2^Q}{4},
\label{eq:VSS32}
\end{align}
where ${\cal E}_{3/2}$ is the eigenenergy obtained from the Schr\"odinger equations \eqref{eq:Sch32}, the first subscript corresponds to $k$ and the second subscript corresponds to $J$. 

In Eqs.~\eqref{eq:VSS12} and \eqref{eq:VSS32}, there are three pentaquark states with  $J^{P}=\left(1/2\right)^-$ and $J^{P}=\left(3/2\right)^-$. 
If the eigenenergies ${\cal E}_{1/2}$ and ${\cal E}_{3/2}$ are nearly degenerate, then the mixing of states with the same $J^{P}$ quantum numbers cannot be treated as a perturbation.  
To correctly determine the physical spectrum or eigenstates, it is necessary to diagonalize the $3\times3$ submatrix of $V_{SS}$ in Eq.~\eqref{eq:VSS}  for $J^{P}=\left(1/2\right)^-$ and $J^{P}=\left(3/2\right)^-$. 
As we show in Eqs.~\eqref{eq:resultscenario11}, \eqref{eq:resultscenario12}, \eqref{eq:resultscenario21}, and \eqref{eq:resultscenario22}, the resulting physical eigenstates (pentaquark states) turn out to be superpositions of states with heavy quark-antiquark pair spin $s=0$ and $s=1$. 

The $3\times3$ submatrix for $J^{P}=\left(1/2\right)^-$ is given by
\renewcommand*\arraystretch{1.5}
\begin{align}
 M_{\left(1/2\right)^-}=\begin{pmatrix} E^{s=0}_{\frac{1}{2}\frac{1}{2}} & -\frac{2\Delta_1^Q}{3\sqrt{3}}-\frac{\Delta_2^Q}{4\sqrt{3}} & -\frac{2\Delta_1^Q}{3\sqrt{6}}-\frac{\Delta_2^Q}{\sqrt{6}} \\ -\frac{2\Delta_1^Q}{3\sqrt{3}}-\frac{\Delta_2^Q}{4\sqrt{3}} & E^{s=1}_{\frac{1}{2}\frac{1}{2}} & \frac{\sqrt{2}\Delta_1^Q}{9}-\frac{\Delta_2^Q}{3\sqrt{2}}\\  -\frac{2\Delta_1^Q}{3\sqrt{6}}-\frac{\Delta_2^Q}{\sqrt{6}} & \frac{\sqrt{2}\Delta_1^Q}{9}-\frac{\Delta_2^Q}{3\sqrt{2}} & E^{s=1}_{\frac{3}{2}\frac{1}{2}}
 \end{pmatrix},
 \label{eq:M12}
\end{align}
where the rows and columns correspond to $|s=0, k=1/2 \rangle_{J^{P}=\left(1/2\right)^-}$, $|s=1, k=1/2 \rangle_{J^{P}=\left(1/2\right)^-}$, and $|s=1, k=3/2 \rangle_{J^{P}=\left(1/2\right)^-}$.  Similarly, for $J^{P}= \left(3/2\right)^-$, the resulting matrix is given by
\renewcommand*\arraystretch{1.5}
\begin{align}
 M_{\left(3/2\right)^-}=\begin{pmatrix} E^{s=1}_{\frac{1}{2}\frac{3}{2}} & \frac{\Delta_1^Q}{3\sqrt{3}}+\frac{\Delta_2^Q}{2\sqrt{3}} & \frac{\sqrt{5}\Delta_1^Q}{9}-\frac{\sqrt{5}\Delta_2^Q}{6} \\ \frac{\Delta_1^Q}{3\sqrt{3}}+\frac{\Delta_2^Q}{2\sqrt{3}} & E^{s=0}_{\frac{3}{2}\frac{3}{2}} & -\frac{\sqrt{5}\Delta_1^Q}{3\sqrt{3}}+\frac{\sqrt{5}\Delta_2^Q}{4\sqrt{3}}\\ \frac{\sqrt{5}\Delta_1^Q}{9}-\frac{\sqrt{5}\Delta_2^Q}{6} & -\frac{\sqrt{5}\Delta_1^Q}{3\sqrt{3}}+\frac{\sqrt{5}\Delta_2^Q}{4\sqrt{3}} & E^{s=1}_{\frac{3}{2}\frac{3}{2}}
 \end{pmatrix},
 \label{eq:M32}
\end{align}
where the rows and columns correspond to $|s=1, k=1/2 \rangle_{J^{P}=\left(3/2\right)^-}$, $|s=0, k=3/2 \rangle_{J^{P}=\left(3/2\right)^-}$, and $|s=1, k=3/2 \rangle_{J^{P}=\left(3/2\right)^-}$. 
We denote the eigenvalues of the matrices $M_{\left(1/2\right)^-}$ and $M_{\left(3/2\right)^-}$ by $E^{i}_{M_{1/2}}$ and $E^{i}_{M_{3/2}}$ ($i=1,2,3$), respectively. 
The eigenvectors of the matrices $M_{\left(1/2\right)^-}$ and $M_{\left(3/2\right)^-}$ are identified with the physical pentaquark states, which are superpositions of the  multiplet states  (Table~\ref{tab:QQbarpenta}) with well-defined heavy quark-antiquark pair spin $s$ and LDF spin $k$. 
The state with $J^{P}=\left(5/2\right)^-$ corresponds to only $s=1$, $k=3/2$ and $l=3/2$. 
Note that $E^{i}_{M_{1/2}}$ and $E^{i}_{M_{3/2}}$ depend on the adjoint baryon masses $\Lambda_{\left(1/2\right)^+}$ and $\Lambda_{\left(3/2\right)^+}$ through the eigenenergies ${\cal E}_{1/2}$ and ${\cal E}_{3/2}$ obtained from the Schr\"odinger equations \eqref{eq:Sch12} and \eqref{eq:Sch32}.

After incorporating spin-dependent corrections, the masses of the pentaquark states are given as follows. 
For $J^{P}=\left(1/2\right)^-$ the masses are $E_{\Sigma_c\bar{D}}+E^{i}_{M_{1/2}}$ in the charm sector and $E_{\Sigma_b\bar{B}}+E^{i}_{M_{1/2}}$ in the bottom sector, while, for $J^{P}=\left(3/2\right)^-$ they are  $E_{\Sigma_c\bar{D}}+E^{i}_{M_{3/2}}$ in the charm sector and $E_{\Sigma_b\bar{B}}+E^{i}_{M_{3/2}}$ in the bottom sector.
Finally, for $J^{P}=\left(5/2\right)^-$,  the masses are given by $E_{\Sigma_c\bar{D}}+E^{s=1}_{\frac{3}{2}\frac{5}{2}}$ in the charm sector and  $E_{\Sigma_b\bar{B}}+E^{s=1}_{\frac{3}{2}\frac{5}{2}}$  in the bottom sector.

In the next subsection, we focus only on the charm pentaquarks. 
We discuss different plausible scenarios for identifying states with the experimentally observed charm pentaquarks $P_{c\bar{c}}$ listed in Table~\ref{tab:exoticPenta}.

\subsection{Identification with experimental $P_{c\bar{c}}$ states}
\label{subsec:identification}

\subsubsection{Scenario 0: ${\cal E}_{1/2}={\cal E}_{3/2}=0$}
For illustration, we first discuss the simplest scenario where the adjoint baryon masses $\Lambda_{\left(1/2\right)^+}$ and $\Lambda_{\left(3/2\right)^+}$ are chosen in such a way to set the 
eigenenergies of the Schr\"odinger equations~\eqref{eq:Sch12} and \eqref{eq:Sch32} to zero: ${\cal E}_{1/2}={\cal E}_{3/2}=0$. 
This scenario was considered by Voloshin to study the decay properties of $P_{c\bar{c}}$ states in the molecular picture in Ref.~\cite{Voloshin:2019aut}.  
Without including the spin-dependent corrections from $V_{SS}$ given in Eq.~\eqref{eq:VSS}, the states are exactly at the spin-isospin-averaged $\Sigma_c\bar{D}$ threshold. 
The critical values of the adjoint baryon masses (in the RS-scheme\footnote{
We use the RS masses, $m_c^{\mathrm{RS}}$ and $m_b^{\mathrm{RS}}$, defined at the renormalon subtraction scale $\nu_f=1\,\mathrm{GeV}$ and given in Sec.~\ref{subsubsec:parameterization}.}) 
that yield ${\cal E}_{1/2}={\cal E}_{3/2}=0$ are $\Lambda_{\left(1/2\right)^+, \mathrm{RS}}=1.149$~GeV and $\Lambda_{\left(3/2\right)^+, \mathrm{RS}}=1.230$~GeV.

Including the spin-dependent corrections from $V_{SS}$, the eigenvalues $E^{i}_{M_{1/2}}$ and $E^{i}_{M_{3/2}}$ in this scenario are
\begin{align}
 E^i_{M_{1/2}} &= \left(-\frac{2\Delta_1^c}{3}-\frac{3\Delta_2^c}{4}, \, -\frac{2\Delta_1^c}{3}+\frac{\Delta_2^c}{4}, \, \frac{\Delta_1^c}{3}+\frac{3\Delta_2^c}{4} \right)=\left(-149.1, -7.8, 56.9\right)~\mathrm{MeV},\nonumber\\
 E^i_{M_{3/2}} &= \left(\frac{\Delta_1^c}{3}-\frac{3\Delta_2^c}{4}, \, -\frac{2\Delta_1^c}{3}+\frac{\Delta_2^c}{4}, \, \frac{\Delta_1^c}{3}+\frac{3\Delta_2^c}{4} \right)=\left(-84.4, -7.8, 56.9\right)~\mathrm{MeV}, 
 \label{eq:E12E320}
\end{align}
where the first, second, and third entries in the parentheses correspond to $i=1, 2, 3$, respectively. 
The masses of the states are obtained after adding $E_{\Sigma_c\bar{D}}$ to the eigenvalues $E^i_{M_{1/2}}$ and $E^i_{M_{3/2}}$ in Eq.~\eqref{eq:E12E320}. 
This implies that for $J^{P}=\left(1/2\right)^-$, there is a state exactly at each of the $\Sigma_c\bar{D}$, $\Sigma_c\bar{D}^*$ and $\Sigma_c^*\bar{D}^*$ thresholds, 
while for $J^{P}=\left(3/2\right)^-$, there is a state exactly at each of the $\Sigma_c^*\bar{D}$, $\Sigma_c\bar{D}^*$ and $\Sigma_c^*\bar{D}^*$ thresholds. 
The state with mass $4.321$~GeV at the $\Sigma_c\bar{D}$ threshold is identified with $P_{c\bar{c}}\left(4312\right)^+$, 
the state with mass $4.385$~GeV at the $\Sigma_c^*\bar{D}$ threshold is identified with $P_{c\bar{c}}\left(4380\right)^+$, 
and the two states with mass $4.462$~GeV at the $\Sigma_c\bar{D}^*$ threshold are identified with $P_{c\bar{c}}\left(4440\right)^+$ and $P_{c\bar{c}}\left(4457\right)^+$, 
whose $J^{P}$ quantum numbers could be both $\left(1/2\right)^-$ and $\left(3/2\right)^-$. 
There are three states at the $\Sigma_c^*\bar{D}^*$ threshold with mass $4.527$~GeV and $J^{P}=\left(1/2\right)^-, \left(3/2\right)^-$, and  $\left(5/2\right)^-$.

However, the experimental states listed in Table~\ref{tab:exoticPenta} have masses slightly below the $\Sigma_c\bar{D}$, $\Sigma_c^*\bar{D}$, and $\Sigma_c\bar{D}^*$ thresholds after including spin-corrections. 
The eigenvalues $E^{i}_{M_{1/2}}$ and $E^{i}_{M_{3/2}}$ ($i=1,2,3$) of the $3\times3$ matrices $M_{1/2}$ and $M_{3/2}$ are functions of the eigenenergies ${\cal E}_{1/2}$ and ${\cal E}_{3/2}$. To reproduce the masses of the four observed pentaquark states within experimental uncertainties (see Table~\ref{tab:exoticPenta}), we identify two optimal choices for these eigenenergies:
$\left({\cal E}_{1/2}, {\cal E}_{3/2}\right)=\left(-23~\mathrm{MeV}, -1~\mathrm{MeV}\right)$ and $\left({\cal E}_{1/2}, {\cal E}_{3/2}\right)=\left(-0.5~\mathrm{MeV}, -14~\mathrm{MeV}\right)$, which we denote as \textit{scenario 1} and \textit{scenario 2}, respectively. 
In the following, we discuss both scenarios.

\subsubsection{Scenario 1: ${\cal E}_{1/2}=-23$~\rm{MeV} and ${\cal E}_{3/2}=-1$~\rm{MeV}}

\begin{table}[t!]
\begin{center}
\small{\renewcommand{\arraystretch}{1.5}
\scriptsize
\begin{tabular}{||c|c|c||}
\hline
   $\begin{array}{c} P_{c\bar{c}}\\{\rm state}\end{array}$       &   Mass (MeV)   &$J^{P}$      \\
   \hline
 $\bm{P_{c\bar{c}}\left(4312\right)^+}$  & $4312$ & $\left(1/2\right)^-$ \\
  
 $\bm{P_{c\bar{c}}\left(4380\right)^+}$  & $4376$ & $\left(3/2\right)^-$ \\
 
$\bm{P_{c\bar{c}}\left(4440\right)^+}$  & $4444$ & $\left(1/2\right)^-$ \\

$\bm{P_{c\bar{c}}\left(4457\right)^+}$  & $4458$ & $\left(3/2\right)^-$ \\

$P_{c\bar{c}}\left(4507\right)^+$  & $4507$ & $\left(1/2\right)^-$\\

$P_{c\bar{c}}\left(4515\right)^+$  & $4515$ & $\left(3/2\right)^-$\\

$P_{c\bar{c}}\left(4526\right)^+$  & $4526$ & $\left(5/2\right)^-$\\
    \hline
\end{tabular}
\caption{ The table shows the masses and assigned $J^{P}$ quantum numbers for the charm pentaquark states in scenario 1 listed in order of increasing mass. 
The states in bold are the experimental states shown in Table~\ref{tab:exoticPenta}.  
The last three states are below the $\Sigma_c^*\bar{D}^*$ threshold and have not yet been observed experimentally. 
All states have isospin $I=1/2$.}
\label{tab:scenario1}}
\end{center}
\end{table}

\begin{figure}[t!]
\begin{minipage}{.6\textwidth}
  \includegraphics[width=\linewidth]{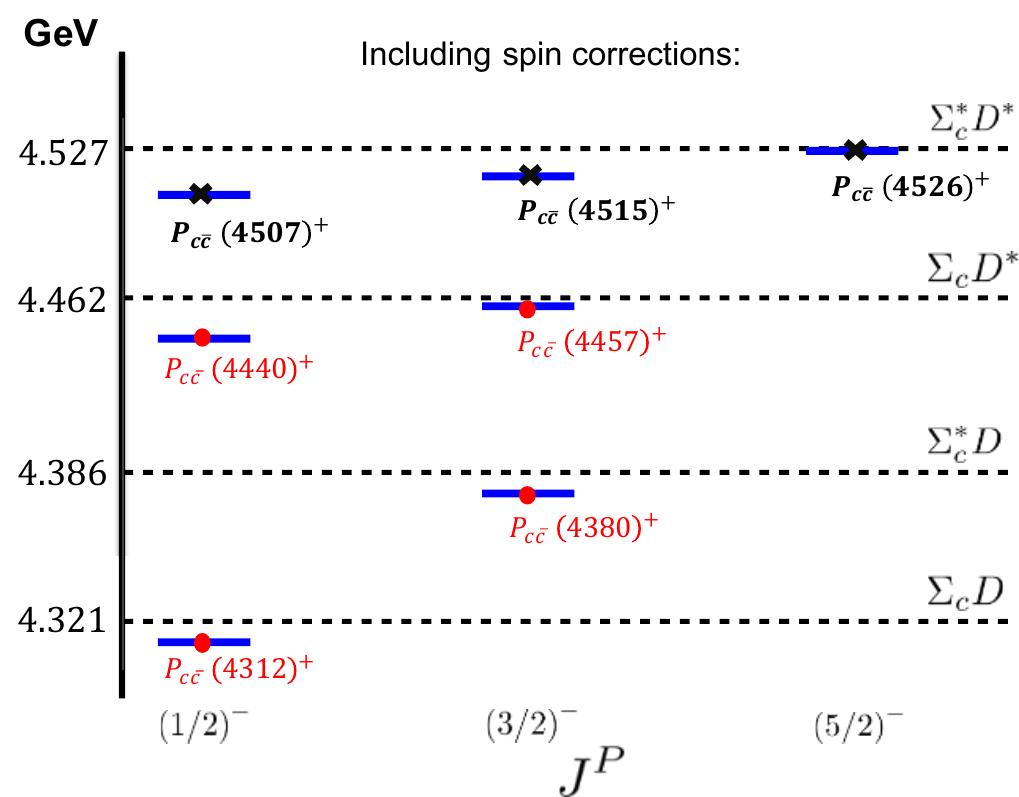}
\end{minipage}\\\vspace{0.5 cm}
\begin{minipage}{.6\textwidth}
  \includegraphics[width=\linewidth]{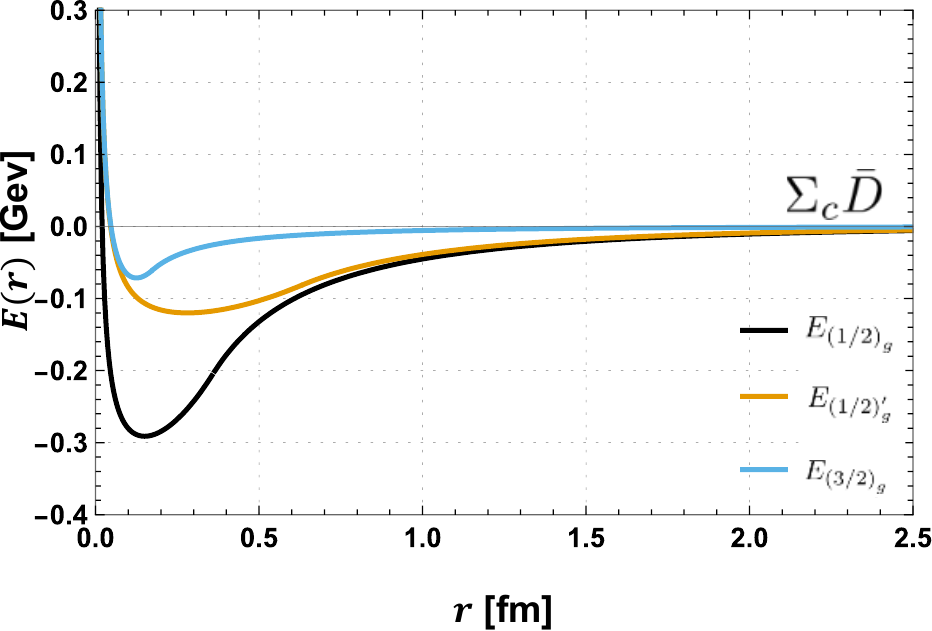}
\end{minipage}
\caption{On the top, we show the spectrum of the lowest $c\bar{c}$ pentaquark multiplet  in Table~\ref{tab:QQbarpenta} after including spin corrections in scenario 1. 
After solving the coupled Schr\"odinger equations \eqref{eq:Sch12} and \eqref{eq:Sch32} and using Eqs.~\eqref{eq:M12} and \eqref{eq:M32} to account for spin-corrections,  
the obtained states are represented as horizontal blue bars. 
The four experimental $P_{c\bar{c}}$ states listed in Table~\ref{tab:exoticPenta} are shown as red dots, while the three unobserved ones are shown as black crosses. 
On the bottom, we show the assumed BO potential curves $E_{\left(1/2\right)_g}$, $E_{\left(1/2\right)_g^\prime}$, and $E_{\left(3/2\right)_g}$ corresponding to the adjoint baryon masses $\Lambda_{\left(1/2\right)^+, \mathrm{RS}}=0.998$~GeV and $\Lambda_{\left(3/2\right)^+, \mathrm{RS}}=1.209$~GeV. 
Isospin of the states is $I=1/2$.}  
\label{fig:scenario1}
\end{figure}

The values of the adjoint baryon masses (in the RS-scheme) that yield $\left({\cal E}_{1/2}, {\cal E}_{3/2}\right)=\left(-23~\mathrm{MeV}, -1~\mathrm{MeV}\right)$ are $\Lambda_{\left(1/2\right)^+, \mathrm{RS}}=0.998$~GeV and $\Lambda_{\left(3/2\right)^+, \mathrm{RS}}=1.209$~GeV. 
Including the spin-dependent corrections from $V_{SS}$, the eigenvalues $E^{i}_{M_{1/2}}$, $E^{i}_{M_{3/2}}$, and $E^{s=1}_{\frac{3}{2}\frac{5}{2}}$ are
\begin{align}
E^i_{M_{1/2}} &=\left(-158.1, -25.6, 36.7\right)~\mathrm{MeV},\qquad\qquad  E^{s=1}_{\frac{3}{2}\frac{5}{2}} = 55.9~\mathrm{MeV},\nonumber\\ 
E^i_{M_{3/2}} &=\left(-93.7, -11.4, 44.8\right)~\mathrm{MeV},
 \label{eq:E12E321}
\end{align}
where the first, second, and third entries in the parentheses correspond to $i=1, 2, 3$, respectively. 
The masses  of the states, obtained after adding $E_{\Sigma_c\bar{D}}$ to the results in Eq.~\eqref{eq:E12E321}, and the assigned $J^{P}$ quantum numbers are listed in Table.~\ref{tab:scenario1}. 
The identification with the experimental states is shown in Fig.~\ref{fig:scenario1}.

The eigenvectors of the matrices $M_{1/2}$ and $M_{3/2}$ in Eqs.~\eqref{eq:M12} and \eqref{eq:M32} allow to express the pentaquark states in Table~\ref{tab:scenario1} in terms of the multiplet states listed in Table~\ref{tab:QQbarpenta}, which have well-defined heavy-quark-antiquark spin $s$ and LDF angular momentum $k$:
\begin{align}
 \begin{pmatrix}
  | P_{c\bar{c}}\left(4312\right)^+\rangle\\
  | P_{c\bar{c}}\left(4440\right)^+\rangle\\
  | P_{c\bar{c}}\left(4507\right)^+\rangle
 \end{pmatrix} &=
 \begin{pmatrix}
  0.541 & 0.327 & 0.775 \\
  -0.207 & -0.842 & 0.499 \\
  -0.815 & 0.430 & 0.388
 \end{pmatrix}
 \begin{pmatrix}
  |s=0, k=1/2 \rangle_{J^{P}=\left(1/2\right)^-}\\
  |s=1, k=1/2 \rangle_{J^{P}=\left(1/2\right)^-}\\
  |s=1, k=3/2 \rangle_{J^{P}=\left(1/2\right)^-}
 \end{pmatrix},
  \label{eq:resultscenario11}
  \end{align}
  \begin{align}
 \begin{pmatrix}
  | P_{c\bar{c}}\left(4380\right)^+\rangle\\
  | P_{c\bar{c}}\left(4457\right)^+\rangle\\
  | P_{c\bar{c}}\left(4515\right)^+\rangle
 \end{pmatrix} &=
 \begin{pmatrix}
  0.646 & -0.484 & 0.589 \\
  -0.361 & 0.486 & 0.796 \\
  0.672 & 0.727 & -0.139
 \end{pmatrix}
 \begin{pmatrix}
  |s=1, k=1/2 \rangle_{J^{P}=\left(3/2\right)^-}\\
  |s=0, k=3/2 \rangle_{J^{P}=\left(3/2\right)^-}\\
  |s=1, k=3/2\rangle_{J^{P}=\left(3/2\right)^-}
 \end{pmatrix},
  \label{eq:resultscenario12}
 \end{align}
  \begin{align}
 |P_{c\bar{c}}\left(4526\right)^+\rangle &=  |s=1, k=3/2 \rangle_{J^{P}=\left(5/2\right)^-}.
 \label{eq:resultscenario13}
\end{align}

\subsubsection{Scenario 2: ${\cal E}_{1/2}=-0.5$~\rm{MeV} and ${\cal E}_{3/2}=-14$~\rm{MeV}}
\begin{figure}[h!]
\begin{minipage}{.6\textwidth}
  \includegraphics[width=1\linewidth]{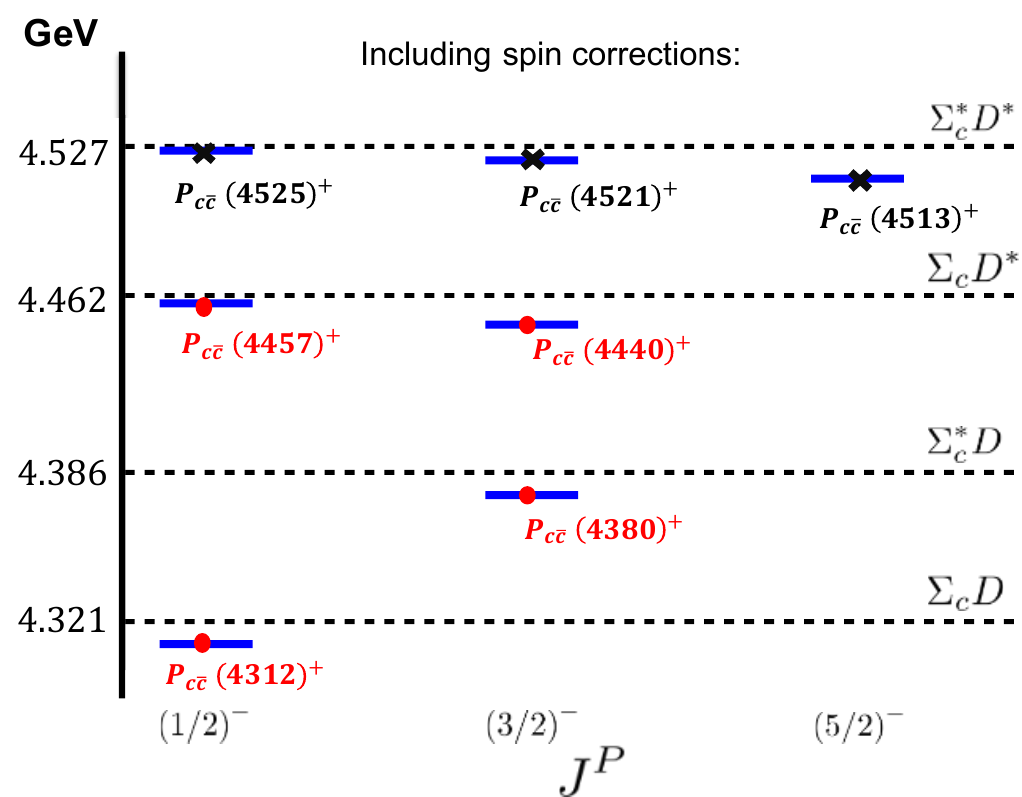}
\end{minipage}\\
\vspace{0.5 cm}
\begin{minipage}{.6\textwidth}
  \includegraphics[width=1\linewidth]{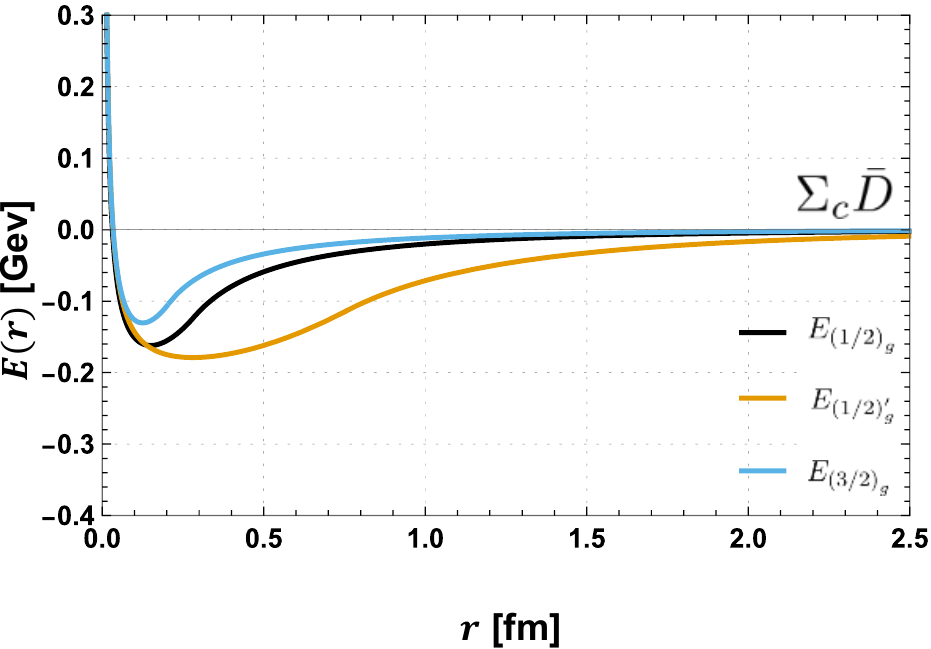}
\end{minipage}
\caption{On the top, we show the spectrum of the lowest $c\bar{c}$ pentaquark multiplet in Table~\ref{tab:QQbarpenta} after including spin corrections in scenario 2. 
The masses are computed from the adjoint baryon masses $\Lambda_{1/2, \mathrm{RS}}=1.125$~GeV and $\Lambda_{3/2, \mathrm{RS}}=1.152$~GeV. 
The labeling is the same as in Fig.~\ref{fig:scenario1}. 
On the bottom, we show the assumed BO potential curves $E_{\left(1/2\right)_g}$, $E_{\left(1/2\right)_g^\prime}$ and $E_{\left(3/2\right)_g}$ for scenario 2 
corresponding to the adjoint baryon masses $\Lambda_{\left(1/2\right)^+, \mathrm{RS}}=1.125$~GeV and $\Lambda_{\left(3/2\right)^+, \mathrm{RS}}=1.152$~GeV. 
Isospin of the states is $I=1/2$.}  
\label{fig:scenario2}
\end{figure}

The values of the adjoint baryon masses (in the RS-scheme) that yield $\left({\cal E}_{1/2}, {\cal E}_{3/2}\right)=\left(-0.5~\mathrm{MeV}, -14~\mathrm{MeV}\right)$ are $\Lambda_{\left(1/2\right)^+, \mathrm{RS}}=1.125$~GeV and $\Lambda_{\left(3/2\right)^+, \mathrm{RS}}=1.152$~GeV. Including the spin-dependent corrections from $V_{SS}$, the eigenvalues $E^{i}_{M_{1/2}}$, $E^{i}_{M_{3/2}}$, and $E^{s=1}_{\frac{3}{2}\frac{5}{2}}$ are
\begin{align}
E^i_{M_{1/2}} &=\left(-158.8, -11.1, 55.0\right)~\mathrm{MeV},\qquad\qquad  E^{s=1}_{\frac{3}{2}\frac{5}{2}} = 42.9~\mathrm{MeV},\nonumber\\ 
E^i_{M_{3/2}} &=\left(-94.2, -20.4, 50.8\right)~\mathrm{MeV},
 \label{eq:E12E322}
\end{align}
where the first, second, and third entries in the parentheses correspond to $i=1, 2, 3$, respectively. 
The masses  of the states, obtained after adding $E_{\Sigma_c\bar{D}}$ to the results in Eq.~\eqref{eq:E12E322}, and the assigned $J^{P}$ quantum numbers are listed in Table.~\ref{tab:scenario2}. 
The identification with the experimental states is shown in Fig.~\ref{fig:scenario2}.

\begin{table}[h!]
\begin{center}
\small{\renewcommand{\arraystretch}{1.5}
\scriptsize
\begin{tabular}{||c|c|c||}
\hline
   $\begin{array}{c} P_{c\bar{c}}\\{\rm state}\end{array}$       &   Mass (MeV)   &$J^{P}$      \\
   \hline
 $\bm{P_{c\bar{c}}\left(4312\right)^+}$  & $4311$ & $\left(1/2\right)^-$ \\
  
 $\bm{P_{c\bar{c}}\left(4380\right)^+}$  & $4376$ & $\left(3/2\right)^-$ \\
 
$\bm{P_{c\bar{c}}\left(4440\right)^+}$  & $4449$ & $\left(3/2\right)^-$ \\

$\bm{P_{c\bar{c}}\left(4457\right)^+}$  & $4459$ & $\left(1/2\right)^-$ \\

$P_{c\bar{c}}\left(4513\right)^+$  & $4513$ & $\left(5/2\right)^-$\\

$P_{c\bar{c}}\left(4521\right)^+$  & $4521$ & $\left(3/2\right)^-$\\

$P_{c\bar{c}}\left(4525\right)^+$  & $4525$ & $\left(1/2\right)^-$\\
    \hline
\end{tabular}
\caption{ The table shows the masses and assigned $J^{P}$ quantum numbers for the charm pentaquark states in scenario 2 listed in order of increasing mass. 
The states in bold are the experimental states shown in Table~\ref{tab:exoticPenta}. 
The last three states are below the $\Sigma_c^*\bar{D}^*$ threshold and have not been experimentally observed so far.
All states have isospin $I=1/2$.}
\label{tab:scenario2}}
\end{center}
\end{table}

Similar to Eqs.~\eqref{eq:resultscenario11}--\eqref{eq:resultscenario13}, we can express the pentaquark states in Table~\ref{tab:scenario2} as:
\begin{align}
 \begin{pmatrix}
  | P_{c\bar{c}}\left(4312\right)^+\rangle\\
  | P_{c\bar{c}}\left(4457\right)^+\rangle\\
  | P_{c\bar{c}}\left(4525\right)^+\rangle
 \end{pmatrix} &=
 \begin{pmatrix}
  0.475 & 0.267 & 0.838 \\
  -0.331 & -0.829 & 0.452 \\
  -0.815 & 0.492 & 0.305
 \end{pmatrix}
 \begin{pmatrix}
  |s=0, k=1/2 \rangle_{J^{P}=\left(1/2\right)^-}\\
  |s=1, k=1/2 \rangle_{J^{P}=\left(1/2\right)^-}\\
  |s=1, k=3/2 \rangle_{J^{P}=\left(1/2\right)^-}
 \end{pmatrix},
 \label{eq:resultscenario21}
 \end{align}
 \begin{align}
 \begin{pmatrix}
  | P_{c\bar{c}}\left(4380\right)^+\rangle\\
  | P_{c\bar{c}}\left(4440\right)^+\rangle\\
  | P_{c\bar{c}}\left(4521\right)^+\rangle
 \end{pmatrix} &=
 \begin{pmatrix}
  0.536 & -0.506 & 0.676 \\
  -0.314 & 0.624 & 0.716 \\
  0.784 & 0.596 & -0.176
 \end{pmatrix}
 \begin{pmatrix}
  |s=1, k=1/2 \rangle_{J^{P}=\left(3/2\right)^-}\\
  |s=0, k=3/2 \rangle_{J^{P}=\left(3/2\right)^-}\\
  |s=1, k=3/2 \rangle_{J^{P}=\left(3/2\right)^-}
 \end{pmatrix},
 \label{eq:resultscenario22}\\
 |P_{c\bar{c}}\left(4513\right)^+\rangle &=  |s=1, k=3/2 \rangle_{J^{P}=\left(5/2\right)^-}.
 \label{eq:resultscenario23}
\end{align}
Contrary to scenario 1, to the states $P_{c\bar{c}}\left(4440\right)$ and $P_{c\bar{c}}\left(4457\right)$ are assigned $J^{P}=\left(3/2\right)^-$ and  $J^{P}=\left(1/2\right)^-$, respectively. Moreover, the pentaquark states that are slightly below the $\Sigma_c^*\bar{D}^*$ threshold have slightly different masses. 
In both scenarios, the pentaquark states with $J^{P}=\left(1/2\right)^-$ and $J^{P}=\left(3/2\right)^-$ are superpositions of heavy-quark pair spin singlet, $s=0$, and triplet, $s=1$, states reflecting a violation of the heavy quark spin symmetry. 
Only the pentaquark state with $J^{P}=\left(5/2\right)^-$ has a well-defined heavy-quark spin configuration with $s=1$.

\section{Decays}
\label{sec:decays}
In this section, we examine the decays of the pentaquark states obtained in scenario 1 (see Eqs. \eqref{eq:resultscenario11}–\eqref{eq:resultscenario13}) and scenario 2 (see Eqs. \eqref{eq:resultscenario21}–\eqref{eq:resultscenario23}), focusing on their decays into $J/\psi$, $\eta_c$, $\Lambda_c\bar{D}$ and $\Lambda_c\bar{D}^*$. 
We compute the semi-inclusive decay rates of pentaquark states to $J/\psi$ and $\eta_c$ following Ref.~\cite{Brambilla:2022hhi}, 
and we evaluate the decay width ratios to $\Lambda_c\bar{D}$ and $\Lambda_c\bar{D}^*$ thresholds following Ref.~\cite{Braaten:2024stn}.

\subsection{Semi-inclusive decays to $J/\psi$ and $\eta_c$.}
\label{subsec:Semi-inlcusive}

Our aim is to compute the semi-inclusive decay rates of a pentaquark $P_{c\bar{c}}$ decaying into quarkonium states $J/\psi$ and $\eta_c$: $P_{c\bar{c}}\rightarrow J/\psi + X$ and $P_{c\bar{c}}\rightarrow \eta_c + X$, where $X$ denotes light hadrons. 
The energy transfer in the transition $P_{c\bar{c}}\rightarrow Q_n + X$, with $Q_n=\{J/\psi, \eta_c\}$ is $\Delta E$,  which is simply the mass difference between the initial pentaquark and the final quarkonium state. 
The energy gap $\Delta E$  is larger than $1$~GeV, and thus satisfies the condition $\Delta E \gg \Lambda_{\mathrm{QCD}}$.  
The $Q\bar{Q}$ pairs in pentaquarks are in a color octet state, while the $Q\bar{Q}$ pairs in $J/\psi$ or $\eta_c$ are in a color singlet state. 
With such a large energy gap $\Delta E$, the gluon emitted by the heavy quarks in the transition from an octet to a singlet state can be treated in weak-coupling perturbation theory 
and semi-inclusive decay widths may be computed along the lines of Refs.~\cite{Brambilla:2022hhi, Oncala:2017hop,TarrusCastella:2021pld}.

In BOEFT, all modes associated with high energy scales down to and including $\Lambda_{\mathrm {QCD}}$ are integrated out, 
which means that  gluons of energy and momentum of order $\Delta E$ should also be integrated out. 
This leads to an imaginary contribution to the pentaquark BO potential related to the semi-inclusive decay rate by the optical theorem. 
In the context of quarkonium hybrids, the semi-inclusive decay rates of hybrids to low-lying quarkonium states at leading and subleading power in the inverse of
the heavy-quark mass were computed in Refs~\cite{Brambilla:2022hhi, Oncala:2017hop, TarrusCastella:2021pld} using the  spectator gluon approximation: 
the LDF (low-energy gluons) that constitute the hybrid do not interact with the high-energy gluons emitted in the transition from octet to singlet that carry the large energy $\Delta E\gg \Lambda_{\mathrm{QCD}}$. 
The leading order contribution comes from the spin-conserving semi-inclusive decay induced by the  chromoelectric-dipole (E1) coupling of the gluon with the  color octet and singlet quark-antiquark pairs, 
while the subleading order contribution comes from the spin-flipping semi-inclusive decay induced by the chromomagnetic-dipole (M1) coupling of the gluon with the  color octet and singlet quark-antiquark pairs \cite{Brambilla:2022hhi}. 
With spin-conserving (or spin-flipping) decays we mean that the spin of the heavy quark-antiquark $Q\bar{Q}$ pair remains the same (or changes) from the initial to the final state.

\begin{figure}[ht]
\begin{center}
\includegraphics[width=0.4\textwidth,angle=0,clip]{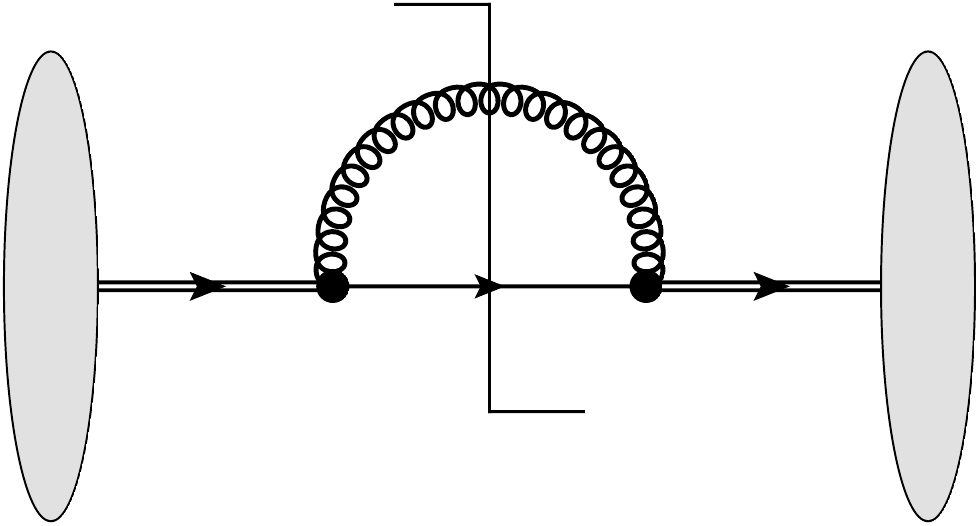}
\caption{The gray blobs stand for a pentaquark state $P_{c\bar{c}}$, and the single and double lines for a $Q\bar{Q}$ pair in a color singlet and octet state, respectively. 
The singlet is associated with the quarkonium state. 
The curly line stands for the energetic gluon emitted in the transition and the black dots for the chromomagnetic-dipole (M1) couplings. 
The vertical line is the cut. 
The imaginary part of this self-energy diagram gives the semi-inclusive widths of the transitions $P_{c\bar{c}}\rightarrow J/\psi + X$ and $P_{c\bar{c}}\rightarrow \eta_c + X$, where $X$ denotes light hadrons. The light quark degrees of freedom that are part of the pentaquark are treated as spectators and are not displayed. }
\label{fig:diagrams}
\end{center}
\end{figure}

In the context of the present study on pentaquarks, we use the results from Ref.~\cite{Brambilla:2022hhi} and the spectator light quark approximation: the LDF (three light quarks) in the  pentaquark do not interact with the high-energy gluon of energy $\Delta E$. 
The transition  $P_{c\bar{c}}\rightarrow Q_n + X$, with $Q_n=\{J/\psi, \eta_c\}$ proceeds via spin-flipping semi-inclusive decay, as both the initial pentaquark and the final quarkonium states have zero orbital angular momentum $\left(L_Q=0\right)$.\footnote{
Spin-conserving semi-inclusive decays of $P_{c\bar{c}}$ would instead produce P-wave quarkonium states.}
We use the spin-averaged masses for pentaquark and quarkonium to estimate the decay rates for both scenarios 1 and 2. 
The spin-flipping semi-inclusive decay width is given by \cite{Brambilla:2022hhi, TarrusCastella:2021pld}
\begin{align}
\Gamma(H_m\to Q_n)=\frac{2\,\alpha_{\rm s}\left(\Delta E\right)}{9}\,T^{\alpha j}\,(T^{\alpha j})^\dagger\,\Delta E^3\,,
\label{eq:Gamma_spincons}
\end{align}
with $T^{\alpha j}$ the matrix element  
\begin{align}
  T^{\alpha j}&=
       \frac{1}{m_Q}   \left[\int d^3{\bm r}\,\Psi_{k^P, {P}=-}^{(N), \alpha\dagger}\left({\bm r}\right)\,
\Phi^{Q\bar{Q}}_{(n)}(\bm{r})\right]\langle \chi_{P_{c\bar{c}}}|\left(S_1^{ j}-S_2^{j}\right)|\chi_{Q_n}\rangle\,,
\label{eq:matrix_SO-spin}
\end{align}
where $k^P=\left(1/2\right)^+$ or $\left(3/2\right)^+$, $\alpha$ denotes the pentaquark vector index,  ${\bm S}_1$ and ${\bm S}_2$ are the spin vectors of the heavy quark and heavy antiquark and $|\chi_{P_{c\bar{c}}}\rangle$ and $|\chi_{Q_n}\rangle$ denote the $Q\bar{Q}$ pair spin state for pentaquark and quarkonium, respectively. 
The spin-matrix elements are computed in Appendix~D of Ref.~\cite{Brambilla:2022hhi}.
The pentaquark wavefunctions are given by Eq.~\eqref{eq:Pwf-1/2} for $k^P=\left(1/2\right)^+$ and Eq.~\eqref{eq:Pwf-3/2} for $k^P=\left(3/2\right)^+$, 
while $\Phi^{Q\bar{Q}}_{(n)}$ denotes the quarkonium wavefunction with quantum numbers $ n \equiv \{n, J, m_j, l, s\}$.\footnote{
The quarkonium wavefunctions are obtained from solving the Schr\"odinger equation with the potential
$$
E_{\Sigma_g^+}(r)=
\begin{cases}
V^{\mathrm{RS}}_s(r, \nu_f),&\, r< R_{\Sigma_g^+} \\
\sigma r+ V_0^{'}, &\, r\ge R_{\Sigma_g^+}
\end{cases},
$$
where we use the RS singlet potential $V^{\mathrm{RS}}_s(r)$ up to order $\alpha^3_s$ in perturbation theory (see, e.g., \cite{Segovia:2018qzb}), 
$\sigma=0.218\,\mathrm{GeV}^2$ is the string tension from the lattice determination in \cite{Brambilla:2022het}, 
and $\nu_f=1$~GeV is the renormalon subtraction scale. 
The constant  $V_0'=-0.651$~GeV and the matching radius $R_{\Sigma_g^+}=0.339$~fm are determined from demanding continuity up to the first derivative.
To compute the energy difference $\Delta E$, we use the spin-averaged quarkonium masses from PDG \cite{ParticleDataGroup:2024cfk}.}

The $P_{c\bar{c}}$ states are superpositions of both heavy quark spin $s=0$ and $s=1$ states for $J^{P}=\left(1/2\right)^-$ and $J^{P}=\left(3/2\right)^-$. 
The non-vanishing of the matrix element~\eqref{eq:matrix_SO-spin} constrains the spin-$0$ component of $P_{c\bar{c}}$ to decay into the spin-$1$ final state $J/\psi$, 
and the spin-$1$ component of $P_{c\bar{c}}$ to decay into the spin-$0$ final state $\eta_c\left(1S\right)$. 
For the spin-$0$ component, the spin-flipping rate in Eq.~\eqref{eq:Gamma_spincons} is multiplied by a factor $3$ corresponding to the $3$ polarizations of $J/\psi$. 
As evident from Eqs.~\eqref{eq:resultscenario11}, \eqref{eq:resultscenario12}, \eqref{eq:resultscenario21}, and \eqref{eq:resultscenario22}, for the spin-$1$ component, both $k^P=\left(1/2\right)^+$ and $k^P=\left(3/2\right)^+$ contribute. 
The results for the semi-inclusive decay rates are shown in Table~\ref{tab:Gamma_exclusive_spin-flipping}.

\begin{table}[t!]
\begin{center}
\small{\renewcommand{\arraystretch}{1.5}
\scriptsize
\begin{minipage}{.49\linewidth}
\begin{tabular}{|c||c|}
\hline
$P_{c\bar{c}}\left[J^{P}\,\right]\,\left(\mathrm{mass}\right)\longrightarrow J/\psi \left[1^{--}\,\right]$ &  $\begin{array}{c}\Gamma \,\,({\rm MeV})\end{array}$\\
\hline
\multicolumn{2}{|c|} {Scenario 1}\\
\hline
$\bm{P_{c\bar{c}}\left[\,\left(1/2\right)^{-}\,\right]\,\left(4312\right)}$    & 31 $^{+11}_{-9}$ $^{+14}_{-7}$   \\

$\bm{P_{c\bar{c}}\left[\,\left(3/2\right)^{-}\,\right]\,\left(4380\right)}$    & 4 $^{+1}_{-1}$ $^{+2}_{-1}$ \\

$\bm{P_{c\bar{c}}\left[\,\left(1/2\right)^{-}\,\right]\,\left(4440\right)}$    & 5 $^{+2}_{-1}$ $^{+2}_{-1}$ \\

$\bm{P_{c\bar{c}}\left[\,\left(3/2\right)^{-}\,\right]\,\left(4457\right)}$    & 4 $^{+1}_{-1}$ $^{+2}_{-1}$ \\

$P_{c\bar{c}}\left[\,\left(1/2\right)^{-}\,\right]\,\left(4507\right)$    & 71 $^{+26}_{-21}$ $^{+32}_{-16}$ \\

$P_{c\bar{c}}\left[\,\left(3/2\right)^{-}\,\right]\,\left(4515\right)$    & 8 $^{+3}_{-2}$ $^{+4}_{-2}$ \\
\hline
\multicolumn{2}{|c|} {Scenario 2}\\
\hline
$\bm{P_{c\bar{c}}\left[\,\left(1/2\right)^{-}\,\right]\,\left(4312\right)}$    & 3 $^{+1}_{-1}$ $^{+1}_{-1}$   \\

$\bm{P_{c\bar{c}}\left[\,\left(3/2\right)^{-}\,\right]\,\left(4380\right)}$    & 16 $^{+6}_{-5}$ $^{+7}_{-4}$ \\

$\bm{P_{c\bar{c}}\left[\,\left(3/2\right)^{-}\,\right]\,\left(4440\right)}$    & 23 $^{+9}_{-7}$ $^{+11}_{-5}$ \\

$\bm{P_{c\bar{c}}\left[\,\left(1/2\right)^{-}\,\right]\,\left(4457\right)}$    & 1.3 $^{+0.5}_{-0.4}$ $^{+0.6}_{-0.3}$ \\

$P_{c\bar{c}}\left[\,\left(3/2\right)^{-}\,\right]\,\left(4521\right)$    & 22 $^{+8}_{-6}$ $^{+10}_{-5}$ \\

$P_{c\bar{c}}\left[\,\left(1/2\right)^{-}\,\right]\,\left(4525\right)$    & 8 $^{+3}_{-2}$ $^{+4}_{-2}$ \\
\hline
\end{tabular}
\end{minipage}
\begin{minipage}{.49\linewidth}
\begin{tabular}{|c||c|}
\hline
$P_{c\bar{c}}\left[J^{P}\,\right]\,\left(\mathrm{mass}\right)\longrightarrow \eta_c\left(1S\right) \left[0^{-+}\,\right]$ &  $\begin{array}{c}\Gamma \,\,({\rm MeV})\end{array}$\\
\hline
\multicolumn{2}{|c|} {Scenario 1}\\
\hline
$\bm{P_{c\bar{c}}\left[\,\left(1/2\right)^{-}\,\right]\,\left(4312\right)}$    & 7 $^{+2}_{-1}$ $^{+2}_{-1}$   \\

$\bm{P_{c\bar{c}}\left[\,\left(3/2\right)^{-}\,\right]\,\left(4380\right)}$    & 17 $^{+5}_{-4}$ $^{+7}_{-3}$ \\

$\bm{P_{c\bar{c}}\left[\,\left(1/2\right)^{-}\,\right]\,\left(4440\right)}$    & 26 $^{+9}_{-7}$ $^{+11}_{-6}$ \\

$\bm{P_{c\bar{c}}\left[\,\left(3/2\right)^{-}\,\right]\,\left(4457\right)}$    & 8 $^{+2}_{-2}$ $^{+3}_{-1}$ \\

$P_{c\bar{c}}\left[\,\left(1/2\right)^{-}\,\right]\,\left(4507\right)$    & 7 $^{+2}_{-2}$ $^{+3}_{-2}$ \\

$P_{c\bar{c}}\left[\,\left(3/2\right)^{-}\,\right]\,\left(4515\right)$    & 16 $^{+6}_{-5}$ $^{+7}_{-4}$ \\

$P_{c\bar{c}}\left[\,\left(5/2\right)^{-}\,\right]\,\left(4526\right)$    & 5 $^{+2}_{-1}$ $^{+2}_{-1}$ \\
\hline
\multicolumn{2}{|c|} {Scenario 2}\\
\hline
$\bm{P_{c\bar{c}}\left[\,\left(1/2\right)^{-}\,\right]\,\left(4312\right)}$    & 15 $^{+5}_{-4}$ $^{+6}_{-3}$   \\

$\bm{P_{c\bar{c}}\left[\,\left(3/2\right)^{-}\,\right]\,\left(4380\right)}$    & 11 $^{+3}_{-3}$ $^{+4}_{-2}$ \\

$\bm{P_{c\bar{c}}\left[\,\left(3/2\right)^{-}\,\right]\,\left(4440\right)}$    & 10 $^{+4}_{-3}$ $^{+5}_{-2}$ \\

$\bm{P_{c\bar{c}}\left[\,\left(1/2\right)^{-}\,\right]\,\left(4457\right)}$    & 7 $^{+2}_{-1}$ $^{+2}_{-1}$ \\

$P_{c\bar{c}}\left[\,\left(5/2\right)^{-}\,\right]\,\left(4513\right)$    & 21 $^{+7}_{-6}$ $^{+9}_{-5}$ \\

$P_{c\bar{c}}\left[\,\left(3/2\right)^{-}\,\right]\,\left(4521\right)$    & 3 $^{+1}_{-1}$ $^{+1}_{-1}$ \\

$P_{c\bar{c}}\left[\,\left(1/2\right)^{-}\,\right]\,\left(4525\right)$    & 3 $^{+1}_{-1}$ $^{+1}_{-0.5}$ \\
\hline
\end{tabular}
\end{minipage}
\caption{Semi-inclusive decay rates of pentaquark states decaying into $J/\psi$ or $\eta_c$: $P_{c\bar{c}}\rightarrow Q_n + X$, with $Q_n=\{J/\psi, \eta_c\}$, where $X$ denotes light hadrons.
The decay rates are computed from Eqs.~\eqref{eq:Gamma_spincons} and \eqref{eq:matrix_SO-spin}.
The pentaquark states are denoted by $P_{c\bar{c}}\left[J^{P}\,\right]\,\left(\mathrm{mass}\right)$, 
where the masses are in MeV. We have used the spin-averaged masses for both pentaquark and quarkonium states to estimate the decay rates. 
For both scenarios 1 and 2 discussed in Sec.~\ref{subsec:identification}, we show the decay rates to $J/\psi$ on the left and decay rates to $\eta_c\left(1S\right)$ on the right.
The first error comes from assuming that the adjoint baryon masses have an uncertainty of the same order as the gluelump masses for hybrids, which is  $\pm\, 0.15\,\mathrm{GeV}$ \cite{Bali:2003jq}. 
The second error bar is from varying the scale of $\alpha_{\rm s}$ from $\Delta E/2$ to $2\,\Delta E$.
For the decay rates shown, $\Delta E \gtrsim 1.3$~GeV  and $\alpha_{\rm s}(\Delta E) \lesssim 0.32$. 
The states in bold are the experimental states shown in Table~\ref{tab:exoticPenta}. 
The pentaquark state  $P_{c\bar{c}}\left[J^{P}=\left(5/2\right)^-\right]$ decays only to $\eta_c\left(1S\right)$.}
\label{tab:Gamma_exclusive_spin-flipping}
}
\end{center}
\end{table}
\normalsize

The sum of the semi-inclusive decay rates to  $J/\psi$ and $\eta_c\left(1S\right)$ in Table~\ref{tab:Gamma_exclusive_spin-flipping} provides a lower bound for the total decay width of the pentaquark state. 
Particularly, for the $P_{c\bar{c}}\left(4312\right)^+$ state, the experimentally measured total width is $10\pm5$~MeV (see Table~\ref{tab:exoticPenta}) \cite{ParticleDataGroup:2024cfk}. 
Our estimate of the lower bound on the total width for $P_{c\bar{c}}\left(4312\right)^+$ is $38^{+18}_{-12}$~MeV in scenario 1 and $18^{+8}_{-5}$~MeV in scenario 2. 
The central value of the estimate from scenario 1 is around $3.8$ times the experimental value, while the estimate from scenario 2 is consistent within errors with the experimental determination -- thus favoring scenario 2. 
Additionally,  we get the ratios of the decay rates to $\eta_c\left(1S\right)$ and $J/\psi$  as
\begin{align}
&\frac{\Gamma\left(P_{c\bar{c}}\left(4312\right)^+\rightarrow \eta_c\left(1S\right) + X\right)}{\Gamma\left(P_{c\bar{c}}\left(4312\right)^+\rightarrow J/\psi + X\right)}= 0.23^{+0.16}_{-0.09}\qquad\mathrm{(scenario\,\,1)},\nonumber\\
&\frac{\Gamma\left(P_{c\bar{c}}\left(4312\right)^+\rightarrow \eta_c\left(1S\right) + X\right)}{\Gamma\left(P_{c\bar{c}}\left(4312\right)^+\rightarrow J/\psi + X\right)}= 5.0^{+3.5}_{-2.9}\qquad\mathrm{(scenario\,\,2)}.
    \label{eq:Pcc4312sce12}
\end{align}
For comparison, in Ref.~\cite{Voloshin:2019aut, Sakai:2019qph} using the heavy quark spin symmetry, the analytic ratio of the exclusive S-wave decay widths for $P_{c\bar{c}}\left(4312\right)^+\rightarrow \eta_c\left(1S\right) + p$ and $P_{c\bar{c}}\left(4312\right)^+\rightarrow J/\psi + p$, where $p$ is the proton, was predicted to be 
\begin{equation}
\frac{\Gamma\left(P_{c\bar{c}}\left(4312\right)^+\rightarrow \eta_c\left(1S\right) + p\right)}{\Gamma\left(P_{c\bar{c}}\left(4312\right)^+\rightarrow J/\psi + p\right)}= 3.0.
    \label{eq:Pcc4312ana}
\end{equation}
Although the central value of our semi-inclusive ratio from scenario 2 in Eq.~\eqref{eq:Pcc4312sce12} is about $1.7$ times the value in Eq.~\eqref{eq:Pcc4312ana}, the two results are compatible within the quoted uncertainties.

For the $P_{c\bar{c}}\left(4380\right)^+$ state, the experimentally measured total width is $210\pm90$~MeV (see Table~\ref{tab:exoticPenta}) \cite{ParticleDataGroup:2024cfk}, which implies that it is a very broad state. 
Our estimate of the lower bound on the total width is $21^{+9}_{-5}$~MeV in scenario 1 and $27^{+10}_{-7}$~MeV in scenario 2 -- both significantly below the central experimental value. 
This discrepancy suggests that the state identified in our analysis might be narrower than the experimentally observed one.
However, a definitive conclusion requires narrowing the large uncertainty affecting the measured width and 
including in the theoretical determination the decay channels to $\Lambda_c\bar{D}^*$ and $\Lambda_c\bar{D}$. 
For the $P_{c\bar{c}}\left(4440\right)^+$ and $P_{c\bar{c}}\left(4457\right)^+$ states, the experimentally measured total widths are $21^{+10}_{-11}$~MeV and $6.4^{+6.0}_{-2.8}$~MeV, respectively (see Table~\ref{tab:exoticPenta}) \cite{ParticleDataGroup:2024cfk}. 
Our estimates of the lower bounds in both scenarios are consistent with the experimental determinations within errors.

For the pentaquark state with $J^{P}=\left(5/2\right)^-$, $P_{c\bar{c}}\left(4526\right)^+$ in scenario 1 or $P_{c\bar{c}}\left(4513\right)^+$ in scenario 2, the decay proceeds only to an $\eta_c\left(1S\right)$ final state, as shown in Table~\ref{tab:Gamma_exclusive_spin-flipping}. 
This is because the state has a well-defined heavy-quark spin configuration $s=1$.

\subsection{Ratios of decays to $\Lambda_c\bar{D}$ and $\Lambda_c\bar{D}^*$.}
\label{subsec:branching ratio}
In this section, our aim is to compute the ratios of the pentaquark decays to the $\Lambda_c\bar{D}$ and $\Lambda_c\bar{D}^*$ thresholds: $P_{c\bar{c}}\rightarrow \Lambda_c\bar{D}$ and $P_{c\bar{c}}\rightarrow \Lambda_c\bar{D}^*$. 
In the BO formalism, these decays proceed through a coupling or mixing potential that  arises between BO potentials sharing the same quantum numbers.
In our current work, the pentaquark BO potentials can mix or couple with BO potentials that asymptotically connect to the spin-isospin averaged $\Lambda_c\bar{D}$ threshold -- provided they share the same BO quantum numbers.  
The pentaquark BO quantum numbers are $\left(1/2\right)_g$ and $\{\left(1/2\right)_g^{\prime}, \left(3/2\right)_g\}$ corresponding to adjoint baryons $k^P=\left(1/2\right)^+$ and $k^P=\left(3/2\right)^+$, respectively, as shown in Table~\ref{tab:QQbarpenta}. 
The $\Lambda_c\bar{D}$ threshold has BO quantum number $\left(1/2\right)_g$ corresponding to total light-quark spin parity $k^P=\left(1/2\right)^+$ with $k^P_{qq}=0^+$ (first row in Table~\ref{tab:qqq}). 
Consequently, relevant for the present case is the mixing of the pentaquark BO potentials with quantum numbers $\left(1/2\right)_g$ and $\left(1/2\right)_g^{\prime}$ with the  BO potential $\left(1/2\right)_g$ connecting with the $\Lambda_c\bar{D}$ and $\Lambda_c\bar{D}^*$ thresholds.
Including this mixing modifies the radial Schr\"odinger equations~\eqref{eq:Sch12} and \eqref{eq:Sch32}, with the mixing potentials appearing as off-diagonal terms in the potential matrix. 
Following Ref.~\cite{Berwein:2024ztx}, the Schr\"odinger equation \eqref{eq:Sch12} gets modified into the two-channel equation 
\begin{align}
 &\hspace{-2.0 cm}\left[-\frac{1}{m_Q r^2}\,\partial_rr^2\partial_r+\frac{1}{m_Qr^2}\begin{pmatrix} \left(l-\frac{1}{2}\right)\left(l+\frac{1}{2}\right) & 0 \\ 0 & \left(l-\frac{1}{2}\right)\left(l+\frac{1}{2}\right) \end{pmatrix} \right.\nonumber\\
	&\hspace{3.0 cm}+\left. \begin{pmatrix} E_{(1/2)_g} & V_1^{\rm mix} \\ V_1^{\rm mix} & E^\Lambda_{(1/2)_g} \end{pmatrix}\right]\hspace{-2pt}\begin{pmatrix} \psi_{\left(1/2\right)}^{(N)} \\ \psi_{\Lambda}^{(N)}\end{pmatrix}
	=\mathcal{E}_{1/2}\begin{pmatrix} \psi_{1/2, \sigma_P}^{(N)} \\ \psi_{\Lambda}^{(N)}\end{pmatrix}\,,
	\label{eq:Sch12-2}
\end{align}
and the Schr\"odinger equation \eqref{eq:Sch32} gets modified into the three-channel equation
\begin{align}
 &\hspace{-0.5 cm}\left[-\frac{1}{m_Q r^2}\,\partial_rr^2\partial_r+\frac{1}{m_Qr^2}\begin{pmatrix} l(l-1)+\frac{9}{4} & -\sqrt{3l(l+1)-\frac{9}{4}} & 0\\ -\sqrt{3l(l+1)-\frac{9}{4}} & l(l+1)-\frac{3}{4} & 0\\ 0 & 0 & \left(l+\frac{1}{2}\right)\left(l+\frac{3}{2}\right)\end{pmatrix} \right. \nonumber\\
	&\hspace{3.0 cm}+\left. \begin{pmatrix} E_{(1/2)_g^{\prime}} & 0 & V_2^{\rm mix}\\ 0 & E_{(3/2)_g} & 0 \\ V_2^{\rm mix} & 0 & E^\Lambda_{(1/2)_g} \end{pmatrix}\right]\hspace{-2pt}\begin{pmatrix} \psi_{\left(1/2\right)^{+\prime}}^{(N)} \\ \psi_{\left(3/2\right)^+}^{(N)}\\ \psi_{\Lambda}^{(N)}\end{pmatrix}
	=\mathcal{E}_{3/2}\begin{pmatrix} \psi_{\left(1/2\right)^{+\prime}}^{(N)} \\ \psi_{\left(3/2\right)^+}^{(N)} \\ \psi_{\Lambda}^{(N)}\end{pmatrix}\,,
	\label{eq:Sch32-2}
\end{align}
where $E^{\Lambda}_{\left(1/2\right)_g}$ is the BO potential that asymptotically connects to the spin-isospin-averaged $\Lambda_c/\Lambda_b-\bar{D}/\bar{B}$ threshold from above, as shown in Fig.~\ref{fig:illustration}, without supporting bound states. 
The mixing is unknown, since it has not yet been measured in lattice QCD.
We assume that it is much smaller than the energy gap of order $200$~MeV that separates the spin-isospin-averaged $\Sigma_c/\Sigma_b-\bar{D}/\bar{B}$ and $\Lambda_c/\Lambda_b-\bar{D}/\bar{B}$ thresholds. 
Under this assumption, the mixing has only a minimal effect on the spectrum obtained from the Schr\"odinger equations without mixing~\eqref{eq:Sch12} and \eqref{eq:Sch32}. 
The decay widths of pentaquark states to (spin-isospin-averaged) $\Lambda_c\bar{D}$ thresholds, which proceeds through the mixing potentials, can be obtained from solving the coupled-channel Schr\"odinger equations \eqref{eq:Sch12-2} and \eqref{eq:Sch32-2}. 
However due to the lack of lattice QCD information on the mixing potentials $V_1^{\rm mix}$ and $V_2^{\rm mix}$, 
we can only provide analytical results for the ratios of the decay widths.

The pentaquark states $P_{c\bar{c}}$ with quantum numbers $J^{P}$ in scenario 1 are listed in Table~\ref{tab:scenario1} and those in scenario 2 are listed in Table~\ref{tab:scenario2}. 
Their expressions in terms of the multiplets listed in Table~\ref{tab:QQbarpenta}, characterized by a well-defined heavy-quark-antiquark spin $s$ and LDF angular momentum $k$, are shown in Eqs.~\eqref{eq:resultscenario11}-\eqref{eq:resultscenario13} and \eqref{eq:resultscenario21}-\eqref{eq:resultscenario23}.  
The S-wave $\Lambda_c\bar{D}$ threshold has $J^{P}=\left(1/2\right)^-$, while the S-wave $\Lambda_c\bar{D}^*$ threshold has $J^{P}=\{\left(1/2\right)^-, \left(3/2\right)^-\}$. 
Therefore, for the pentaquark states $P_{c\bar{c}}\left[J^{P}=\left(1/2\right)^-\right]$ in both scenarios 1 and 2, the decay to S-wave $\Lambda_c\bar{D}$ and $\Lambda_c\bar{D}^*$ is allowed, while, for $P_{c\bar{c}}\left[J^{P}=\left(3/2\right)^-\right]$ in both scenarios 1 and 2, only the decay to S-wave $\Lambda_c\bar{D}^*$ is allowed, while the decay to $\Lambda_c\bar{D}$ requires a D-wave transition. 
Similarly, the pentaquark states $P_{c\bar{c}}\left[J^{P}=\left(5/2\right)^-\right]$ in both scenarios 1 and 2 can decay to both $\Lambda_c\bar{D}$ and $\Lambda_c\bar{D}^*$ only through D-wave transitions. 
Since D-wave decays involve higher angular momentum barriers, they are expected to be suppressed compared to S-wave decays.

Following Ref.~\cite{Braaten:2024stn}, the mixing or coupling potential governing the transitions $P_{c\bar{c}}\left[J^{P}\right]\rightarrow \Lambda_c\bar{D}, \Lambda_c\bar{D}^*$ can be written as
\begin{multline}
V^{J^{P}}_{s,l,\eta}\Bigl(k^P, L_Q \to \bigl[\bigl(\tfrac{1}{2}^+, 0^+\bigr) J_1, \bigl(\tfrac{1}{2}^-, \tfrac{1}{2}^+\bigr) J_2\bigr] J^{\prime}, L_Q^\prime\Bigr) = \sqrt{2}\,(-1)^{1/2+s + L_Q^\prime + J} \sqrt{\tilde{s} \tilde{l}\tilde{J}_1\tilde{J}_2 \tilde{J^{\prime}} } \\
\times 
\begin{Bmatrix}
s  & \tfrac{1}{2} & J^{\prime} \\
L_Q^\prime & J & l \\
\end{Bmatrix}
\begin{Bmatrix}
\frac{1}{2} & \frac{1}{2} & s \\
0 & \tfrac{1}{2} & \tfrac{1}{2} \\
J_1 & J_2 & J^{\prime} \\
\end{Bmatrix}
G_{l,\eta}^{P}\bigl(k^P, L_Q \to  \tfrac{1}{2}^+, L_Q^\prime\bigr),
\label{eq:mixpot}
\end{multline}
where $J_1$ and $J_2$ are the spins of the  $\Lambda_c$ baryon and $\bar{D}$ or $\bar{D}^*$ mesons respectively, $J_1=1/2$ for $\Lambda_c$ and $J_2=0$ or $1$ for $\bar{D}$ or $\bar{D}^*$, $J^{\prime}$ is the quantum number of ${\bm J}^{\prime} = {\bm J}_1+{\bm J}_2$, whose values are $J^{\prime}=1/2$ for $\Lambda_c\bar{D}$ and $J^{\prime}=\{1/2, 3/2\}$ for $\Lambda_c\bar{D}^*$, 
$L_Q^{\prime}$ is the quantum number denoting the relative angular momentum between the baryon $\Lambda_c$ and the meson $\bar{D}$ or $\bar{D}^*$, whose values are $L_Q^{\prime} =0$ for S-wave and $L_Q^{\prime} =2$ for D-wave, and $\tilde{J}=2J+1$. 
The curly brackets  $\left\{\begin{smallmatrix} j_1 & j_2 & j_3 \\ j_4 & j_5 & j_6\end{smallmatrix}\right\}$ and $\left\{\begin{smallmatrix} j_1 & j_2 & j_3 \\ j_4 & j_5 & j_6 \\ j_7 & j_8 & j_9 \end{smallmatrix}\right\}$ are the Wigner 6-$j$ symbol and Wigner 9-$j$ symbol, respectively. 
The quantum numbers in the parenthesis and square brackets on the left-hand side of Eq.~\eqref{eq:mixpot} have been summed over and the resulting quantum numbers are written to the right.\footnote{
In the notation $\bigl[\bigl(\tfrac{1}{2}^+, 0^+\bigr) J_1, \bigl(\tfrac{1}{2}^-, \tfrac{1}{2}^+\bigr) J_2\bigr] J^{\prime}$ appearing in Eq.~\eqref{eq:mixpot}, 
the first parenthesis identifies the $\Lambda_c$, whose heavy quark spin $1/2$ and light quark spin $0$ have been summed to give the total spin $J_1$, which is written to the right of the parenthesis.  
Similarly, the second parenthesis identifies the $\bar{D}$ or $\bar{D}^*$, whose heavy quark spin $1/2$ and light quark spin $1/2$ have been summed to give the total spin $J_2$, which is written to the right of the parenthesis. 
The sum of $J_1$ and $J_2$ gives $J^{\prime}$, which is the quantum number of the S-wave $\Lambda_c\bar{D}$ or $\Lambda_c\bar{D}^*$ system written to the right of the square bracket.}
The superscript on $V$ in Eq.~\eqref{eq:mixpot} are quantum numbers $J^{P}$ which is exactly conserved between the pentaquark states $P_{c\bar{c}}$ and $\Lambda_c\bar{D}$ or $\Lambda_c\bar{D}^*$ thresholds. 
The transition amplitude $G^{P}_{l,\eta}$, which depends on the combined angular momentum $l$, can be expressed as
\begin{multline}
G_{l,\eta}^{P}\bigl(k^P, L_Q \to (0^+, \tfrac{1}{2}^+) \tfrac{1}{2}, L_Q^\prime\bigr) =
(-1)^{L_Q + L_Q^\prime} \\
\times\sum_{\lambda=\mp1/2}
\cg{k}{\lambda}{l}{-\lambda}{L_Q}{0}
\cg{\tfrac{1}{2}}{\lambda}{l}{-\lambda}{L_Q^\prime}{0}
g_{\lambda,\eta}\bigl(k^P \to (0^+, \tfrac{1}{2}^+)\bigr),
\label{eq:mixpot1}
\end{multline}
where $\cg{j1}{m1}{j2}{m2}{j}{m}$ denotes Clebsch--Gordon coefficients, $\lambda$ is the eigenvalue of ${\bm K}\cdot\hat{{\bm r}}$ that for $\Lambda_c\bar{D}$ or $\Lambda_c\bar{D}^*$ is $\lambda=\pm 1/2$, corresponding to total light-quark spin parity $\left(1/2\right)^+$. 
The quantity $g_{\lambda,\eta}$ is the transition amplitude between the light quark state in the pentaquark and the spin-isospin averaged  $\Lambda_c\bar{D}$ threshold in the static limit.  
This quantity can, in principle, be computed in lattice QCD, although it is currently unknown.\footnote{
The transition rate has been computed only in the case of string breaking (avoided crossing) for the case of quarkonium and lowest meson-antimeson thresholds in lattice QCD in Ref.~\cite{Bali:2005fu, Bulava:2024jpj}.} 
Since $g_{\lambda, \eta}$ depends solely on the light-quark states in the pentaquark and the spin-isospin averaged  $\Lambda_c\bar{D}$ threshold, it is the same for all pentaquark states with the same light-quark quantum numbers $k^P$.
Using Eq.~\eqref{eq:mixpot1}, we get for the S-wave transitions:
\begin{align}
 G_{\tfrac{1}{2},+}^{-}\bigl(\tfrac{1}{2}^+, 0 \to (0^+, \tfrac{1}{2}^+) \tfrac{1}{2}, 0\bigr)&=\frac{1}{2}\left[g_{1/2,+}\bigl(\tfrac{1}{2}^+ \to (0^+, \tfrac{1}{2}^+)\bigr)+g_{-1/2,+}\bigl(\tfrac{1}{2}^+ \to (0^+, \tfrac{1}{2}^+)\bigr)\right],\nonumber\\
 G_{\tfrac{3}{2},+}^{-}\bigl(\tfrac{3}{2}^+, 0 \to (0^+, \tfrac{1}{2}^+) \tfrac{1}{2}, 0\bigr)&=0.
 \label{eq:mixpot2}
\end{align}
This result implies that, among the components in Eqs.~\eqref{eq:resultscenario11}–\eqref{eq:resultscenario13} and \eqref{eq:resultscenario21}–\eqref{eq:resultscenario23}, those involving LDF with $k=3/2$, i.e. $|s=1, k=3/2 \rangle_{J^{P}=\left(1/2\right)^-}$, $|s=0, k=3/2 \rangle_{J^{P}=\left(3/2\right)^-}$ and so on, do not contribute to the S-wave decays of $P_{c\bar{c}}$ states to $\Lambda_c\bar{D}$ and $\Lambda_c\bar{D}^*$. 
A direct consequence is that $\Gamma\left[P_{c\bar{c}}\left[J^{P}=\left(5/2\right)^-\right]\rightarrow \Lambda_c\bar{D}, \Lambda_c\bar{D}^*\right]=0$ for S-wave transitions, 
in agreement with the earlier discussion that such decays proceed only via D-wave and are thus suppressed. 

Only the transition amplitude $G^-_{1/2, +}$ in Eq.~\eqref{eq:mixpot2} contributes to the decays of $P_{c\bar{c}}$ states to S-wave $\Lambda_c\bar{D}$ and $\Lambda_c\bar{D}^*$ states. 
Therefore, it cancels out in the ratio of decay rates. 
Following Ref.~\cite{Braaten:2024stn}, the ratio of decay widths for the pentaquark states $P_{c\bar{c}}$ with $J^{P}=\left(1/2\right)^-$ is given by
\begin{align}
 \frac{\Gamma\left(P_{c\bar{c}}\left[J^{P}=\left(1/2\right)^-\right]\rightarrow \Lambda_c\bar{D}\right)/v_{\Lambda_c\bar{D}}}{\Gamma\left(P_{c\bar{c}}\left[J^{P}=\left(1/2\right)^-\right]\rightarrow \Lambda_c\bar{D}^*\right)/v_{\Lambda_c\bar{D}^*}}=\frac{\bigg|a\,V^{\tfrac{1}{2}^-}_{0,\tfrac{1}{2},+}\left(\Lambda_c\bar{D}; J^{\prime}=\tfrac{1}{2}\right)+ b\,V^{\tfrac{1}{2}^-}_{1,\tfrac{1}{2},+}\left(\Lambda_c\bar{D}; J^{\prime}=\tfrac{1}{2}\right)\bigg|^2}{\displaystyle\sum_{J^{\prime}=1/2}^{3/2}\bigg|a\,V^{\tfrac{1}{2}^-}_{0,\tfrac{1}{2},+}\left(\Lambda_c\bar{D}^*; J^{\prime}\right)+ b\,V^{\tfrac{1}{2}^-}_{1,\tfrac{1}{2},+}\left(\Lambda_c\bar{D}^*; J^{\prime}\right)\bigg|^2},
 \label{eq:Gamma1/2}
\end{align}
where the phase space factor for each final state channel with orbital angular momentum $L_Q^\prime$ is proportional to $v_{\alpha}^{2 L_Q^\prime + 1}$, $\alpha$ being either $\Lambda_c\bar{D}^*$ or $\Lambda_c\bar{D}$ and $v_\alpha$ the velocity of each of the outgoing heavy hadrons in the center-of-mass frame. 
The velocity $v_\alpha$ is determined through energy and momentum conservation by the masses of the initial pentaquark $P_{c\bar{c}}$, final baryon $\Lambda_c$ and meson $\bar{D}^*$ or $\bar{D}$. 
The factors $a$ and $b$ in Eq.~\eqref{eq:Gamma1/2} are the elements of the $3\times3$ matrices in Eqs.~\eqref{eq:resultscenario11} and \eqref{eq:resultscenario21}: $a = {}_{J^{P}=\left(1/2\right)^-}\langle s=0, k=1/2 {\big |} P_{c\bar{c}}\left[J^{P}=\left(1/2\right)^-\right]\rangle$ and $b = {}_{J^{P}=\left(1/2\right)^-}\langle s=1, k=1/2 {\big |}P_{c\bar{c}}\left[J^{P}=\left(1/2\right)^-\right]\rangle$. 
For simplicity, in Eq.~\eqref{eq:Gamma1/2}, we have used a compact notation for the coupling potential in Eq.~\eqref{eq:mixpot}:
\begin{align}
 V^{\tfrac{1}{2}^-}_{0,\tfrac{1}{2},+}\left(\Lambda_c\bar{D}^*; J^{\prime}\right) &= V^{\tfrac{1}{2}^-}_{0,\tfrac{1}{2},+}\Bigl(\tfrac{1}{2}^+, 0 \to \bigl[\bigl(\tfrac{1}{2}^+, 0^+\bigr)\frac{1}{2}, \bigl(\tfrac{1}{2}^-, \tfrac{1}{2}^+\bigr) 1\bigr] J^{\prime}, 0\Bigr),\nonumber\\
 V^{\tfrac{1}{2}^-}_{1,\tfrac{1}{2},+}\left(\Lambda_c\bar{D}^*; J^{\prime}\right) &= V^{\tfrac{1}{2}^-}_{1,\tfrac{1}{2},+}\Bigl(\tfrac{1}{2}^+, 0 \to \bigl[\bigl(\tfrac{1}{2}^+, 0^+\bigr)\frac{1}{2}, \bigl(\tfrac{1}{2}^-, \tfrac{1}{2}^+\bigr) 1\bigr] J^{\prime}, 0\Bigr),\nonumber\\
 V^{\tfrac{1}{2}^-}_{0,\tfrac{1}{2},+}\left(\Lambda_c\bar{D}; J^{\prime}=\tfrac{1}{2}\right)&=V^{\tfrac{1}{2}^-}_{0,\tfrac{1}{2},+}\Bigl(\tfrac{1}{2}^+, 0 \to \bigl[\bigl(\tfrac{1}{2}^+, 0^+\bigr)\frac{1}{2}, \bigl(\tfrac{1}{2}^-, \tfrac{1}{2}^+\bigr) 0\bigr] \tfrac{1}{2}, 0\Bigr).
\end{align}
Similarly, the ratio of decay widths for the pentaquark states $P_{c\bar{c}}$ with $J^{P}=\left(3/2\right)^-$ can be obtained by setting $a=0$ and $b = {}_{J^{P}=\left(3/2\right)^-}\langle s=1, k=1/2 {\big |} P_{c\bar{c}}\left[J^{P}=\left(3/2\right)^-\right]\rangle$ in Eq.~\eqref{eq:Gamma1/2}, the matrix elements given by Eqs.~\eqref{eq:resultscenario12} and \eqref{eq:resultscenario22}, and substituting $V^{\tfrac{1}{2}^-}_{1,\tfrac{1}{2},+}\left(\Lambda_c\bar{D}; J^{\prime}=\tfrac{1}{2}\right)$ with $V^{\tfrac{3}{2}^-}_{1,\tfrac{1}{2},+}\left(\Lambda_c\bar{D}; J^{\prime}=\tfrac{1}{2}\right)$ and $V^{\tfrac{1}{2}^-}_{1,\tfrac{1}{2},+}\left(\Lambda_c\bar{D}^*; J^{\prime}\right)$ with $V^{\tfrac{3}{2}^-}_{1,\tfrac{1}{2},+}\left(\Lambda_c\bar{D}^*; J^{\prime}\right)$.

\begin{figure}[ht]
\begin{center}
\includegraphics[width=0.7\textwidth,angle=0,clip]{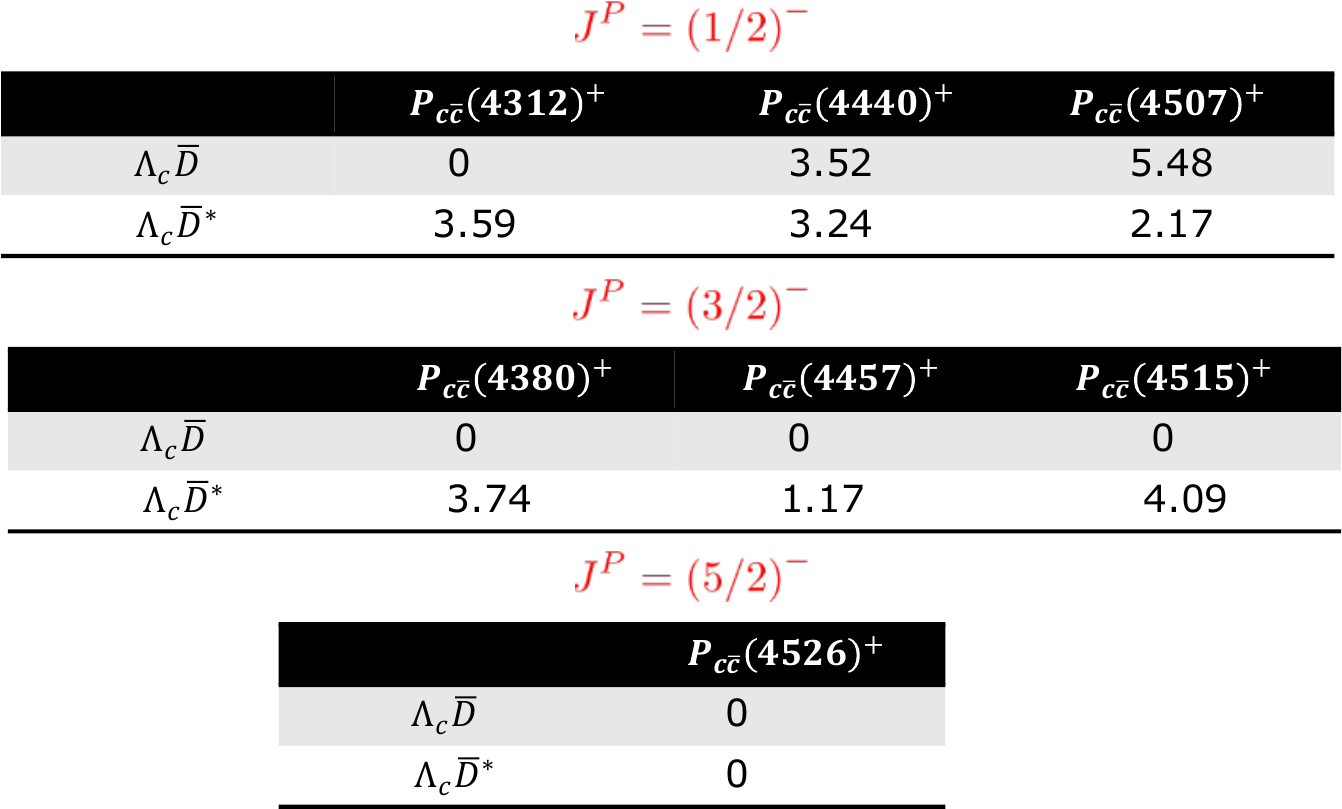}
\caption{Scenario 1: The rows marked $\Lambda_c\bar{D}$ show $\bigg|a\,V^{\tfrac{1}{2}^-}_{0,\tfrac{1}{2},+}\left(\Lambda_c\bar{D}; J^{\prime}=\tfrac{1}{2}\right)+ b\,V^{\tfrac{1}{2}^-}_{1,\tfrac{1}{2},+}\left(\Lambda_c\bar{D}; J^{\prime}=\tfrac{1}{2}\right)\bigg|^2$ $/G_{\tfrac{1}{2},+}^{-}\bigl(\tfrac{1}{2}^+, 0 \to (0^+, \tfrac{1}{2}^+) \tfrac{1}{2}, 0\bigr)^2$ and the rows marked $\Lambda_c\bar{D}^*$ show $\displaystyle\sum_{J^{\prime}=1/2}^{3/2}\bigg|a\,V^{\tfrac{1}{2}^-}_{0,\tfrac{1}{2},+}\left(\Lambda_c\bar{D}^*; J^{\prime}\right)+ b\,V^{\tfrac{1}{2}^-}_{1,\tfrac{1}{2},+}\left(\Lambda_c\bar{D}^*; J^{\prime}\right)\bigg|^2$ $/ G_{\tfrac{1}{2},+}^{-}\bigl(\tfrac{1}{2}^+, 0 \to (0^+, \tfrac{1}{2}^+) \tfrac{1}{2}, 0\bigr)^2$, i.e. the numerator and denominator of the ratio in the right-hand side of Eq.~\eqref{eq:Gamma1/2}, for the different states listed in the columns.
The ratios of the different entries provide all the decay widths ratios that follow from Eq.~\eqref{eq:Gamma1/2}.}
\label{fig:decayLambdacD-1}
\end{center}
\end{figure}

\begin{figure}[ht]
\begin{center}
\includegraphics[width=0.7\textwidth,angle=0,clip]{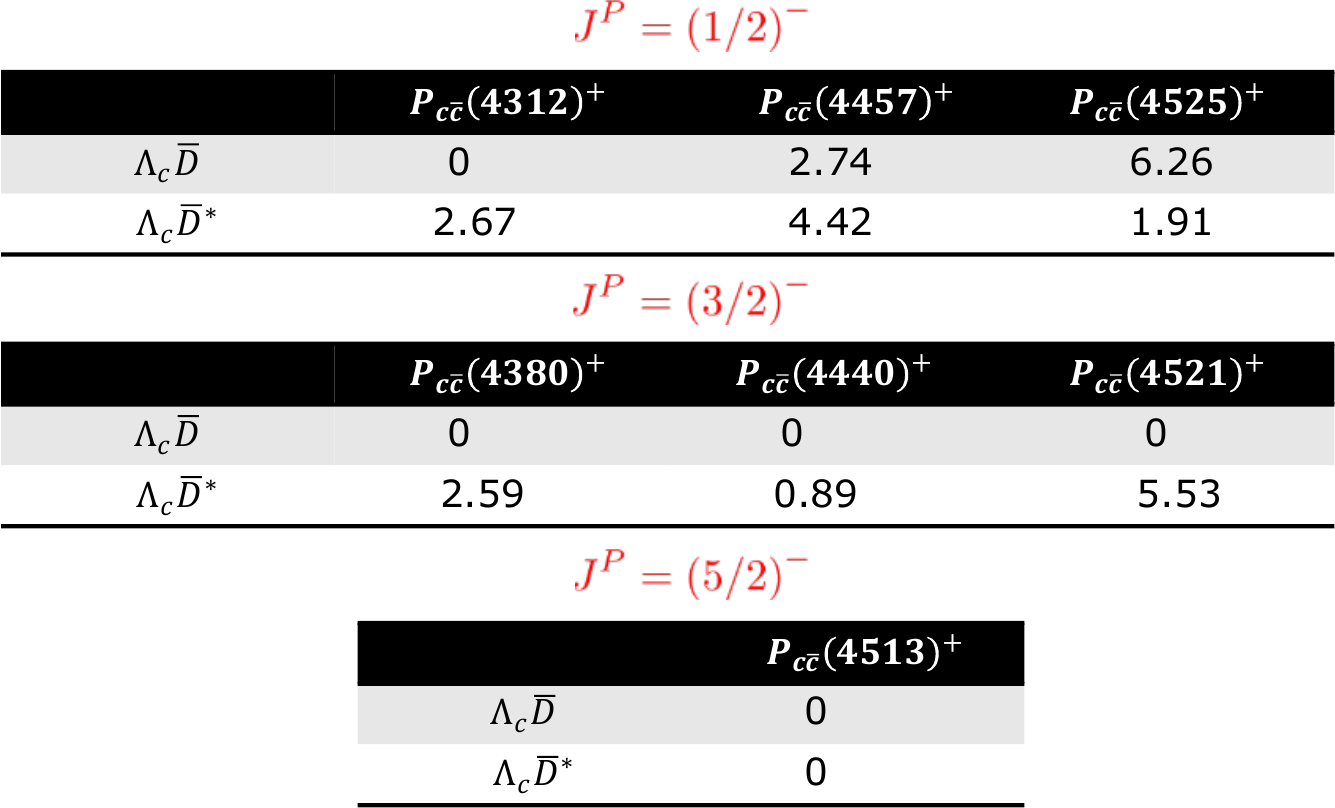}
\caption{Scenario 2: Same as in Fig.~\ref{fig:decayLambdacD-1}.}
\label{fig:decayLambdacD-2}
\end{center}
\end{figure}

The results for the numerator and denominator of the ratio in the right-hand side of Eq.~\eqref{eq:Gamma1/2} for different 
pentaquarks $P_{c\bar{c}}$ decaying into S-wave $\Lambda_c\bar{D}$ and $\Lambda_c\bar{D}^*$ are shown in Fig.~\ref{fig:decayLambdacD-1} for scenario 1 and in Fig.~\ref{fig:decayLambdacD-2} for scenario 2. 
For a given pentaquark state, $P_{c\bar{c}}$, from the entries in Figs.~\ref{fig:decayLambdacD-1} and \ref{fig:decayLambdacD-2}, it follows that 
\begin{align}
&\frac{\Gamma^{(1)}\left(P_{c\bar{c}}\left(4440\right)^+\rightarrow \Lambda_c\bar{D}\right)/v_{\Lambda_c\bar{D}}}{\Gamma^{(1)}\left(P_{c\bar{c}}\left(4440\right)^+\rightarrow \Lambda_c\bar{D}^*\right)/v_{\Lambda_c\bar{D}^*}}= 1.09,\qquad\frac{\Gamma^{(2)}\left(P_{c\bar{c}}\left(4440\right)^+\rightarrow \Lambda_c\bar{D}\right)/v_{\Lambda_c\bar{D}}}{\Gamma^{(2)}\left(P_{c\bar{c}}\left(4440\right)^+\rightarrow \Lambda_c\bar{D}^*\right)/v_{\Lambda_c\bar{D}^*}}= 0,\nonumber\\
&\frac{\Gamma^{(1)}\left(P_{c\bar{c}}\left(4457\right)^+\rightarrow \Lambda_c\bar{D}\right)/v_{\Lambda_c\bar{D}}}{\Gamma^{(1)}\left(P_{c\bar{c}}\left(4457\right)^+\rightarrow \Lambda_c\bar{D}^*\right)/v_{\Lambda_c\bar{D}^*}}=0,\qquad \frac{\Gamma^{(2)}\left(P_{c\bar{c}}\left(4457\right)^+\rightarrow \Lambda_c\bar{D}\right)/v_{\Lambda_c\bar{D}}}{\Gamma^{(2)}\left(P_{c\bar{c}}\left(4457\right)^+\rightarrow \Lambda_c\bar{D}^*\right)/v_{\Lambda_c\bar{D}^*}}= 0.62,\nonumber\\
&\frac{\Gamma^{(1)}\left(P_{c\bar{c}}\left(4507\right)^+\rightarrow \Lambda_c\bar{D}\right)/v_{\Lambda_c\bar{D}}}{\Gamma^{(1)}\left(P_{c\bar{c}}\left(4507\right)^+\rightarrow \Lambda_c\bar{D}^*\right)/v_{\Lambda_c\bar{D}^*}}= 2.53,\qquad \frac{\Gamma^{(2)}\left(P_{c\bar{c}}\left(4525\right)^+\rightarrow \Lambda_c\bar{D}\right)/v_{\Lambda_c\bar{D}}}{\Gamma^{(2)}\left(P_{c\bar{c}}\left(4525\right)^+\rightarrow \Lambda_c\bar{D}^*\right)/v_{\Lambda_c\bar{D}^*}}= 3.28,
\label{eq:Br}
\end{align}
where the superscripts $(1)$ or $(2)$ denote results in scenario 1 and 2, respectively. 
We recall that the state $P_{c\bar{c}}\left(4440\right)^+$ is assigned $J^{P}=\left(1/2\right)^-$ in scenario 1, while in scenario 2 it has $J^{P}=\left(3/2\right)^-$; 
the reverse holds for $P_{c\bar{c}}\left(4457\right)^+$. 
The state  $P_{c\bar{c}}\left(4507\right)^+$ in scenario 1 and the state $P_{c\bar{c}}\left(4525\right)^+$ in scenario 2 are close to the $\Sigma_c^*\bar{D}^*$ threshold with $J^{P}=\left(1/2\right)^-$.

In Ref.~\cite{Voloshin:2019aut}, considering that the states $P_{c\bar{c}}\left(4440\right)^+$ or $P_{c\bar{c}}\left(4457\right)^+$ are exactly at the $\Sigma_c\bar{D}^*$ threshold with quantum numbers  $J^{P}=\left(1/2\right)^-$ (scenario 0 in Sec.~\ref{subsec:identification}), the analytically predicted ratio of S-wave decay rates turns out to be 
\begin{equation}
    \frac{\Gamma\left(P_{c\bar{c}}\left[J^{P}=\left(1/2\right)^-\right]\rightarrow \Lambda_c\bar{D}\right)/v_{\Lambda_c\bar{D}}}{\Gamma\left(P_{c\bar{c}}\left[J^{P}=\left(1/2\right)^-\right]\rightarrow \Lambda_c\bar{D}^*\right)/v_{\Lambda_c\bar{D}^*}}= \frac{3}{4}.
    \label{eq:VoloshinPc1}
\end{equation}
Our results in~\eqref{eq:Br} yield a ratio of $1.09$ in scenario 1 and $0.62$ in scenario 2. 
The result of Eq.~\eqref{eq:VoloshinPc1} is therefore closer to the one of scenario 2, where the $P_{c\bar{c}}\left(4440\right)^+$ and $P_{c\bar{c}}\left(4457\right)^+$  are assigned with $J^{P}=\left(3/2\right)^-$ and $J^{P}=\left(1/2\right)^-$, respectively. 

Considering scenario 2 in Fig.~\ref{fig:decayLambdacD-2}, for $J^{P}=\left(1/2\right)^-$ states, we observe that the $P_{c\bar{c}}\left(4312\right)^+$ only decays into $\Lambda_c\bar{D}^*$, which is consistent with the prediction from the molecular picture~\cite{Voloshin:2019aut}.\footnote{
In the molecular picture, the $P_{c\bar{c}}\left(4312\right)^+$ is a $\Sigma_c\bar{D}$ molecule. 
There is no $DD\pi$ vertex to account for the transition to $\Lambda_c\bar{D}$~\cite{Voloshin:2019aut}.} 
For $P_{c\bar{c}}\left(4457\right)^+$, the relative partial decay width to a $\Lambda_c\bar{D}^*$ final state is about $1.66$ times larger than that of $P_{c\bar{c}}\left(4312\right)^+$, 
which implies that $P_{c\bar{c}}\left(4457\right)^+$ exhibits a slightly broader width into the $\Lambda_c\bar{D}^*$ channel, consistent with its higher mass and larger available phase space.\footnote{
The total S-wave width of $P_{c\bar{c}}\left(4457\right)^+$ is around $2.7$ times larger than the total S-wave width of $P_{c\bar{c}}\left(4312\right)^+$. 
However, their measured total widths are similar (see Table~\ref{tab:exoticPenta}).}  
For $P_{c\bar{c}}\left(4525\right)^+$, the dominant S-wave decay is to $\Lambda_c\bar{D}$, which is $2.29$ times larger than that of $P_{c\bar{c}}\left(4457\right)^+$ into the same channel. 
The large total S-wave width suggests that the $P_{c\bar{c}}\left(4525\right)^+$ is likely a relatively broad resonance compared to lower-lying states such as $P_{c\bar{c}}\left(4312\right)^+$ and $P_{c\bar{c}}\left(4457\right)^+$. 
Also, accounting for the decays of $P_{c\bar{c}}\left(4525\right)^+$ to $J/\psi$ and $\eta_c\left(1S\right)$ from Table~\ref{tab:Gamma_exclusive_spin-flipping}, the broad width may explain why this state has not yet been observed experimentally.  
 
For $J^{P}=\left(3/2\right)^-$ states, we observe that all the states only decay to S-wave $\Lambda_c\bar{D}^*$, as discussed at the beginning of this section. 
For $P_{c\bar{c}}\left(4380\right)^+$, the relative partial decay rate for decays into $\Lambda_c\bar{D}^*$ is of the same order  as that of $P_{c\bar{c}}\left(4312\right)^+$ ($2.59$ vs $2.67$ in Fig.~\ref{fig:decayLambdacD-2}), which suggests that $P_{c\bar{c}}\left(4380\right)^+$ has a total width comparable to the one of $P_{c\bar{c}}\left(4312\right)^+$. 
Also including decays to $J/\psi$ and $\eta_c\left(1S\right)$ from Table~\ref{tab:Gamma_exclusive_spin-flipping}, our analysis suggests that this state is narrower than the broad structure seen in experiments, in agreement with Refs.~\cite{Du:2019pij, Du:2021fmf}. 
For $P_{c\bar{c}}\left(4440\right)^+$, the relative partial decay rate for decays into $\Lambda_c\bar{D}^*$ is about $3$ times smaller than that of  $P_{c\bar{c}}\left(4380\right)^+$. 
Moreover, from Table~\ref{tab:Gamma_exclusive_spin-flipping}, we see that this state may have significant decay modes to $J/\psi$ and $\eta_c\left(1S\right)$.
For $P_{c\bar{c}}\left(4521\right)^+$, the relative partial decay rate for decays into $\Lambda_c\bar{D}^*$  is $2.14$ times larger than that of $P_{c\bar{c}}\left(4380\right)^+$, 
which suggests that this state is likely a relatively broad resonance compared to lower-lying states such as $P_{c\bar{c}}\left(4380\right)^+$ and $P_{c\bar{c}}\left(4440\right)^+$. 
Also accounting for the decays of $P_{c\bar{c}}\left(4525\right)^+$ to $J/\psi$ and $\eta_c\left(1S\right)$ from Table~\ref{tab:Gamma_exclusive_spin-flipping}, the broad nature of the resonance could be the reason why this state has not been seen in experiments.

The $J^{P}=\left(5/2\right)^-$ state $P_{c\bar{c}}\left(4513\right)^+$ can decay into $\Lambda_c\bar{D}$ and $\Lambda_c\bar{D}^*$ only via D-wave transitions, 
which are suppressed compared to S-wave decays, 
indicating that the state may have a narrower width compared to other pentaquark states in Fig.~\ref{fig:decayLambdacD-2}. 
Moreover, from Table~\ref{tab:Gamma_exclusive_spin-flipping}, we observe that this state only decays to $\eta_c\left(1S\right)$ with a width of $21$~MeV (in scenario 2). 
The narrow width for decays into  $\Lambda_c\bar{D}$, $\Lambda_c\bar{D}^*$ and the fact that the state decays to $\eta_c\left(1S\right)$ and not to $J/\psi$ clearly makes the detection of $P_{c\bar{c}}\left(4513\right)^+$ a challenging enterprise.

\section{Predictions for bottom pentaquarks $P_{b\bar{b}}$}
\label{sec:Pbb}
As discussed in the previous section, scenario 2 is favored based on the analyses of the charm pentaquark semi-inclusive decay widths into $J/\psi$ and $\eta_c\left(1S\right)$, and decay patterns into S-wave $\Lambda_c\bar{D}$ and $\Lambda_c\bar{D}^*$ thresholds. 
In this section, we provide predictions for the bottom pentaquark states $P_{b\bar{b}}$ adopting scenario 2.

\begin{figure}[ht!]
 \centering
  \includegraphics[width=0.7\linewidth]{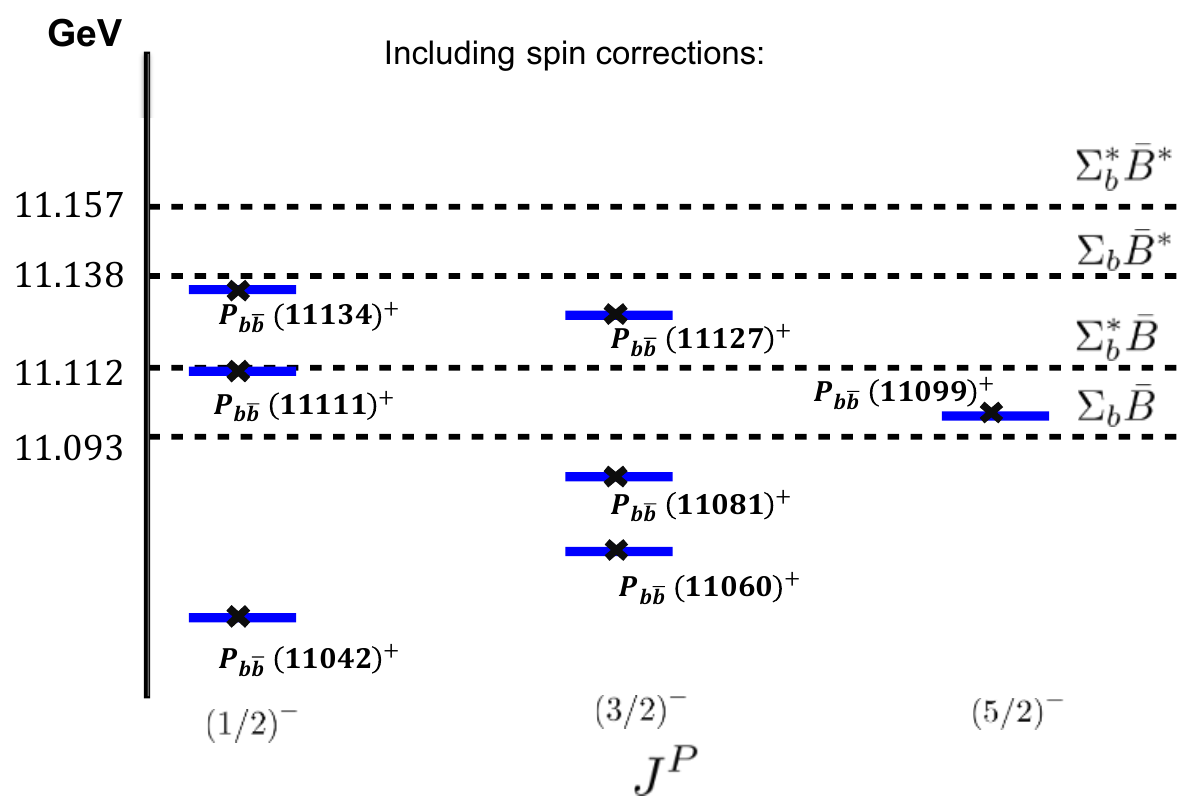}
\caption{Predicted spectrum of the lowest-lying bottom pentaquark states $P_{b\bar{b}}$ and their assigned $J^{P}$ quantum numbers in scenario 2, 
which the analysis of charm pentaquarks favors. 
The masses of the states, in units of MeV, are indicated in the parentheses next to each state.  
The states, after solving the coupled Schr\"odinger equations \eqref{eq:Sch12} and \eqref{eq:Sch32} and using Eqs.~\eqref{eq:M12} and \eqref{eq:M32} to account for spin-corrections, are represented by horizontal blue bars; the black crosses on the horizontal bars denote that the states have not been seen yet in experiments. 
The masses are computed from the adjoint baryon masses $\Lambda_{\left(1/2\right)^+, \mathrm{RS}}=1.125$~GeV and $\Lambda_{\left(3/2\right)^+, \mathrm{RS}}=1.152$~GeV.  
Isospin is $I=1/2$ for all the states.}  
\label{fig:Pbbscenario2}
\end{figure}

The values of the adjoint baryon masses (in the RS-scheme) tuned on the charm pentaquark case are  $\Lambda_{\left(1/2\right)^+, \mathrm{RS}}=1.125$~GeV and $\Lambda_{\left(3/2\right)^+, \mathrm{RS}}=1.152$~GeV. 
Solving the Schr\"odinger equations~\eqref{eq:Sch12} and \eqref{eq:Sch32} with these inputs for the adjoint baryon masses appearing in the potentials~\eqref{eq:QQbar_V} and with $m_b^{\mathrm{RS}}=4.863$~GeV yields $\left({\cal E}_{1/2}, {\cal E}_{3/2}\right)=\left(-21~\mathrm{MeV}, -59~\mathrm{MeV}\right)$. 
Including the spin-dependent corrections from $V_{SS}$, the eigenvalues $E^{i}_{M_{1/2}}$, $E^{i}_{M_{3/2}}$, and $E^{s=1}_{\frac{3}{2}\frac{5}{2}}$ corresponding to the $J^{P}=\left(1/2\right)^-$, $\left(3/2\right)^-$, and $\left(5/2\right)^-$ states are
\begin{align}
 E^i_{M_{1/2}} &=\left(-97.7, -28.4, -5.5\right)~\mathrm{MeV},\qquad\qquad  E^{s=1}_{\frac{3}{2}\frac{5}{2}} = -41.0~\mathrm{MeV},\nonumber\\ 
E^i_{M_{3/2}} &=\left(-79.1, -58.1,- 12.7\right)~\mathrm{MeV},
 \label{eq:E12E322Pbb}
\end{align}
where the first, second, and third entries in the parentheses correspond to $i=1, 2, 3$ respectively. 
The masses  of the states, obtained after adding $E_{\Sigma_b\bar{B}}=11.140$~GeV to the results in Eq.~\eqref{eq:E12E322Pbb}, and the assigned $J^{P}$ quantum numbers are shown in Fig.~\ref{fig:Pbbscenario2}. 
The eigenvectors of the matrices $M_{1/2}$ and $M_{3/2}$ in Eqs.~\eqref{eq:M12} and \eqref{eq:M32} allow to express the pentaquark states as superpositions of the multiplet states listed in Table~\ref{tab:QQbarpenta}, which have well-defined heavy-quark-antiquark spin $s$ and LDF angular momentum $k$:
\begin{align}
 \begin{pmatrix}
  | P_{b\bar{b}}\left(11042\right)^+\rangle\\
  | P_{b\bar{b}}\left(11111\right)^+\rangle\\
  | P_{b\bar{b}}\left(11134\right)^+\rangle
 \end{pmatrix} &=
 \begin{pmatrix}
  0.318 & 0.153 & 0.936 \\
  -0.522 & -0.796 & 0.307 \\
  -0.791 & 0.586 & 0.173
 \end{pmatrix}
 \begin{pmatrix}
  |s=0, k=1/2 \rangle_{J^{P}=\left(1/2\right)^-}\\
  |s=1, k=1/2 \rangle_{J^{P}=\left(1/2\right)^-}\\
  |s=1, k=3/2 \rangle_{J^{P}=\left(1/2\right)^-}
 \end{pmatrix},
 \label{eq:Pbbresultscenario21}
 \end{align}
 \begin{align}
 \begin{pmatrix}
  | P_{b\bar{b}}\left(11060\right)^+\rangle\\
  | P_{b\bar{b}}\left(11081\right)^+\rangle\\
  | P_{b\bar{b}}\left(11127\right)^+\rangle
 \end{pmatrix} &=
 \begin{pmatrix}
  0.309 & -0.503 & 0.807 \\
  -0.177 & 0.804 & 0.568 \\
  0.934 & 0.319 & -0.160
 \end{pmatrix}
 \begin{pmatrix}
  |s=1, k=1/2 \rangle_{J^{P}=\left(3/2\right)^-}\\
  |s=0, k=3/2 \rangle_{J^{P}=\left(3/2\right)^-}\\
  |s=1, k=3/2  \rangle_{J^{P}=\left(3/2\right)^-}
 \end{pmatrix},
 \label{eq:Pbbresultscenario22}
 \end{align}
 \begin{align}
 |P_{b\bar{b}}\left(11099\right)^+\rangle &=  |s=1, k=3/2  \rangle_{J^{P}=\left(5/2\right)^-}.
 \label{eq:Pbbresultscenario23}
\end{align}

\begin{table}[t!]
\begin{center}
\small{\renewcommand{\arraystretch}{1.5}
\scriptsize
\begin{minipage}{.49\linewidth}
\begin{tabular}{|c||c|}
\hline
$P_{b\bar{b}}\left[J^{P}\,\right]\,\left(\mathrm{mass}\right)\longrightarrow \Upsilon(1S) \left[1^{--}\,\right]$ &  $\begin{array}{c}\Gamma \,\,({\rm MeV})\end{array}$\\
\hline
\multicolumn{2}{|c|} {Scenario 2}\\
\hline
$P_{b\bar{b}}\left[\,\left(1/2\right)^{-}\,\right]\,\left(11042\right)$    & 1 $^{+0.4}_{-0.3}$ $^{+0.5}_{-0.3}$   \\

$P_{b\bar{b}}\left[\,\left(3/2\right)^{-}\,\right]\,\left(11060\right)$    & 2 $^{+1}_{-1}$ $^{+1}_{-0.5}$ \\

$P_{b\bar{b}}\left[\,\left(3/2\right)^{-}\,\right]\,\left(11081\right)$    & 6 $^{+2}_{-2}$ $^{+2}_{-2}$ \\

$P_{b\bar{b}}\left[\,\left(1/2\right)^{-}\,\right]\,\left(11111\right)$    & 3 $^{+1}_{-1}$ $^{+1}_{-1}$ \\

$P_{b\bar{b}}\left[\,\left(3/2\right)^{-}\,\right]\,\left(11127\right)$    & 1 $^{+0.3}_{-0.2}$ $^{+0.4}_{-0.2}$ \\

$P_{b\bar{b}}\left[\,\left(1/2\right)^{-}\,\right]\,\left(11134\right)$    & 8 $^{+2}_{-2}$ $^{+3}_{-2}$ \\
\hline
\end{tabular}
\end{minipage}
\begin{minipage}{.49\linewidth}
\begin{tabular}{|c||c|}
\hline
$P_{b\bar{b}}\left[J^{P}\,\right]\,\left(\mathrm{mass}\right)\longrightarrow \eta_b\left(1S\right) \left[0^{-+}\,\right]$ &  $\begin{array}{c}\Gamma \,\,({\rm MeV})\end{array}$\\
\hline
\multicolumn{2}{|c|} {Scenario 2}\\
\hline
$P_{b\bar{b}}\left[\,\left(1/2\right)^{-}\,\right]\,\left(11042\right)$    & 3 $^{+1}_{-1}$ $^{+1}_{-1}$   \\

$P_{b\bar{b}}\left[\,\left(3/2\right)^{-}\,\right]\,\left(11060\right)$    & 2 $^{+1}_{-1}$ $^{+1}_{-0.4}$ \\

$P_{b\bar{b}}\left[\,\left(3/2\right)^{-}\,\right]\,\left(11081\right)$    & 1 $^{+0.3}_{-0.3}$ $^{+0.4}_{-0.2}$ \\

$P_{b\bar{b}}\left[\,\left(5/2\right)^{-}\,\right]\,\left(11099\right)$    & 3 $^{+1}_{-1}$ $^{+1}_{-1}$ \\

$P_{b\bar{b}}\left[\,\left(1/2\right)^{-}\,\right]\,\left(11111\right)$    & 3 $^{+1}_{-1}$ $^{+1}_{-1}$ \\

$P_{b\bar{b}}\left[\,\left(3/2\right)^{-}\,\right]\,\left(11127\right)$    & 4 $^{+1}_{-1}$ $^{+1}_{-1}$ \\

$P_{b\bar{b}}\left[\,\left(1/2\right)^{-}\,\right]\,\left(11134\right)$    & 1.5 $^{+0.4}_{-0.3}$ $^{+0.6}_{-0.2}$ \\
\hline
\end{tabular}
\end{minipage}
\caption{Predictions of the semi-inclusive decay rates of pentaquark states $P_{b\bar{b}}$ decaying into $\Upsilon\left(1S\right)$ or $\eta_b\left(1S\right)$ in scenario 2: $P_{b\bar{b}}\rightarrow Q_n + X$, with $Q_n=\{\Upsilon(1S), \eta_b\left(1S\right)\}$, where $X$ denotes light hadrons.
The decay rates are computed from Eqs.~\eqref{eq:Gamma_spincons} and \eqref{eq:matrix_SO-spin}.
The pentaquark states are denoted by $P_{b\bar{b}}\left[J^{P}\,\right]\,\left(\mathrm{mass}\right)$, 
where the masses are in MeV. We have used the spin-averaged masses for both pentaquark and quarkonium states to estimate the decay rates. 
We show the decay rates to $\Upsilon\left(1S\right)$ on the left and the decay rates to $\eta_b\left(1S\right)$ and on the right. 
The pentaquark state $P_{b\bar{b}}\left[J^{P}=\left(5/2\right)^-\right]$ decays only to $\eta_b$.
}
\label{tab:PbbGamma_exclusive_spin-flipping}
}
\end{center}
\end{table}
\normalsize

Our predictions for the spin-flipping semi-inclusive decay widths into $\Upsilon\left(1S\right)$ and $\eta_b \left(1S\right)$ of the pentaquark states $P_{b\bar{b}}$ shown in Fig.~\ref{fig:Pbbscenario2}, computed using Eqs.~\eqref{eq:Gamma_spincons} and \eqref{eq:matrix_SO-spin} along with Eqs.~\eqref{eq:Pbbresultscenario21}--\eqref{eq:Pbbresultscenario23}, are summarized in Table~\ref{tab:PbbGamma_exclusive_spin-flipping}. 
The transition is spin-flipping as the non-vanishing of the matrix element in Eq.~\eqref{eq:matrix_SO-spin} constrains the spin-$0$ component of $P_{b\bar{b}}$ to decay into the spin-$1$ final state $\Upsilon(1S)$ and the spin-$1$ component of $P_{b\bar{b}}$ to decay into the spin-$0$ final state $\eta_b\left(1S\right)$. 
The sum of the decay widths to $\Upsilon(1S)$ and $\eta_b\left(1S\right)$ sets a lower bound on the total decay width of the $P_{b\bar{b}}$ state. 
Compared to the semi-inclusive decays in the charm sector given in Table~\ref{tab:Gamma_exclusive_spin-flipping}, 
the semi-inclusive rates in the bottom sector are smaller due to $1/m_Q$ suppression from the heavier bottom mass. 
The pentaquark state $P_{b\bar{b}}\left[J^{P}=\left(5/2\right)^-\right](11099)$ only decays to $\eta_b$  as this state has a well-defined heavy-quark spin $s=1$, as seen from Eq.~\eqref{eq:Pbbresultscenario23}.

\begin{figure}[ht]
\begin{center}
\includegraphics[width=0.7\textwidth,angle=0,clip]{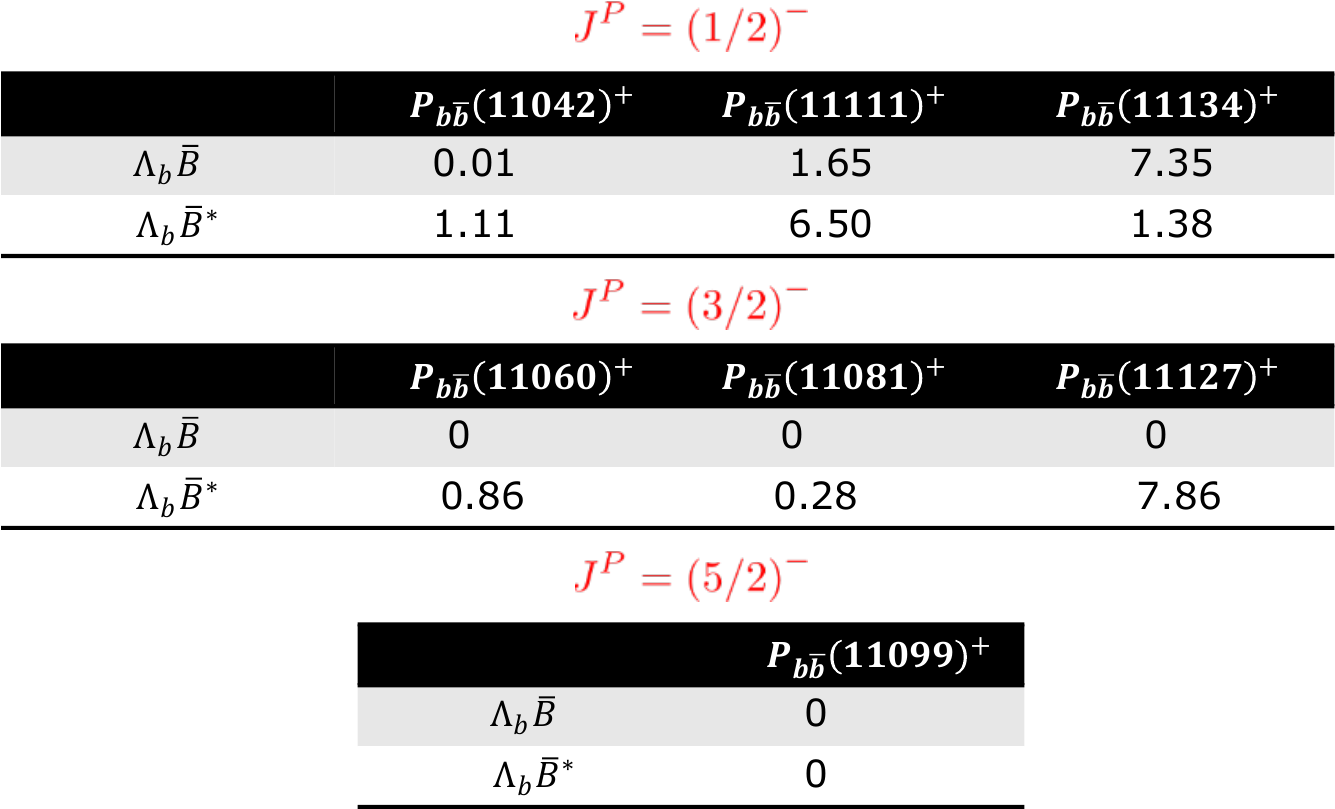}
\caption{Same as in Fig.~\ref{fig:decayLambdacD-2} but for $P_{b\bar{b}}$ states decaying into S-wave $\Lambda_b\bar{B}$ and $\Lambda_b\bar{B}^*$ in scenario 2.}
\label{fig:PbbdecayLambdabB-2}
\end{center}
\end{figure}

The results for the numerator and denominator of the ratio in the right-hand side of Eq.~\eqref{eq:Gamma1/2} obtained using Eqs.~\eqref{eq:Pbbresultscenario21}--\eqref{eq:Pbbresultscenario23} for different pentaquarks $P_{b\bar{b}}$ decaying into S-wave $\Lambda_b\bar{B}$ and $\Lambda_b\bar{B}^*$ 
are shown in Fig.~\ref{fig:PbbdecayLambdabB-2} for scenario 2.
For a given pentaquark state, $P_{b\bar{b}}$, from the entries in Fig.~\ref{fig:PbbdecayLambdabB-2}, it follows that 
\begin{align}
&\frac{\Gamma\left(P_{b\bar{b}}\left(11111\right)^+\rightarrow \Lambda_b\bar{B}\right)/v_{\Lambda_b\bar{B}}}{\Gamma\left(P_{b\bar{b}}\left(11111\right)^+\rightarrow \Lambda_b\bar{B}^*\right)/v_{\Lambda_b\bar{B}^*}}= 0.25, \quad \frac{\Gamma\left(P_{b\bar{b}}\left(11134\right)^+\rightarrow \Lambda_b\bar{B}\right)/v_{\Lambda_b\bar{B}}}{\Gamma\left(P_{b\bar{b}}\left(11134\right)^+\rightarrow \Lambda_b\bar{B}^*\right)/v_{\Lambda_b\bar{B}^*}}= 5.30.
\label{eq:PbbBr}
\end{align}
If the state $P_{b\bar{b}}\left(11111\right)^+$ with $J^{P}=\left(1/2\right)^-$ were exactly at the $\Sigma_b\bar{B}^*$ threshold (scenario 0 in Sec.~\ref{subsec:identification}), then the ratio of decay widths would be given by Eq.~\eqref{eq:VoloshinPc1}. 
Our results yield a ratio of $0.25$ (see Eq.~\eqref{eq:PbbBr}), which is $3$ times smaller than Eq.~\eqref{eq:VoloshinPc1}. 
This deviation reflects the fact that bottom pentaquark states are more deeply bound, as it follows from the values of the binding energies ${\cal E}_{1/2}$ and ${\cal E}_{3/2}$. 

Considering Fig.~\ref{fig:PbbdecayLambdabB-2}, for $J^{P}=\left(1/2\right)^-$ states, we observe that $P_{b\bar{b}}\left(11042\right)^+$ decays most often into $\Lambda_b\bar{B}^*$. 
For $P_{b\bar{b}}\left(11111\right)^+$, which has dominant decay into $\Lambda_b\bar{B}^*$, the relative partial decay width to a $\Lambda_b\bar{B}^*$ final state is about $5.86$ times larger than that of $P_{b\bar{b}}\left(11042\right)^+$ implying that the state has much broader width into the $\Lambda_c\bar{B}^*$ channel.  
For $P_{b\bar{b}}\left(11134\right)^+$, the dominant S-wave decay is into $\Lambda_b\bar{B}$, whose width is $4.45$ times larger than that of $P_{b\bar{b}}\left(11111\right)^+$ into the same channel. 
The large total S-wave width for both $P_{b\bar{b}}\left(11111\right)^+$ and $P_{b\bar{b}}\left(11134\right)^+$ suggests that both states are likely to manifest as relatively broad resonances compared to  $P_{b\bar{b}}\left(11042\right)^+$. 
For $J^{P}=\left(3/2\right)^-$ states, we observe that all the states decay only to S-wave $\Lambda_b\bar{B}^*$ as discussed at the beginning of this section. 
For $P_{b\bar{b}}\left(11060\right)^+$, the relative partial decay rate for decays into $\Lambda_b\bar{B}^*$ is slightly smaller ($0.7$ times) than that of  $P_{b\bar{b}}\left(11042\right)^+$. 
For $P_{b\bar{b}}\left(11081\right)^+$, the relative partial decay rate for decays into $\Lambda_b\bar{B}^*$ is about $3$ times smaller than that of  $P_{b\bar{b}}\left(11060\right)^+$. 
Moreover, accounting for the decays to $\Upsilon\left(1S\right)$ and $\eta_b\left(1S\right)$ in Table~\ref{tab:PbbGamma_exclusive_spin-flipping}, both $P_{b\bar{b}}\left(11060\right)^+$ and $P_{b\bar{b}}\left(11081\right)^+$ are possibly narrow pentaquark states. 
For $P_{b\bar{b}}\left(11127\right)^+$, the relative partial decay rate for decays into $\Lambda_b\bar{B}^*$  is around $9$ times larger than that of $P_{b\bar{b}}\left(11060\right)^+$, which suggests that this state will likely be  a relatively much broader resonance compared to  $P_{b\bar{b}}\left(11060\right)^+$ and $P_{b\bar{b}}\left(11081\right)^+$. 
The $J^{P}=\left(5/2\right)^-$ state $P_{b\bar{b}}\left(11099\right)^+$ can decay into $\Lambda_b\bar{B}$ and $\Lambda_b\bar{B}^*$ only via D-wave transitions, which are suppressed compared to S-wave decays indicating that the state will have a narrower width compared to other pentaquark states in Fig.~\ref{fig:PbbdecayLambdabB-2}. 
Accounting for the decay to  $\eta_b\left(1S\right)$ from Table~\ref{tab:PbbGamma_exclusive_spin-flipping}, further supports that this state may probably be a narrow resonance.

\section{Summary and conclusions}
\label{sec:conclusion}
In this work, we have presented a QCD based description of quarkonium  pentaquark $Q\bar{Q}qqq$ states both in the charm and bottom sector. 
The description is based on the Born--Oppenheimer nonrelativistic effective field theory framework (BOEFT), whose underlying principle is the systematic factorization of the dynamics of the heavy quarks and the three light quarks based on the hierarchy $\Lambda_{\mathrm{QCD}}\gg m_Qv^2$. 
The BOEFT leads to coupled-channel Schr\"odinger equations (Eqs.~\eqref{eq:Sch12} and \eqref{eq:Sch32}) for describing the pentaquarks, as derived in Ref.~\cite{Berwein:2024ztx}. 
These equations require pentaquark BO potentials (static energies) as inputs. 
They appear in the diagonal entries of the potential matrix in the Schr\"odinger Eqs.~\eqref{eq:Sch12} and \eqref{eq:Sch32}. 
In Sec.~\ref{subsec:Pstatic}, we write down the pentaquark BO potentials at short distances.
At short distances, they are the sum of a repulsive color octet potential, adjoint baryon masses $\Lambda_\kappa$ and non-perturbative terms of ${\cal O}\left(r^2\right)$. 
Currently, there are no lattice QCD        computations available for either the adjoint baryon spectrum or the pentaquark BO potentials. 
To facilitate future lattice studies, we give the expressions of the light quark operators for the lowest adjoint baryons with quantum numbers $\kappa=\{\left(1/2\right)^+,\, I=1/2\}$ and $\{\left(3/2\right)^+,\, I=1/2\}$ in Eqs.~\eqref{eq:OpQQbarqqq} and \eqref{eq:OpQQbarqqq-1} and the expression of the gauge-invariant interpolating operators in Eq.~\eqref{eq:intop}.
The two-point correlation functions of the interpolating operators can be expressed in terms of generalized Wilson loops, computed in lattice QCD and eventually used to extract the pentaquark BO potentials.

In this work, we propose to identify the four experimentally observed hidden-charm pentaquark states $P_{c\bar{c}}$ listed in Table~\ref{tab:exoticPenta} with states in the lowest multiplet of the BO potentials $\left(1/2\right)_g$ and $\{\left(1/2\right)_g^\prime, \left(3/2\right)_g\}$.
These potentials connect smoothly to the $\Sigma_c\bar{D}$ threshold at long distances due to BO quantum number conservation \cite{Berwein:2024ztx} and match to the spectrum of the adjoint baryons $\left(1/2\right)^+$ and $\left(3/2\right)^+$, respectively, at short distances. 
The form of the potentials at intermediate distances is unknown, as they have not been computed in lattice QCD.
At leading order in the $1/m_Q$ expansion, solving the Schr\"odinger Eqs.~\eqref{eq:Sch12} and \eqref{eq:Sch32} with these BO potentials yields a degenerate (spin-averaged) set of states independent of the heavy quark spin ${\bm S}$ (see Table~\ref{tab:QQbarpenta}). 
Since, the BO potentials for the $Q\bar{Q}$ pentaquark at large $r$ evolve smoothly into the heavy-light baryon-antimeson thresholds in the static limit and because the spin-dependent interaction that leads to spin-splitting in the heavy-light baryon-antimeson threshold appears at order $1/m_Q$,
in Sec.~\ref{subsec:spin}, we account for the spin splittings in the pentaquark multiplet in Table~\ref{tab:QQbarpenta} by choosing a spin interaction that resembles the one responsible for the spin interaction  
in the heavy-light baryon-antimeson states (see Eq. \eqref{eq:VSS}).
Besides the four states already observed, our assumed form (at intermediate distances) for the $(1/2)_g$ BO potential connecting with the $\Sigma_c\bar{D}$ threshold at large distances and the $(1/2)^+$ adjoint baryon at short distances supports three additional states, which are near the $\Sigma_c^*\bar{D}^*$ threshold.    
Seven states are also predicted by the molecular picture based on the heavy-quark spin symmetry in Refs.~\cite{Liu:2019tjn,Du:2019pij,Xiao:2019aya,Du:2021fmf}. 
In the absence of lattice data, however, other choices for the behaviour of the BO potentials at intermediate distances are possible.
If the $(1/2)_g$ BO potential connecting with the $\Lambda_c\bar{D}$ threshold would support bound states, then three additional low-lying pentaquark states in the charm sector could exist. 
Ten low-lying pentaquark states, some having positive parity, are predicted in compact pentaquark models such as in \cite{Ali:2019npk}.  
Because there is no experimental evidence of states near the $\Lambda_c\bar{D}$ threshold, we have assumed that the BO potential that connects to that threshold falls off monotonically 
from above like in Fig.~\ref{fig:illustration} and hence does not support bound states. 
Another possibility is that also the $(1/2)_g$ BO potential connecting to the $\Sigma_c\bar{D}$ threshold at large distances and the $(1/2)^+$ adjoint baryon at short distances does not support bound states.
This case, which brings the number of predicted low-lying pentaquarks to four, has been investigated in \cite{Alasiri:2025roh}. 
In BOEFT, all the ground-state pentaquark states have negative parity. 
The positive parity pentaquark states arise from orbital excitations with heavy-quark orbital angular momentum $L_Q=1$ (P-wave) and therefore lie at higher energies than the ground-state pentaquarks.

By treating the adjoint baryon masses $\Lambda_{\left(1/2\right)^+}$ and $\Lambda_{\left(3/2\right)^+}$ as free parameters fitted on the masses of the four observed $P_{c\bar{c}}$ states after including spin corrections, we find two viable scenarios, scenario 1 and scenario 2, whose spectra and $J^{P}$ assignments are summarized in Tables~\ref{tab:scenario1} and \ref{tab:scenario2}, and in Figs.~\ref{fig:scenario1} and \ref{fig:scenario2}. 
Notably, in scenario 2, our predicted masses for the three unobserved $P_{c\bar{c}}$ states  show good agreement with the molecular model predictions in Refs.~\cite{Du:2019pij,Du:2021fmf}, even though BOEFT makes no a priori assumption of a molecular configuration. 
The agreement emerges from the complex QCD dynamics encoded in the solutions of the Schr\"odinger equations \eqref{eq:Sch12} and \eqref{eq:Sch32}. 
In particular, the BOEFT can be closer or farther
away from the molecular picture results depending on the values of the adjoint
baryon masses and the corresponding eigenergies coming from the diagonalization of the matrices in Eqs.~\eqref{eq:M12} and \eqref{eq:M32}. 
It is therefore of crucial importance to calculate
on the lattice the adjoint baryon masses $\Lambda_{\left(1/2\right)^+}$ and $\Lambda_{\left(3/2\right)^+}$.
After including spin corrections, the resulting $P_{c\bar{c}}$ states are superpositions of multiplet states (Table~\ref{tab:QQbarpenta}) with the heavy quark-antiquark pair in both spin $s=0$ and $s=1$ configurations, as given by Eqs.~\eqref{eq:resultscenario11}-\eqref{eq:resultscenario13} and Eqs.~\eqref{eq:resultscenario21}-\eqref{eq:resultscenario23} for scenario 1 and 2, respectively. 
In Sec.~\ref{subsec:Semi-inlcusive}, we compute semi-inclusive decay rates into $J/\psi$ and $\eta_c(1S)$ following Ref.~\cite{Brambilla:2022hhi}. 
Spin-flip transitions mediate the decays: the spin-$0$ component of $P_{c\bar{c}}$ decays to $J/\psi$ and the spin-$1$ component  of $P_{c\bar{c}}$ decays to $\eta_c\left(1S\right)$. 
Spin-conserving transitions of $P_{c\bar{c}}$ would instead lead to P-wave quarkonium states, and hence do not contribute.
The sum of these rates provides a lower bound on the total width of the pentaquark state. 
Results for both scenarios are summarized in Table~\ref{tab:Gamma_exclusive_spin-flipping}. 
In particular, for $P_{c\bar{c}}\left(4312\right)^+$, which has $J^{P}=\left(1/2\right)^-$ in both scenarios, 
comparing our lower bound estimate with the total width favors scenario~2.  
Scenario 2 assigns $P_{c\bar{c}}\left(4440\right)^+$ to $J^{P}=\left(3/2\right)^-$ and $P_{c\bar{c}}\left(4457\right)^+$ to $J^{P}=\left(1/2\right)^-$. 
This $J^{P}$ assignment agrees with the one in Refs.~\cite{Liu:2019zvb,Yalikun:2021bfm,Du:2019pij,Du:2021fmf,Yamaguchi:2019seo, Li:2025ejt}. 
In scenario 2, the state $P_{c\bar{c}}\left(4513\right)^+$ with $J^{P}=\left(5/2\right)^-$ decays only to $\eta_c(1S)$.

In Sec.~\ref{subsec:branching ratio}, we present model-independent predictions for the decay ratios of $P_{c\bar{c}}$ states into S-wave $\Lambda_c\bar{D}$ and $\Lambda_c\bar{D}^*$ following Ref.~\cite{Braaten:2024stn}. 
These transitions are governed by a single transition amplitude (see Eqs.~\eqref{eq:mixpot1} and \eqref{eq:mixpot2}), which, in principle, is calculable in lattice QCD. 
The transition amplitude cancels in the decay width ratios, as shown in Eq. \eqref{eq:Gamma1/2}.
We find that the $P_{c\bar{c}}$ states with $J^{P}=\left(3/2\right)^-$ decay only to S-wave $\Lambda_c\bar{D}^*$ 
and that the decay to $\Lambda_c\bar{D}$ requires a D-wave transition. 
Also the $P_{c\bar{c}}$ states with $J^{P}=\left(5/2\right)^-$ can decay to both $\Lambda_c\bar{D}$ and $\Lambda_c\bar{D}^*$ only through a D-wave. 
Our results are summarized in Eq.~\eqref{eq:Br} and Figs.~\ref{fig:decayLambdacD-1} and \ref{fig:decayLambdacD-2}. 
Again, we find that scenario 2 is favored.
For $P_{c\bar{c}}\left(4380\right)^+$, to which a now obsolete LHCb analysis attributed a broad width \cite{LHCb:2015yax}, we find that the decay rate into S-wave $\Lambda_c\bar{D}^*$ is of the same order  as that of $P_{c\bar{c}}\left(4312\right)^+$ (see Fig.~\ref{fig:decayLambdacD-2}). 
Also including the decay widths to $J/\psi$ and $\eta_c\left(1S\right)$ from Table~\ref{tab:Gamma_exclusive_spin-flipping}, our analysis suggests that $P_{c\bar{c}}\left(4380\right)^+$ is a narrow state, 
in agreement with the studies in Refs.~\cite{Du:2019pij, Du:2021fmf}.

In Sec.~\ref{sec:Pbb}, based on scenario 2, we report our predictions for the bottom pentaquark $P_{b\bar{b}}$ states. 
The adjoint baryon masses and the BO static potentials are independent of the heavy quark mass.
Unlike in the molecular model, which cannot be easily extended from the charm to the bottom sector,  in the BOEFT it is enough to solve the Schr\"odinger equations \eqref{eq:Sch12} and \eqref{eq:Sch32} using the bottom quark mass and the BO static potentials in Eq.~\eqref{eq:QQbar_V}, with the adjoint baryon masses $\Lambda_{\left(1/2\right)^+}$ and $\Lambda_{\left(3/2\right)^+}$ fixed in the charm sector. 
The results  for the masses of the lowest seven $P_{b\bar{b}}$ states along with their $J^{P}$ quantum number assignments are shown in Fig.~\ref{fig:Pbbscenario2}. 
The states are more deeply bound than in the charm sector as reflected by the values of the binding energies.
Our predictions for the semi-inclusive decays into $\Upsilon(1S)$ and $\eta_b(1S)$ are given in Table~\ref{tab:PbbGamma_exclusive_spin-flipping}, while model-independent results for the relative decay rates into S-wave $\Lambda_b\bar{B}$ and $\Lambda_b\bar{B}^*$ are shown in Fig.~\ref{fig:PbbdecayLambdabB-2}. 
We find that for the pentaquark state $P_{b\bar{b}}\left(11111\right)^+$ with $J^P=\left(1/2\right)^-$, the ratio of the decay rates into S-wave $\Lambda_b\bar{B}$ and $\Lambda_b\bar{B}^*$ is different from the molecular picture prediction in Ref.~\cite{Voloshin:2019aut}.

Finally, we emphasize that definitive, model-independent predictions for hidden-heavy pentaquarks, both $P_{c\bar{c}}$ and $P_{b\bar{b}}$, need several lattice QCD inputs,  most importantly the adjoint baryon spectrum and the pentaquark BO potentials.
The computation of the adjoint baryon spectrum, and in particular the adjoint baryon masses $\Lambda_{\left(1/2\right)^+}$ and $\Lambda_{\left(3/2\right)^+}$, may confirm or correct our choices for these quantities.
The computation of the BO potentials would establish whether the ground-state multiplet contains ten, seven or just four $P_{c\bar{c}}$ and $P_{b\bar{b}}$ states. 
Obviously, the experimental confirmation or discovery of new $P_{c\bar{c}}$ and $P_{b\bar{b}}$ states is the most awaited and needed information to constrain the different scenarios for hidden heavy pentaquarks in the framework of the BOEFT presented in this and related works. 

Notably, the LHCb collaboration recently reported the observation of a new pentaquark state $P_{c\bar{c}}\left(4337\right)^+$ in the decay $B_s^0\to J/\psi p \bar{p}$, seen as a structure in the $J/\psi p$ invariant-mass distribution with mass $4337^{+7}_{-4}$~MeV and width $29^{+30}_{-18}$~MeV \cite{LHCb:2021chn}. 
If this state has $J^{P}=\left(1/2\right)^+$, which currently has the highest statistical significance of $3.7\sigma$ \cite{LHCb:2021chn}, then within the BOEFT framework it could correspond to a P-wave resonance state in the BO potential $\left(1/2\right)_g$ that dip slightly below the $\Lambda_c\bar{D}$ threshold, then cross it again and asymptotically connects to the $\Lambda_c\bar{D}$ threshold from above.    
Additionally, no $P_{c\bar{c}}$ states with quartet isospin $I=3/2$ have yet been observed experimentally. 
In the BOEFT, the $I=3/2$ pentaquark states are described by a single channel Schr\"odinger equation with BO potential $\left(1/2\right)_g$ matching to an adjoint baryon with quantum number $k^P=\left(1/2\right)^+$ and $I=3/2$ at short distances  \cite{Berwein:2024ztx}. 
A comprehensive investigation of these states will be pursued in future work.

The nonstrange and strange charm pentaquark states $P_{c\bar{c}}$ and $P_{cs}$ will be searched and studied through the hadro-production channels  $pp\rightarrow pp J/\psi$ and $pp\rightarrow pK^+\Lambda J/\psi$ at the upcoming CBM experiment at the GSI-FAIR facility \cite{Messchendorp:2025vzi}.
The direct observation of existing and new pentaquark states through hadro-production channels has the potential to provide new valuable insights into the formation and internal structure of these states.

\section*{Acknowledgements}
This work has been supported by the DFG cluster of excellence ORIGINS funded by the Deutsche Forschungsgemeinschaft under Germany’s Excellence Strategy-EXC-2094-390783311. N.~B. acknowledges the Advanced ERC grant ERC-2023-ADG-Project EFT-XYZ.

\bibliography{Pentaquark}

\begin{thebibliography}{139}%
\makeatletter
\providecommand \@ifxundefined [1]{%
 \@ifx{#1\undefined}
}%
\providecommand \@ifnum [1]{%
 \ifnum #1\expandafter \@firstoftwo
 \else \expandafter \@secondoftwo
 \fi
}%
\providecommand \@ifx [1]{%
 \ifx #1\expandafter \@firstoftwo
 \else \expandafter \@secondoftwo
 \fi
}%
\providecommand \natexlab [1]{#1}%
\providecommand \enquote  [1]{``#1''}%
\providecommand \bibnamefont  [1]{#1}%
\providecommand \bibfnamefont [1]{#1}%
\providecommand \citenamefont [1]{#1}%
\providecommand \href@noop [0]{\@secondoftwo}%
\providecommand \href [0]{\begingroup \@sanitize@url \@href}%
\providecommand \@href[1]{\@@startlink{#1}\@@href}%
\providecommand \@@href[1]{\endgroup#1\@@endlink}%
\providecommand \@sanitize@url [0]{\catcode `\\12\catcode `\$12\catcode
  `\&12\catcode `\#12\catcode `\^12\catcode `\_12\catcode `\%12\relax}%
\providecommand \@@startlink[1]{}%
\providecommand \@@endlink[0]{}%
\providecommand \url  [0]{\begingroup\@sanitize@url \@url }%
\providecommand \@url [1]{\endgroup\@href {#1}{\urlprefix }}%
\providecommand \urlprefix  [0]{URL }%
\providecommand \Eprint [0]{\href }%
\providecommand \doibase [0]{https://doi.org/}%
\providecommand \selectlanguage [0]{\@gobble}%
\providecommand \bibinfo  [0]{\@secondoftwo}%
\providecommand \bibfield  [0]{\@secondoftwo}%
\providecommand \translation [1]{[#1]}%
\providecommand \BibitemOpen [0]{}%
\providecommand \bibitemStop [0]{}%
\providecommand \bibitemNoStop [0]{.\EOS\space}%
\providecommand \EOS [0]{\spacefactor3000\relax}%
\providecommand \BibitemShut  [1]{\csname bibitem#1\endcsname}%
\let\auto@bib@innerbib\@empty
\bibitem [{\citenamefont {Gell-Mann}(1964)}]{GellMann:1964nj}%
  \BibitemOpen
  \bibfield  {author} {\bibinfo {author} {\bibfnamefont {M.}~\bibnamefont
  {Gell-Mann}},\ }\bibfield  {title} {\bibinfo {title} {{A Schematic Model of
  Baryons and Mesons}},\ }\href {https://doi.org/10.1016/S0031-9163(64)92001-3}
  {\bibfield  {journal} {\bibinfo  {journal} {Phys. Lett.}\ }\textbf {\bibinfo
  {volume} {8}},\ \bibinfo {pages} {214} (\bibinfo {year} {1964})}\BibitemShut
  {NoStop}%
\bibitem [{\citenamefont {Zweig}(1964)}]{Zweig:1964CERN}%
  \BibitemOpen
  \bibfield  {author} {\bibinfo {author} {\bibfnamefont {G.}~\bibnamefont
  {Zweig}},\ }\bibfield  {title} {\bibinfo {title} {Developments in the quark
  theory of hadrons},\ }\href@noop {} {\bibfield  {journal} {\bibinfo
  {journal} {CERN Report No.8182/TH.401, CERN Report No.8419/TH.412}\ }
  (\bibinfo {year} {1964})}\BibitemShut {NoStop}%
\bibitem [{\citenamefont {Jaffe}\ and\ \citenamefont
  {Johnson}(1976)}]{Jaffe:1975fd}%
  \BibitemOpen
  \bibfield  {author} {\bibinfo {author} {\bibfnamefont {R.~L.}\ \bibnamefont
  {Jaffe}}\ and\ \bibinfo {author} {\bibfnamefont {K.}~\bibnamefont
  {Johnson}},\ }\bibfield  {title} {\bibinfo {title} {{Unconventional States of
  Confined Quarks and Gluons}},\ }\href
  {https://doi.org/10.1016/0370-2693(76)90423-8} {\bibfield  {journal}
  {\bibinfo  {journal} {Phys. Lett. B}\ }\textbf {\bibinfo {volume} {60}},\
  \bibinfo {pages} {201} (\bibinfo {year} {1976})}\BibitemShut {NoStop}%
\bibitem [{\citenamefont {Brambilla}\ \emph
  {et~al.}(2020{\natexlab{a}})\citenamefont {Brambilla}, \citenamefont
  {Eidelman}, \citenamefont {Hanhart}, \citenamefont {Nefediev}, \citenamefont
  {Shen}, \citenamefont {Thomas}, \citenamefont {Vairo},\ and\ \citenamefont
  {Yuan}}]{Brambilla:2019esw}%
  \BibitemOpen
  \bibfield  {author} {\bibinfo {author} {\bibfnamefont {N.}~\bibnamefont
  {Brambilla}}, \bibinfo {author} {\bibfnamefont {S.}~\bibnamefont {Eidelman}},
  \bibinfo {author} {\bibfnamefont {C.}~\bibnamefont {Hanhart}}, \bibinfo
  {author} {\bibfnamefont {A.}~\bibnamefont {Nefediev}}, \bibinfo {author}
  {\bibfnamefont {C.-P.}\ \bibnamefont {Shen}}, \bibinfo {author}
  {\bibfnamefont {C.~E.}\ \bibnamefont {Thomas}}, \bibinfo {author}
  {\bibfnamefont {A.}~\bibnamefont {Vairo}},\ and\ \bibinfo {author}
  {\bibfnamefont {C.-Z.}\ \bibnamefont {Yuan}},\ }\bibfield  {title} {\bibinfo
  {title} {{The $XYZ$ states: experimental and theoretical status and
  perspectives}},\ }\href {https://doi.org/10.1016/j.physrep.2020.05.001}
  {\bibfield  {journal} {\bibinfo  {journal} {Phys. Rept.}\ }\textbf {\bibinfo
  {volume} {873}},\ \bibinfo {pages} {1} (\bibinfo {year}
  {2020}{\natexlab{a}})},\ \Eprint {https://arxiv.org/abs/1907.07583}
  {arXiv:1907.07583 [hep-ex]} \BibitemShut {NoStop}%
\bibitem [{\citenamefont {Choi}\ \emph {et~al.}(2003)\citenamefont {Choi} \emph
  {et~al.}}]{Belle:2003nnu}%
  \BibitemOpen
  \bibfield  {author} {\bibinfo {author} {\bibfnamefont {S.~K.}\ \bibnamefont
  {Choi}} \emph {et~al.} (\bibinfo {collaboration} {Belle}),\ }\bibfield
  {title} {\bibinfo {title} {{Observation of a narrow charmonium-like state in
  exclusive $B^\pm \to K^\pm \pi^+ \pi^- J/\psi$ decays}},\ }\href
  {https://doi.org/10.1103/PhysRevLett.91.262001} {\bibfield  {journal}
  {\bibinfo  {journal} {Phys. Rev. Lett.}\ }\textbf {\bibinfo {volume} {91}},\
  \bibinfo {pages} {262001} (\bibinfo {year} {2003})},\ \Eprint
  {https://arxiv.org/abs/hep-ex/0309032} {arXiv:hep-ex/0309032} \BibitemShut
  {NoStop}%
\bibitem [{\citenamefont {Esposito}\ \emph {et~al.}(2017)\citenamefont
  {Esposito}, \citenamefont {Pilloni},\ and\ \citenamefont
  {Polosa}}]{Esposito:2016noz}%
  \BibitemOpen
  \bibfield  {author} {\bibinfo {author} {\bibfnamefont {A.}~\bibnamefont
  {Esposito}}, \bibinfo {author} {\bibfnamefont {A.}~\bibnamefont {Pilloni}},\
  and\ \bibinfo {author} {\bibfnamefont {A.~D.}\ \bibnamefont {Polosa}},\
  }\bibfield  {title} {\bibinfo {title} {{Multiquark Resonances}},\ }\href
  {https://doi.org/10.1016/j.physrep.2016.11.002} {\bibfield  {journal}
  {\bibinfo  {journal} {Phys. Rept.}\ }\textbf {\bibinfo {volume} {668}},\
  \bibinfo {pages} {1} (\bibinfo {year} {2017})},\ \Eprint
  {https://arxiv.org/abs/1611.07920} {arXiv:1611.07920 [hep-ph]} \BibitemShut
  {NoStop}%
\bibitem [{\citenamefont {Olsen}\ \emph {et~al.}(2018)\citenamefont {Olsen},
  \citenamefont {Skwarnicki},\ and\ \citenamefont {Zieminska}}]{Olsen:2017bmm}%
  \BibitemOpen
  \bibfield  {author} {\bibinfo {author} {\bibfnamefont {S.~L.}\ \bibnamefont
  {Olsen}}, \bibinfo {author} {\bibfnamefont {T.}~\bibnamefont {Skwarnicki}},\
  and\ \bibinfo {author} {\bibfnamefont {D.}~\bibnamefont {Zieminska}},\
  }\bibfield  {title} {\bibinfo {title} {{Nonstandard heavy mesons and baryons:
  Experimental evidence}},\ }\href
  {https://doi.org/10.1103/RevModPhys.90.015003} {\bibfield  {journal}
  {\bibinfo  {journal} {Rev. Mod. Phys.}\ }\textbf {\bibinfo {volume} {90}},\
  \bibinfo {pages} {015003} (\bibinfo {year} {2018})},\ \Eprint
  {https://arxiv.org/abs/1708.04012} {arXiv:1708.04012 [hep-ph]} \BibitemShut
  {NoStop}%
\bibitem [{\citenamefont {Guo}\ \emph {et~al.}(2018)\citenamefont {Guo},
  \citenamefont {Hanhart}, \citenamefont {Mei\ss{}ner}, \citenamefont {Wang},
  \citenamefont {Zhao},\ and\ \citenamefont {Zou}}]{Guo:2017jvc}%
  \BibitemOpen
  \bibfield  {author} {\bibinfo {author} {\bibfnamefont {F.-K.}\ \bibnamefont
  {Guo}}, \bibinfo {author} {\bibfnamefont {C.}~\bibnamefont {Hanhart}},
  \bibinfo {author} {\bibfnamefont {U.-G.}\ \bibnamefont {Mei\ss{}ner}},
  \bibinfo {author} {\bibfnamefont {Q.}~\bibnamefont {Wang}}, \bibinfo {author}
  {\bibfnamefont {Q.}~\bibnamefont {Zhao}},\ and\ \bibinfo {author}
  {\bibfnamefont {B.-S.}\ \bibnamefont {Zou}},\ }\bibfield  {title} {\bibinfo
  {title} {{Hadronic molecules}},\ }\href
  {https://doi.org/10.1103/RevModPhys.90.015004} {\bibfield  {journal}
  {\bibinfo  {journal} {Rev. Mod. Phys.}\ }\textbf {\bibinfo {volume} {90}},\
  \bibinfo {pages} {015004} (\bibinfo {year} {2018})},\ \bibinfo {note}
  {[Erratum: Rev.Mod.Phys. 94, 029901 (2022)]},\ \Eprint
  {https://arxiv.org/abs/1705.00141} {arXiv:1705.00141 [hep-ph]} \BibitemShut
  {NoStop}%
\bibitem [{\citenamefont {Ali}\ \emph {et~al.}(2017)\citenamefont {Ali},
  \citenamefont {Lange},\ and\ \citenamefont {Stone}}]{Ali:2017jda}%
  \BibitemOpen
  \bibfield  {author} {\bibinfo {author} {\bibfnamefont {A.}~\bibnamefont
  {Ali}}, \bibinfo {author} {\bibfnamefont {J.~S.}\ \bibnamefont {Lange}},\
  and\ \bibinfo {author} {\bibfnamefont {S.}~\bibnamefont {Stone}},\ }\bibfield
   {title} {\bibinfo {title} {{Exotics: Heavy Pentaquarks and Tetraquarks}},\
  }\href {https://doi.org/10.1016/j.ppnp.2017.08.003} {\bibfield  {journal}
  {\bibinfo  {journal} {Prog. Part. Nucl. Phys.}\ }\textbf {\bibinfo {volume}
  {97}},\ \bibinfo {pages} {123} (\bibinfo {year} {2017})},\ \Eprint
  {https://arxiv.org/abs/1706.00610} {arXiv:1706.00610 [hep-ph]} \BibitemShut
  {NoStop}%
\bibitem [{\citenamefont {Chen}\ \emph {et~al.}(2023)\citenamefont {Chen},
  \citenamefont {Chen}, \citenamefont {Liu}, \citenamefont {Liu},\ and\
  \citenamefont {Zhu}}]{Chen:2022asf}%
  \BibitemOpen
  \bibfield  {author} {\bibinfo {author} {\bibfnamefont {H.-X.}\ \bibnamefont
  {Chen}}, \bibinfo {author} {\bibfnamefont {W.}~\bibnamefont {Chen}}, \bibinfo
  {author} {\bibfnamefont {X.}~\bibnamefont {Liu}}, \bibinfo {author}
  {\bibfnamefont {Y.-R.}\ \bibnamefont {Liu}},\ and\ \bibinfo {author}
  {\bibfnamefont {S.-L.}\ \bibnamefont {Zhu}},\ }\bibfield  {title} {\bibinfo
  {title} {{An updated review of the new hadron states}},\ }\href
  {https://doi.org/10.1088/1361-6633/aca3b6} {\bibfield  {journal} {\bibinfo
  {journal} {Rept. Prog. Phys.}\ }\textbf {\bibinfo {volume} {86}},\ \bibinfo
  {pages} {026201} (\bibinfo {year} {2023})},\ \Eprint
  {https://arxiv.org/abs/2204.02649} {arXiv:2204.02649 [hep-ph]} \BibitemShut
  {NoStop}%
\bibitem [{\citenamefont {Brambilla}\ \emph {et~al.}(2011)\citenamefont
  {Brambilla} \emph {et~al.}}]{Brambilla:2010cs}%
  \BibitemOpen
  \bibfield  {author} {\bibinfo {author} {\bibfnamefont {N.}~\bibnamefont
  {Brambilla}} \emph {et~al.},\ }\bibfield  {title} {\bibinfo {title} {{Heavy
  Quarkonium: Progress, Puzzles, and Opportunities}},\ }\href
  {https://doi.org/10.1140/epjc/s10052-010-1534-9} {\bibfield  {journal}
  {\bibinfo  {journal} {Eur. Phys. J. C}\ }\textbf {\bibinfo {volume} {71}},\
  \bibinfo {pages} {1534} (\bibinfo {year} {2011})},\ \Eprint
  {https://arxiv.org/abs/1010.5827} {arXiv:1010.5827 [hep-ph]} \BibitemShut
  {NoStop}%
\bibitem [{\citenamefont {Lebed}(2023)}]{Lebed:2023vnd}%
  \BibitemOpen
  \bibfield  {author} {\bibinfo {author} {\bibfnamefont {R.~F.}\ \bibnamefont
  {Lebed}},\ }\bibfield  {title} {\bibinfo {title} {{Theory of the
  (Heavy-Quark) Exotic Hadrons: A Primer for the Flavor Community}},\ }\href
  {https://doi.org/10.22323/1.445.0028} {\bibfield  {journal} {\bibinfo
  {journal} {PoS}\ }\textbf {\bibinfo {volume} {FPCP2023}},\ \bibinfo {pages}
  {028} (\bibinfo {year} {2023})},\ \Eprint {https://arxiv.org/abs/2308.00781}
  {arXiv:2308.00781 [hep-ph]} \BibitemShut {NoStop}%
\bibitem [{\citenamefont {Aaij}\ \emph {et~al.}(2015)\citenamefont {Aaij} \emph
  {et~al.}}]{LHCb:2015yax}%
  \BibitemOpen
  \bibfield  {author} {\bibinfo {author} {\bibfnamefont {R.}~\bibnamefont
  {Aaij}} \emph {et~al.} (\bibinfo {collaboration} {LHCb}),\ }\bibfield
  {title} {\bibinfo {title} {{Observation of $J/\psi p$ Resonances Consistent
  with Pentaquark States in $\Lambda_b^0 \to J/\psi K^- p$ Decays}},\ }\href
  {https://doi.org/10.1103/PhysRevLett.115.072001} {\bibfield  {journal}
  {\bibinfo  {journal} {Phys. Rev. Lett.}\ }\textbf {\bibinfo {volume} {115}},\
  \bibinfo {pages} {072001} (\bibinfo {year} {2015})},\ \Eprint
  {https://arxiv.org/abs/1507.03414} {arXiv:1507.03414 [hep-ex]} \BibitemShut
  {NoStop}%
\bibitem [{\citenamefont {Aaij}\ \emph {et~al.}(2019)\citenamefont {Aaij} \emph
  {et~al.}}]{LHCb:2019kea}%
  \BibitemOpen
  \bibfield  {author} {\bibinfo {author} {\bibfnamefont {R.}~\bibnamefont
  {Aaij}} \emph {et~al.} (\bibinfo {collaboration} {LHCb}),\ }\bibfield
  {title} {\bibinfo {title} {{Observation of a narrow pentaquark state,
  $P_c(4312)^+$, and of two-peak structure of the $P_c(4450)^+$}},\ }\href
  {https://doi.org/10.1103/PhysRevLett.122.222001} {\bibfield  {journal}
  {\bibinfo  {journal} {Phys. Rev. Lett.}\ }\textbf {\bibinfo {volume} {122}},\
  \bibinfo {pages} {222001} (\bibinfo {year} {2019})},\ \Eprint
  {https://arxiv.org/abs/1904.03947} {arXiv:1904.03947 [hep-ex]} \BibitemShut
  {NoStop}%
\bibitem [{\citenamefont {Aaij}\ \emph {et~al.}(2021)\citenamefont {Aaij} \emph
  {et~al.}}]{LHCb:2020jpq}%
  \BibitemOpen
  \bibfield  {author} {\bibinfo {author} {\bibfnamefont {R.}~\bibnamefont
  {Aaij}} \emph {et~al.} (\bibinfo {collaboration} {LHCb}),\ }\bibfield
  {title} {\bibinfo {title} {{Evidence of a $J/\psi\Lambda$ structure and
  observation of excited $\Xi^-$ states in the $\Xi^-_b \to J/\psi\Lambda K^-$
  decay}},\ }\href {https://doi.org/10.1016/j.scib.2021.02.030} {\bibfield
  {journal} {\bibinfo  {journal} {Sci. Bull.}\ }\textbf {\bibinfo {volume}
  {66}},\ \bibinfo {pages} {1278} (\bibinfo {year} {2021})},\ \Eprint
  {https://arxiv.org/abs/2012.10380} {arXiv:2012.10380 [hep-ex]} \BibitemShut
  {NoStop}%
\bibitem [{\citenamefont {Aaij}\ \emph {et~al.}(2023)\citenamefont {Aaij} \emph
  {et~al.}}]{LHCb:2022ogu}%
  \BibitemOpen
  \bibfield  {author} {\bibinfo {author} {\bibfnamefont {R.}~\bibnamefont
  {Aaij}} \emph {et~al.} (\bibinfo {collaboration} {LHCb}),\ }\bibfield
  {title} {\bibinfo {title} {{Observation of a
  J/\ensuremath{\psi}\ensuremath{\Lambda} Resonance Consistent with a Strange
  Pentaquark Candidate in
  B-\textrightarrow{}J/\ensuremath{\psi}\ensuremath{\Lambda}p\textasciimacron{}
  Decays}},\ }\href {https://doi.org/10.1103/PhysRevLett.131.031901} {\bibfield
   {journal} {\bibinfo  {journal} {Phys. Rev. Lett.}\ }\textbf {\bibinfo
  {volume} {131}},\ \bibinfo {pages} {031901} (\bibinfo {year} {2023})},\
  \Eprint {https://arxiv.org/abs/2210.10346} {arXiv:2210.10346 [hep-ex]}
  \BibitemShut {NoStop}%
\bibitem [{\citenamefont {Navas}\ \emph {et~al.}(2024)\citenamefont {Navas}
  \emph {et~al.}}]{ParticleDataGroup:2024cfk}%
  \BibitemOpen
  \bibfield  {author} {\bibinfo {author} {\bibfnamefont {S.}~\bibnamefont
  {Navas}} \emph {et~al.} (\bibinfo {collaboration} {Particle Data Group}),\
  }\bibfield  {title} {\bibinfo {title} {{Review of particle physics}},\ }\href
  {https://doi.org/10.1103/PhysRevD.110.030001} {\bibfield  {journal} {\bibinfo
   {journal} {Phys. Rev. D}\ }\textbf {\bibinfo {volume} {110}},\ \bibinfo
  {pages} {030001} (\bibinfo {year} {2024})}\BibitemShut {NoStop}%
\bibitem [{\citenamefont {Wu}\ \emph {et~al.}(2010)\citenamefont {Wu},
  \citenamefont {Molina}, \citenamefont {Oset},\ and\ \citenamefont
  {Zou}}]{Wu:2010jy}%
  \BibitemOpen
  \bibfield  {author} {\bibinfo {author} {\bibfnamefont {J.-J.}\ \bibnamefont
  {Wu}}, \bibinfo {author} {\bibfnamefont {R.}~\bibnamefont {Molina}}, \bibinfo
  {author} {\bibfnamefont {E.}~\bibnamefont {Oset}},\ and\ \bibinfo {author}
  {\bibfnamefont {B.~S.}\ \bibnamefont {Zou}},\ }\bibfield  {title} {\bibinfo
  {title} {{Prediction of narrow $N^*$ and $\Lambda^*$ resonances with hidden
  charm above 4 GeV}},\ }\href {https://doi.org/10.1103/PhysRevLett.105.232001}
  {\bibfield  {journal} {\bibinfo  {journal} {Phys. Rev. Lett.}\ }\textbf
  {\bibinfo {volume} {105}},\ \bibinfo {pages} {232001} (\bibinfo {year}
  {2010})}\BibitemShut {NoStop}%
\bibitem [{\citenamefont {Wu}\ \emph {et~al.}(2011)\citenamefont {Wu},
  \citenamefont {Molina}, \citenamefont {Oset},\ and\ \citenamefont
  {Zou}}]{Wu:2010vk}%
  \BibitemOpen
  \bibfield  {author} {\bibinfo {author} {\bibfnamefont {J.-J.}\ \bibnamefont
  {Wu}}, \bibinfo {author} {\bibfnamefont {R.}~\bibnamefont {Molina}}, \bibinfo
  {author} {\bibfnamefont {E.}~\bibnamefont {Oset}},\ and\ \bibinfo {author}
  {\bibfnamefont {B.~S.}\ \bibnamefont {Zou}},\ }\bibfield  {title} {\bibinfo
  {title} {{Dynamically generated $N^{*}$ and $\Lambda^*$ resonances in the
  hidden charm sector around 4.3 GeV}},\ }\href
  {https://doi.org/10.1103/PhysRevC.84.015202} {\bibfield  {journal} {\bibinfo
  {journal} {Phys. Rev. C}\ }\textbf {\bibinfo {volume} {84}},\ \bibinfo
  {pages} {015202} (\bibinfo {year} {2011})}\BibitemShut {NoStop}%
\bibitem [{\citenamefont {Wang}\ \emph {et~al.}(2011)\citenamefont {Wang},
  \citenamefont {Huang}, \citenamefont {Zhang},\ and\ \citenamefont
  {Zou}}]{Wang:2011rga}%
  \BibitemOpen
  \bibfield  {author} {\bibinfo {author} {\bibfnamefont {W.~L.}\ \bibnamefont
  {Wang}}, \bibinfo {author} {\bibfnamefont {F.}~\bibnamefont {Huang}},
  \bibinfo {author} {\bibfnamefont {Z.~Y.}\ \bibnamefont {Zhang}},\ and\
  \bibinfo {author} {\bibfnamefont {B.~S.}\ \bibnamefont {Zou}},\ }\bibfield
  {title} {\bibinfo {title} {{$\Sigma_c \bar{D}$ and $\Lambda_c \bar{D}$ states
  in a chiral quark model}},\ }\href
  {https://doi.org/10.1103/PhysRevC.84.015203} {\bibfield  {journal} {\bibinfo
  {journal} {Phys. Rev. C}\ }\textbf {\bibinfo {volume} {84}},\ \bibinfo
  {pages} {015203} (\bibinfo {year} {2011})}\BibitemShut {NoStop}%
\bibitem [{\citenamefont {Yang}\ \emph {et~al.}(2012)\citenamefont {Yang},
  \citenamefont {Sun}, \citenamefont {He}, \citenamefont {Liu},\ and\
  \citenamefont {Zhu}}]{Yang:2011wz}%
  \BibitemOpen
  \bibfield  {author} {\bibinfo {author} {\bibfnamefont {Z.-C.}\ \bibnamefont
  {Yang}}, \bibinfo {author} {\bibfnamefont {Z.-F.}\ \bibnamefont {Sun}},
  \bibinfo {author} {\bibfnamefont {J.}~\bibnamefont {He}}, \bibinfo {author}
  {\bibfnamefont {X.}~\bibnamefont {Liu}},\ and\ \bibinfo {author}
  {\bibfnamefont {S.-L.}\ \bibnamefont {Zhu}},\ }\bibfield  {title} {\bibinfo
  {title} {{The possible hidden-charm molecular baryons composed of
  anti-charmed meson and charmed baryon}},\ }\href
  {https://doi.org/10.1088/1674-1137/36/1/002} {\bibfield  {journal} {\bibinfo
  {journal} {Chin. Phys. C}\ }\textbf {\bibinfo {volume} {36}},\ \bibinfo
  {pages} {6} (\bibinfo {year} {2012})}\BibitemShut {NoStop}%
\bibitem [{\citenamefont {Wu}\ \emph {et~al.}(2012)\citenamefont {Wu},
  \citenamefont {Lee},\ and\ \citenamefont {Zou}}]{Wu:2012md}%
  \BibitemOpen
  \bibfield  {author} {\bibinfo {author} {\bibfnamefont {J.-J.}\ \bibnamefont
  {Wu}}, \bibinfo {author} {\bibfnamefont {T.~S.~H.}\ \bibnamefont {Lee}},\
  and\ \bibinfo {author} {\bibfnamefont {B.~S.}\ \bibnamefont {Zou}},\
  }\bibfield  {title} {\bibinfo {title} {{Nucleon Resonances with Hidden Charm
  in Coupled-Channel Models}},\ }\href
  {https://doi.org/10.1103/PhysRevC.85.044002} {\bibfield  {journal} {\bibinfo
  {journal} {Phys. Rev. C}\ }\textbf {\bibinfo {volume} {85}},\ \bibinfo
  {pages} {044002} (\bibinfo {year} {2012})}\BibitemShut {NoStop}%
\bibitem [{\citenamefont {Li}\ and\ \citenamefont {Liu}(2014)}]{Li:2014gra}%
  \BibitemOpen
  \bibfield  {author} {\bibinfo {author} {\bibfnamefont {X.-Q.}\ \bibnamefont
  {Li}}\ and\ \bibinfo {author} {\bibfnamefont {X.}~\bibnamefont {Liu}},\
  }\bibfield  {title} {\bibinfo {title} {{A possible global group structure for
  exotic states}},\ }\href {https://doi.org/10.1140/epjc/s10052-014-3198-3}
  {\bibfield  {journal} {\bibinfo  {journal} {Eur. Phys. J. C}\ }\textbf
  {\bibinfo {volume} {74}},\ \bibinfo {pages} {3198} (\bibinfo {year}
  {2014})}\BibitemShut {NoStop}%
\bibitem [{\citenamefont {Chen}\ \emph {et~al.}(2015)\citenamefont {Chen},
  \citenamefont {Liu}, \citenamefont {Li},\ and\ \citenamefont
  {Zhu}}]{Chen:2015loa}%
  \BibitemOpen
  \bibfield  {author} {\bibinfo {author} {\bibfnamefont {R.}~\bibnamefont
  {Chen}}, \bibinfo {author} {\bibfnamefont {X.}~\bibnamefont {Liu}}, \bibinfo
  {author} {\bibfnamefont {X.-Q.}\ \bibnamefont {Li}},\ and\ \bibinfo {author}
  {\bibfnamefont {S.-L.}\ \bibnamefont {Zhu}},\ }\bibfield  {title} {\bibinfo
  {title} {{Identifying exotic hidden-charm pentaquarks}},\ }\href
  {https://doi.org/10.1103/PhysRevLett.115.132002} {\bibfield  {journal}
  {\bibinfo  {journal} {Phys. Rev. Lett.}\ }\textbf {\bibinfo {volume} {115}},\
  \bibinfo {pages} {132002} (\bibinfo {year} {2015})}\BibitemShut {NoStop}%
\bibitem [{\citenamefont {Karliner}\ and\ \citenamefont
  {Rosner}(2015)}]{Karliner:2015ina}%
  \BibitemOpen
  \bibfield  {author} {\bibinfo {author} {\bibfnamefont {M.}~\bibnamefont
  {Karliner}}\ and\ \bibinfo {author} {\bibfnamefont {J.~L.}\ \bibnamefont
  {Rosner}},\ }\bibfield  {title} {\bibinfo {title} {{New Exotic Meson and
  Baryon Resonances from Doubly-Heavy Hadronic Molecules}},\ }\href
  {https://doi.org/10.1103/PhysRevLett.115.122001} {\bibfield  {journal}
  {\bibinfo  {journal} {Phys. Rev. Lett.}\ }\textbf {\bibinfo {volume} {115}},\
  \bibinfo {pages} {122001} (\bibinfo {year} {2015})}\BibitemShut {NoStop}%
\bibitem [{\citenamefont {Chen}\ \emph {et~al.}(2016)\citenamefont {Chen},
  \citenamefont {Chen}, \citenamefont {Liu},\ and\ \citenamefont
  {Zhu}}]{Chen:2016qju}%
  \BibitemOpen
  \bibfield  {author} {\bibinfo {author} {\bibfnamefont {H.-X.}\ \bibnamefont
  {Chen}}, \bibinfo {author} {\bibfnamefont {W.}~\bibnamefont {Chen}}, \bibinfo
  {author} {\bibfnamefont {X.}~\bibnamefont {Liu}},\ and\ \bibinfo {author}
  {\bibfnamefont {S.-L.}\ \bibnamefont {Zhu}},\ }\bibfield  {title} {\bibinfo
  {title} {{The hidden-charm pentaquark and tetraquark states}},\ }\href
  {https://doi.org/10.1016/j.physrep.2016.05.004} {\bibfield  {journal}
  {\bibinfo  {journal} {Phys. Rept.}\ }\textbf {\bibinfo {volume} {639}},\
  \bibinfo {pages} {1} (\bibinfo {year} {2016})},\ \Eprint
  {https://arxiv.org/abs/1601.02092} {arXiv:1601.02092 [hep-ph]} \BibitemShut
  {NoStop}%
\bibitem [{\citenamefont {Liu}\ \emph {et~al.}(2019)\citenamefont {Liu},
  \citenamefont {Pan}, \citenamefont {Peng}, \citenamefont
  {S\'anchez~S\'anchez}, \citenamefont {Geng}, \citenamefont {Hosaka},\ and\
  \citenamefont {Pavon~Valderrama}}]{Liu:2019tjn}%
  \BibitemOpen
  \bibfield  {author} {\bibinfo {author} {\bibfnamefont {M.-Z.}\ \bibnamefont
  {Liu}}, \bibinfo {author} {\bibfnamefont {Y.-W.}\ \bibnamefont {Pan}},
  \bibinfo {author} {\bibfnamefont {F.-Z.}\ \bibnamefont {Peng}}, \bibinfo
  {author} {\bibfnamefont {M.}~\bibnamefont {S\'anchez~S\'anchez}}, \bibinfo
  {author} {\bibfnamefont {L.-S.}\ \bibnamefont {Geng}}, \bibinfo {author}
  {\bibfnamefont {A.}~\bibnamefont {Hosaka}},\ and\ \bibinfo {author}
  {\bibfnamefont {M.}~\bibnamefont {Pavon~Valderrama}},\ }\bibfield  {title}
  {\bibinfo {title} {{Emergence of a complete heavy-quark spin symmetry
  multiplet: seven molecular pentaquarks in light of the latest LHCb
  analysis}},\ }\href {https://doi.org/10.1103/PhysRevLett.122.242001}
  {\bibfield  {journal} {\bibinfo  {journal} {Phys. Rev. Lett.}\ }\textbf
  {\bibinfo {volume} {122}},\ \bibinfo {pages} {242001} (\bibinfo {year}
  {2019})}\BibitemShut {NoStop}%
\bibitem [{\citenamefont {Du}\ \emph {et~al.}(2020)\citenamefont {Du},
  \citenamefont {Baru}, \citenamefont {Guo}, \citenamefont {Hanhart},
  \citenamefont {Mei{\ss}ner}, \citenamefont {Oller},\ and\ \citenamefont
  {Wang}}]{Du:2019pij}%
  \BibitemOpen
  \bibfield  {author} {\bibinfo {author} {\bibfnamefont {M.-L.}\ \bibnamefont
  {Du}}, \bibinfo {author} {\bibfnamefont {V.}~\bibnamefont {Baru}}, \bibinfo
  {author} {\bibfnamefont {F.-K.}\ \bibnamefont {Guo}}, \bibinfo {author}
  {\bibfnamefont {C.}~\bibnamefont {Hanhart}}, \bibinfo {author} {\bibfnamefont
  {U.-G.}\ \bibnamefont {Mei{\ss}ner}}, \bibinfo {author} {\bibfnamefont
  {J.~A.}\ \bibnamefont {Oller}},\ and\ \bibinfo {author} {\bibfnamefont
  {Q.}~\bibnamefont {Wang}},\ }\bibfield  {title} {\bibinfo {title}
  {{Interpretation of the LHCb $P_c$ States as Hadronic Molecules and Hints of
  a Narrow $P_c(4380)$}},\ }\href
  {https://doi.org/10.1103/PhysRevLett.124.072001} {\bibfield  {journal}
  {\bibinfo  {journal} {Phys. Rev. Lett.}\ }\textbf {\bibinfo {volume} {124}},\
  \bibinfo {pages} {072001} (\bibinfo {year} {2020})},\ \Eprint
  {https://arxiv.org/abs/1910.11846} {arXiv:1910.11846 [hep-ph]} \BibitemShut
  {NoStop}%
\bibitem [{\citenamefont {Du}\ \emph {et~al.}(2021)\citenamefont {Du},
  \citenamefont {Baru}, \citenamefont {Guo}, \citenamefont {Hanhart},
  \citenamefont {Mei{\ss}ner}, \citenamefont {Oller},\ and\ \citenamefont
  {Wang}}]{Du:2021fmf}%
  \BibitemOpen
  \bibfield  {author} {\bibinfo {author} {\bibfnamefont {M.-L.}\ \bibnamefont
  {Du}}, \bibinfo {author} {\bibfnamefont {V.}~\bibnamefont {Baru}}, \bibinfo
  {author} {\bibfnamefont {F.-K.}\ \bibnamefont {Guo}}, \bibinfo {author}
  {\bibfnamefont {C.}~\bibnamefont {Hanhart}}, \bibinfo {author} {\bibfnamefont
  {U.-G.}\ \bibnamefont {Mei{\ss}ner}}, \bibinfo {author} {\bibfnamefont
  {J.~A.}\ \bibnamefont {Oller}},\ and\ \bibinfo {author} {\bibfnamefont
  {Q.}~\bibnamefont {Wang}},\ }\bibfield  {title} {\bibinfo {title}
  {{Revisiting the nature of the P$_{c}$ pentaquarks}},\ }\href
  {https://doi.org/10.1007/JHEP08(2021)157} {\bibfield  {journal} {\bibinfo
  {journal} {JHEP}\ }\textbf {\bibinfo {volume} {08}},\ \bibinfo {pages}
  {157}},\ \Eprint {https://arxiv.org/abs/2102.07159} {arXiv:2102.07159
  [hep-ph]} \BibitemShut {NoStop}%
\bibitem [{\citenamefont {Chen}\ \emph
  {et~al.}(2019{\natexlab{a}})\citenamefont {Chen}, \citenamefont {Chen},\ and\
  \citenamefont {Zhu}}]{Chen:2019bip}%
  \BibitemOpen
  \bibfield  {author} {\bibinfo {author} {\bibfnamefont {H.-X.}\ \bibnamefont
  {Chen}}, \bibinfo {author} {\bibfnamefont {W.}~\bibnamefont {Chen}},\ and\
  \bibinfo {author} {\bibfnamefont {S.-L.}\ \bibnamefont {Zhu}},\ }\bibfield
  {title} {\bibinfo {title} {{Possible interpretations of the $P_c(4312)$,
  $P_c(4440)$, and $P_c(4457)$}},\ }\href
  {https://doi.org/10.1103/PhysRevD.100.051501} {\bibfield  {journal} {\bibinfo
   {journal} {Phys. Rev. D}\ }\textbf {\bibinfo {volume} {100}},\ \bibinfo
  {pages} {051501} (\bibinfo {year} {2019}{\natexlab{a}})},\ \Eprint
  {https://arxiv.org/abs/1903.11001} {arXiv:1903.11001 [hep-ph]} \BibitemShut
  {NoStop}%
\bibitem [{\citenamefont {Chen}\ \emph
  {et~al.}(2019{\natexlab{b}})\citenamefont {Chen}, \citenamefont {Sun},
  \citenamefont {Liu},\ and\ \citenamefont {Zhu}}]{Chen:2019asm}%
  \BibitemOpen
  \bibfield  {author} {\bibinfo {author} {\bibfnamefont {R.}~\bibnamefont
  {Chen}}, \bibinfo {author} {\bibfnamefont {Z.-F.}\ \bibnamefont {Sun}},
  \bibinfo {author} {\bibfnamefont {X.}~\bibnamefont {Liu}},\ and\ \bibinfo
  {author} {\bibfnamefont {S.-L.}\ \bibnamefont {Zhu}},\ }\bibfield  {title}
  {\bibinfo {title} {{Strong LHCb evidence supporting the existence of the
  hidden-charm molecular pentaquarks}},\ }\href
  {https://doi.org/10.1103/PhysRevD.100.011502} {\bibfield  {journal} {\bibinfo
   {journal} {Phys. Rev. D}\ }\textbf {\bibinfo {volume} {100}},\ \bibinfo
  {pages} {011502} (\bibinfo {year} {2019}{\natexlab{b}})}\BibitemShut
  {NoStop}%
\bibitem [{\citenamefont {Guo}\ \emph {et~al.}(2019)\citenamefont {Guo},
  \citenamefont {Jing}, \citenamefont {Meissner},\ and\ \citenamefont
  {Sakai}}]{Guo:2019twa}%
  \BibitemOpen
  \bibfield  {author} {\bibinfo {author} {\bibfnamefont {F.-K.}\ \bibnamefont
  {Guo}}, \bibinfo {author} {\bibfnamefont {H.-J.}\ \bibnamefont {Jing}},
  \bibinfo {author} {\bibfnamefont {U.-G.}\ \bibnamefont {Meissner}},\ and\
  \bibinfo {author} {\bibfnamefont {S.}~\bibnamefont {Sakai}},\ }\bibfield
  {title} {\bibinfo {title} {Isospin breaking decays as a diagnosis of the
  hadronic molecular structure of the pc(4457)},\ }\href
  {https://doi.org/10.1103/PhysRevD.99.091501} {\bibfield  {journal} {\bibinfo
  {journal} {Phys. Rev. D}\ }\textbf {\bibinfo {volume} {99}},\ \bibinfo
  {pages} {091501} (\bibinfo {year} {2019})}\BibitemShut {NoStop}%
\bibitem [{\citenamefont {He}(2019)}]{He:2019ify}%
  \BibitemOpen
  \bibfield  {author} {\bibinfo {author} {\bibfnamefont {J.}~\bibnamefont
  {He}},\ }\bibfield  {title} {\bibinfo {title} {{Study of $P_c(4457)$,
  $P_c(4440)$, and $P_c(4312)$ in a quasipotential Bethe-Salpeter equation
  approach}},\ }\href {https://doi.org/10.1140/epjc/s10052-019-6906-1}
  {\bibfield  {journal} {\bibinfo  {journal} {Eur. Phys. J. C}\ }\textbf
  {\bibinfo {volume} {79}},\ \bibinfo {pages} {393} (\bibinfo {year}
  {2019})}\BibitemShut {NoStop}%
\bibitem [{\citenamefont {Guo}\ and\ \citenamefont
  {Oller}(2019)}]{Guo:2019kdc}%
  \BibitemOpen
  \bibfield  {author} {\bibinfo {author} {\bibfnamefont {Z.-H.}\ \bibnamefont
  {Guo}}\ and\ \bibinfo {author} {\bibfnamefont {J.}~\bibnamefont {Oller}},\
  }\bibfield  {title} {\bibinfo {title} {Anatomy of the newly observed
  hidden-charm pentaquark states: Pc(4312), pc(4440) and pc(4457)},\ }\href
  {https://doi.org/10.1016/j.physletb.2019.04.021} {\bibfield  {journal}
  {\bibinfo  {journal} {Phys. Lett. B}\ }\textbf {\bibinfo {volume} {793}},\
  \bibinfo {pages} {144} (\bibinfo {year} {2019})}\BibitemShut {NoStop}%
\bibitem [{\citenamefont {Shimizu}\ \emph {et~al.}(2019)\citenamefont
  {Shimizu}, \citenamefont {Yamaguchi},\ and\ \citenamefont
  {Harada}}]{Shimizu:2019ptd}%
  \BibitemOpen
  \bibfield  {author} {\bibinfo {author} {\bibfnamefont {Y.}~\bibnamefont
  {Shimizu}}, \bibinfo {author} {\bibfnamefont {Y.}~\bibnamefont {Yamaguchi}},\
  and\ \bibinfo {author} {\bibfnamefont {M.}~\bibnamefont {Harada}},\
  }\bibfield  {title} {\bibinfo {title} {{Heavy quark spin multiplet structure
  of $P_c(4312)$, $P_c(4440)$, and $P_c(4457)$}},\ }\href@noop {} {\  (\bibinfo
  {year} {2019})},\ \Eprint {https://arxiv.org/abs/1904.00587}
  {arXiv:1904.00587 [hep-ph]} \BibitemShut {NoStop}%
\bibitem [{\citenamefont {Xiao}\ \emph
  {et~al.}(2019{\natexlab{a}})\citenamefont {Xiao}, \citenamefont {Huang},
  \citenamefont {Dong}, \citenamefont {Geng},\ and\ \citenamefont
  {Chen}}]{Xiao:2019pjg}%
  \BibitemOpen
  \bibfield  {author} {\bibinfo {author} {\bibfnamefont {C.-J.}\ \bibnamefont
  {Xiao}}, \bibinfo {author} {\bibfnamefont {Y.}~\bibnamefont {Huang}},
  \bibinfo {author} {\bibfnamefont {Y.-B.}\ \bibnamefont {Dong}}, \bibinfo
  {author} {\bibfnamefont {L.-S.}\ \bibnamefont {Geng}},\ and\ \bibinfo
  {author} {\bibfnamefont {D.-Y.}\ \bibnamefont {Chen}},\ }\bibfield  {title}
  {\bibinfo {title} {Exploring the molecular scenario of pc(4312), pc(4440),
  and pc(4457)},\ }\href {https://doi.org/10.1103/PhysRevD.100.014022}
  {\bibfield  {journal} {\bibinfo  {journal} {Phys. Rev. D}\ }\textbf {\bibinfo
  {volume} {100}},\ \bibinfo {pages} {014022} (\bibinfo {year}
  {2019}{\natexlab{a}})}\BibitemShut {NoStop}%
\bibitem [{\citenamefont {Xiao}\ \emph
  {et~al.}(2019{\natexlab{b}})\citenamefont {Xiao}, \citenamefont {Nieves},\
  and\ \citenamefont {Oset}}]{Xiao:2019aya}%
  \BibitemOpen
  \bibfield  {author} {\bibinfo {author} {\bibfnamefont {C.~W.}\ \bibnamefont
  {Xiao}}, \bibinfo {author} {\bibfnamefont {J.}~\bibnamefont {Nieves}},\ and\
  \bibinfo {author} {\bibfnamefont {E.}~\bibnamefont {Oset}},\ }\bibfield
  {title} {\bibinfo {title} {{Heavy quark spin symmetric molecular states from
  ${\bar D}^{(*)}\Sigma_c^{(*)}$ and other coupled channels in the light of the
  recent LHCb pentaquarks}},\ }\href
  {https://doi.org/10.1103/PhysRevD.100.014021} {\bibfield  {journal} {\bibinfo
   {journal} {Phys. Rev. D}\ }\textbf {\bibinfo {volume} {100}},\ \bibinfo
  {pages} {014021} (\bibinfo {year} {2019}{\natexlab{b}})}\BibitemShut
  {NoStop}%
\bibitem [{\citenamefont {Wang}\ \emph
  {et~al.}(2020{\natexlab{a}})\citenamefont {Wang}, \citenamefont {Chen},
  \citenamefont {Liu},\ and\ \citenamefont {Liu}}]{Wang:2019nwt}%
  \BibitemOpen
  \bibfield  {author} {\bibinfo {author} {\bibfnamefont {F.-L.}\ \bibnamefont
  {Wang}}, \bibinfo {author} {\bibfnamefont {R.}~\bibnamefont {Chen}}, \bibinfo
  {author} {\bibfnamefont {Z.-W.}\ \bibnamefont {Liu}},\ and\ \bibinfo {author}
  {\bibfnamefont {X.}~\bibnamefont {Liu}},\ }\bibfield  {title} {\bibinfo
  {title} {{Probing new types of $P_c$ states inspired by the interaction
  between $S$-wave charmed baryon and anti-charmed meson in a $\bar T$
  doublet}},\ }\href {https://doi.org/10.1103/PhysRevC.101.025201} {\bibfield
  {journal} {\bibinfo  {journal} {Phys. Rev. C}\ }\textbf {\bibinfo {volume}
  {101}},\ \bibinfo {pages} {025201} (\bibinfo {year}
  {2020}{\natexlab{a}})}\BibitemShut {NoStop}%
\bibitem [{\citenamefont {Meng}\ \emph {et~al.}(2019)\citenamefont {Meng},
  \citenamefont {Wang}, \citenamefont {Wang},\ and\ \citenamefont
  {Zhu}}]{Meng:2019ilv}%
  \BibitemOpen
  \bibfield  {author} {\bibinfo {author} {\bibfnamefont {L.}~\bibnamefont
  {Meng}}, \bibinfo {author} {\bibfnamefont {B.}~\bibnamefont {Wang}}, \bibinfo
  {author} {\bibfnamefont {G.-J.}\ \bibnamefont {Wang}},\ and\ \bibinfo
  {author} {\bibfnamefont {S.-L.}\ \bibnamefont {Zhu}},\ }\bibfield  {title}
  {\bibinfo {title} {{The hidden charm pentaquark states and
  $\Sigma_c\bar{D}^{(*)}$ interaction in chiral perturbation theory}},\ }\href
  {https://doi.org/10.1103/PhysRevD.100.014031} {\bibfield  {journal} {\bibinfo
   {journal} {Phys. Rev. D}\ }\textbf {\bibinfo {volume} {100}},\ \bibinfo
  {pages} {014031} (\bibinfo {year} {2019})}\BibitemShut {NoStop}%
\bibitem [{\citenamefont {Wu}\ \emph {et~al.}(2019)\citenamefont {Wu},
  \citenamefont {Lee},\ and\ \citenamefont {Zou}}]{Wu:2019adv}%
  \BibitemOpen
  \bibfield  {author} {\bibinfo {author} {\bibfnamefont {J.-J.}\ \bibnamefont
  {Wu}}, \bibinfo {author} {\bibfnamefont {T.~S.~H.}\ \bibnamefont {Lee}},\
  and\ \bibinfo {author} {\bibfnamefont {B.-S.}\ \bibnamefont {Zou}},\
  }\bibfield  {title} {\bibinfo {title} {{Nucleon resonances with hidden charm
  in \ensuremath{\gamma}p reactions}},\ }\href
  {https://doi.org/10.1103/PhysRevC.100.035206} {\bibfield  {journal} {\bibinfo
   {journal} {Phys. Rev. C}\ }\textbf {\bibinfo {volume} {100}},\ \bibinfo
  {pages} {035206} (\bibinfo {year} {2019})}\BibitemShut {NoStop}%
\bibitem [{\citenamefont {Voloshin}(2019)}]{Voloshin:2019aut}%
  \BibitemOpen
  \bibfield  {author} {\bibinfo {author} {\bibfnamefont {M.~B.}\ \bibnamefont
  {Voloshin}},\ }\bibfield  {title} {\bibinfo {title} {{Some decay properties
  of hidden-charm pentaquarks as baryon-meson molecules}},\ }\href
  {https://doi.org/10.1103/PhysRevD.100.034020} {\bibfield  {journal} {\bibinfo
   {journal} {Phys. Rev. D}\ }\textbf {\bibinfo {volume} {100}},\ \bibinfo
  {pages} {034020} (\bibinfo {year} {2019})},\ \Eprint
  {https://arxiv.org/abs/1907.01476} {arXiv:1907.01476 [hep-ph]} \BibitemShut
  {NoStop}%
\bibitem [{\citenamefont {Wang}\ and\ \citenamefont
  {Wang}(2020)}]{Wang:2019hyc}%
  \BibitemOpen
  \bibfield  {author} {\bibinfo {author} {\bibfnamefont {Z.-G.}\ \bibnamefont
  {Wang}}\ and\ \bibinfo {author} {\bibfnamefont {X.}~\bibnamefont {Wang}},\
  }\bibfield  {title} {\bibinfo {title} {{Analysis of the strong decays of the
  $P_c(4312)$ as a pentaquark molecular state with QCD sum rules}},\ }\href
  {https://doi.org/10.1088/1674-1137/ababf7} {\bibfield  {journal} {\bibinfo
  {journal} {Chin. Phys. C}\ }\textbf {\bibinfo {volume} {44}},\ \bibinfo
  {pages} {103102} (\bibinfo {year} {2020})}\BibitemShut {NoStop}%
\bibitem [{\citenamefont {Yamaguchi}\ \emph {et~al.}(2020)\citenamefont
  {Yamaguchi}, \citenamefont {Garc{\'\i}a-Tecocoatzi}, \citenamefont
  {Giachino}, \citenamefont {Hosaka}, \citenamefont {Santopinto}, \citenamefont
  {Takeuchi},\ and\ \citenamefont {Takizawa}}]{Yamaguchi:2019seo}%
  \BibitemOpen
  \bibfield  {author} {\bibinfo {author} {\bibfnamefont {Y.}~\bibnamefont
  {Yamaguchi}}, \bibinfo {author} {\bibfnamefont {H.}~\bibnamefont
  {Garc{\'\i}a-Tecocoatzi}}, \bibinfo {author} {\bibfnamefont {A.}~\bibnamefont
  {Giachino}}, \bibinfo {author} {\bibfnamefont {A.}~\bibnamefont {Hosaka}},
  \bibinfo {author} {\bibfnamefont {E.}~\bibnamefont {Santopinto}}, \bibinfo
  {author} {\bibfnamefont {S.}~\bibnamefont {Takeuchi}},\ and\ \bibinfo
  {author} {\bibfnamefont {M.}~\bibnamefont {Takizawa}},\ }\bibfield  {title}
  {\bibinfo {title} {{$P_c$ pentaquarks with chiral tensor and quark
  dynamics}},\ }\href {https://doi.org/10.1103/PhysRevD.101.091502} {\bibfield
  {journal} {\bibinfo  {journal} {Phys. Rev. D}\ }\textbf {\bibinfo {volume}
  {101}},\ \bibinfo {pages} {091502} (\bibinfo {year} {2020})},\ \Eprint
  {https://arxiv.org/abs/1907.04684} {arXiv:1907.04684 [hep-ph]} \BibitemShut
  {NoStop}%
\bibitem [{\citenamefont {Liu}\ \emph {et~al.}(2021)\citenamefont {Liu},
  \citenamefont {Wu}, \citenamefont {S\'anchez~S\'anchez}, \citenamefont
  {Valderrama}, \citenamefont {Geng},\ and\ \citenamefont {Xie}}]{Liu:2019zvb}%
  \BibitemOpen
  \bibfield  {author} {\bibinfo {author} {\bibfnamefont {M.-Z.}\ \bibnamefont
  {Liu}}, \bibinfo {author} {\bibfnamefont {T.-W.}\ \bibnamefont {Wu}},
  \bibinfo {author} {\bibfnamefont {M.}~\bibnamefont {S\'anchez~S\'anchez}},
  \bibinfo {author} {\bibfnamefont {M.~P.}\ \bibnamefont {Valderrama}},
  \bibinfo {author} {\bibfnamefont {L.-S.}\ \bibnamefont {Geng}},\ and\
  \bibinfo {author} {\bibfnamefont {J.-J.}\ \bibnamefont {Xie}},\ }\bibfield
  {title} {\bibinfo {title} {{Spin-parities of the $P_c(4440)$ and $P_c(4457)$
  in the one-boson-exchange model}},\ }\href
  {https://doi.org/10.1103/PhysRevD.103.054004} {\bibfield  {journal} {\bibinfo
   {journal} {Phys. Rev. D}\ }\textbf {\bibinfo {volume} {103}},\ \bibinfo
  {pages} {054004} (\bibinfo {year} {2021})}\BibitemShut {NoStop}%
\bibitem [{\citenamefont {Lin}\ and\ \citenamefont {Zou}(2019)}]{Lin:2019qiv}%
  \BibitemOpen
  \bibfield  {author} {\bibinfo {author} {\bibfnamefont {Y.-H.}\ \bibnamefont
  {Lin}}\ and\ \bibinfo {author} {\bibfnamefont {B.-S.}\ \bibnamefont {Zou}},\
  }\bibfield  {title} {\bibinfo {title} {{Strong decays of the latest LHCb
  pentaquark candidates in hadronic molecule pictures}},\ }\href
  {https://doi.org/10.1103/PhysRevD.100.056005} {\bibfield  {journal} {\bibinfo
   {journal} {Phys. Rev. D}\ }\textbf {\bibinfo {volume} {100}},\ \bibinfo
  {pages} {056005} (\bibinfo {year} {2019})}\BibitemShut {NoStop}%
\bibitem [{\citenamefont {Wang}\ \emph {et~al.}(2019)\citenamefont {Wang},
  \citenamefont {Meng},\ and\ \citenamefont {Zhu}}]{Wang:2019ato}%
  \BibitemOpen
  \bibfield  {author} {\bibinfo {author} {\bibfnamefont {B.}~\bibnamefont
  {Wang}}, \bibinfo {author} {\bibfnamefont {L.}~\bibnamefont {Meng}},\ and\
  \bibinfo {author} {\bibfnamefont {S.-L.}\ \bibnamefont {Zhu}},\ }\bibfield
  {title} {\bibinfo {title} {{Hidden-charm and hidden-bottom molecular
  pentaquarks in chiral effective field theory}},\ }\href
  {https://doi.org/10.1007/JHEP11(2019)108} {\bibfield  {journal} {\bibinfo
  {journal} {JHEP}\ }\textbf {\bibinfo {volume} {11}},\ \bibinfo {pages}
  {108}}\BibitemShut {NoStop}%
\bibitem [{\citenamefont {Gutsche}\ and\ \citenamefont
  {Lyubovitskij}(2019)}]{Gutsche:2019mkg}%
  \BibitemOpen
  \bibfield  {author} {\bibinfo {author} {\bibfnamefont {T.}~\bibnamefont
  {Gutsche}}\ and\ \bibinfo {author} {\bibfnamefont {V.~E.}\ \bibnamefont
  {Lyubovitskij}},\ }\bibfield  {title} {\bibinfo {title} {{Structure and
  decays of hidden heavy pentaquarks}},\ }\href
  {https://doi.org/10.1103/PhysRevD.100.094031} {\bibfield  {journal} {\bibinfo
   {journal} {Phys. Rev. D}\ }\textbf {\bibinfo {volume} {100}},\ \bibinfo
  {pages} {094031} (\bibinfo {year} {2019})}\BibitemShut {NoStop}%
\bibitem [{\citenamefont {Burns}\ and\ \citenamefont
  {Swanson}(2019)}]{Burns:2019iih}%
  \BibitemOpen
  \bibfield  {author} {\bibinfo {author} {\bibfnamefont {T.~J.}\ \bibnamefont
  {Burns}}\ and\ \bibinfo {author} {\bibfnamefont {E.~S.}\ \bibnamefont
  {Swanson}},\ }\bibfield  {title} {\bibinfo {title} {{Molecular interpretation
  of the $P_c$(4440) and $P_c$(4457) states}},\ }\href
  {https://doi.org/10.1103/PhysRevD.100.114033} {\bibfield  {journal} {\bibinfo
   {journal} {Phys. Rev. D}\ }\textbf {\bibinfo {volume} {100}},\ \bibinfo
  {pages} {114033} (\bibinfo {year} {2019})}\BibitemShut {NoStop}%
\bibitem [{\citenamefont {Wang}\ \emph
  {et~al.}(2020{\natexlab{b}})\citenamefont {Wang}, \citenamefont {Xiao},
  \citenamefont {Chen}, \citenamefont {Liu}, \citenamefont {Liu},\ and\
  \citenamefont {Zhu}}]{Wang:2019spc}%
  \BibitemOpen
  \bibfield  {author} {\bibinfo {author} {\bibfnamefont {G.-J.}\ \bibnamefont
  {Wang}}, \bibinfo {author} {\bibfnamefont {L.-Y.}\ \bibnamefont {Xiao}},
  \bibinfo {author} {\bibfnamefont {R.}~\bibnamefont {Chen}}, \bibinfo {author}
  {\bibfnamefont {X.-H.}\ \bibnamefont {Liu}}, \bibinfo {author} {\bibfnamefont
  {X.}~\bibnamefont {Liu}},\ and\ \bibinfo {author} {\bibfnamefont {S.-L.}\
  \bibnamefont {Zhu}},\ }\bibfield  {title} {\bibinfo {title} {{Probing
  hidden-charm decay properties of $P_c$ states in a molecular scenario}},\
  }\href {https://doi.org/10.1103/PhysRevD.102.036012} {\bibfield  {journal}
  {\bibinfo  {journal} {Phys. Rev. D}\ }\textbf {\bibinfo {volume} {102}},\
  \bibinfo {pages} {036012} (\bibinfo {year} {2020}{\natexlab{b}})}\BibitemShut
  {NoStop}%
\bibitem [{\citenamefont {Xu}\ \emph {et~al.}(2020)\citenamefont {Xu},
  \citenamefont {Li}, \citenamefont {Chang},\ and\ \citenamefont
  {Wang}}]{Xu:2020gjl}%
  \BibitemOpen
  \bibfield  {author} {\bibinfo {author} {\bibfnamefont {H.}~\bibnamefont
  {Xu}}, \bibinfo {author} {\bibfnamefont {Q.}~\bibnamefont {Li}}, \bibinfo
  {author} {\bibfnamefont {C.-H.}\ \bibnamefont {Chang}},\ and\ \bibinfo
  {author} {\bibfnamefont {G.-L.}\ \bibnamefont {Wang}},\ }\bibfield  {title}
  {\bibinfo {title} {{Recently observed $P_c$ as molecular states and possible
  mixture of $P_c(4457)$}},\ }\href
  {https://doi.org/10.1103/PhysRevD.101.054037} {\bibfield  {journal} {\bibinfo
   {journal} {Phys. Rev. D}\ }\textbf {\bibinfo {volume} {101}},\ \bibinfo
  {pages} {054037} (\bibinfo {year} {2020})}\BibitemShut {NoStop}%
\bibitem [{\citenamefont {Kuang}\ \emph {et~al.}(2020)\citenamefont {Kuang},
  \citenamefont {Dai}, \citenamefont {Kang},\ and\ \citenamefont
  {Yao}}]{Kuang:2020bnk}%
  \BibitemOpen
  \bibfield  {author} {\bibinfo {author} {\bibfnamefont {S.-Q.}\ \bibnamefont
  {Kuang}}, \bibinfo {author} {\bibfnamefont {L.-Y.}\ \bibnamefont {Dai}},
  \bibinfo {author} {\bibfnamefont {X.-W.}\ \bibnamefont {Kang}},\ and\
  \bibinfo {author} {\bibfnamefont {D.-L.}\ \bibnamefont {Yao}},\ }\bibfield
  {title} {\bibinfo {title} {{Pole analysis on the hadron spectroscopy of
  $\Lambda_b\to J/\Psi p K^-$}},\ }\href
  {https://doi.org/10.1140/epjc/s10052-020-8008-5} {\bibfield  {journal}
  {\bibinfo  {journal} {Eur. Phys. J. C}\ }\textbf {\bibinfo {volume} {80}},\
  \bibinfo {pages} {433} (\bibinfo {year} {2020})}\BibitemShut {NoStop}%
\bibitem [{\citenamefont {Peng}\ \emph {et~al.}(2020)\citenamefont {Peng},
  \citenamefont {Liu}, \citenamefont {S{\'a}nchez~S{\'a}nchez},\ and\
  \citenamefont {Pavon~Valderrama}}]{Peng:2020xrf}%
  \BibitemOpen
  \bibfield  {author} {\bibinfo {author} {\bibfnamefont {F.-Z.}\ \bibnamefont
  {Peng}}, \bibinfo {author} {\bibfnamefont {M.-Z.}\ \bibnamefont {Liu}},
  \bibinfo {author} {\bibfnamefont {M.}~\bibnamefont
  {S{\'a}nchez~S{\'a}nchez}},\ and\ \bibinfo {author} {\bibfnamefont
  {M.}~\bibnamefont {Pavon~Valderrama}},\ }\bibfield  {title} {\bibinfo {title}
  {{Heavy-hadron molecules from light-meson-exchange saturation}},\ }\href
  {https://doi.org/10.1103/PhysRevD.102.114020} {\bibfield  {journal} {\bibinfo
   {journal} {Phys. Rev. D}\ }\textbf {\bibinfo {volume} {102}},\ \bibinfo
  {pages} {114020} (\bibinfo {year} {2020})},\ \Eprint
  {https://arxiv.org/abs/2004.05658} {arXiv:2004.05658 [hep-ph]} \BibitemShut
  {NoStop}%
\bibitem [{\citenamefont {Xiao}\ \emph {et~al.}(2020)\citenamefont {Xiao},
  \citenamefont {Lu}, \citenamefont {Wu},\ and\ \citenamefont
  {Geng}}]{Xiao:2020frg}%
  \BibitemOpen
  \bibfield  {author} {\bibinfo {author} {\bibfnamefont {C.~W.}\ \bibnamefont
  {Xiao}}, \bibinfo {author} {\bibfnamefont {J.~X.}\ \bibnamefont {Lu}},
  \bibinfo {author} {\bibfnamefont {J.~J.}\ \bibnamefont {Wu}},\ and\ \bibinfo
  {author} {\bibfnamefont {L.~S.}\ \bibnamefont {Geng}},\ }\bibfield  {title}
  {\bibinfo {title} {{How to reveal the nature of three or more pentaquark
  states}},\ }\href {https://doi.org/10.1103/PhysRevD.102.056018} {\bibfield
  {journal} {\bibinfo  {journal} {Phys. Rev. D}\ }\textbf {\bibinfo {volume}
  {102}},\ \bibinfo {pages} {056018} (\bibinfo {year} {2020})}\BibitemShut
  {NoStop}%
\bibitem [{\citenamefont {Dong}\ \emph {et~al.}(2021)\citenamefont {Dong},
  \citenamefont {Guo},\ and\ \citenamefont {Zou}}]{Dong:2021juy}%
  \BibitemOpen
  \bibfield  {author} {\bibinfo {author} {\bibfnamefont {X.-K.}\ \bibnamefont
  {Dong}}, \bibinfo {author} {\bibfnamefont {F.-K.}\ \bibnamefont {Guo}},\ and\
  \bibinfo {author} {\bibfnamefont {B.-S.}\ \bibnamefont {Zou}},\ }\bibfield
  {title} {\bibinfo {title} {{A survey of heavy-antiheavy hadronic
  molecules}},\ }\href {https://doi.org/10.13725/j.cnki.pip.2021.02.001}
  {\bibfield  {journal} {\bibinfo  {journal} {Progr. Phys.}\ }\textbf {\bibinfo
  {volume} {41}},\ \bibinfo {pages} {65} (\bibinfo {year} {2021})}\BibitemShut
  {NoStop}%
\bibitem [{\citenamefont {Burns}\ and\ \citenamefont
  {Swanson}(2022)}]{Burns:2021jlu}%
  \BibitemOpen
  \bibfield  {author} {\bibinfo {author} {\bibfnamefont {T.~J.}\ \bibnamefont
  {Burns}}\ and\ \bibinfo {author} {\bibfnamefont {E.~S.}\ \bibnamefont
  {Swanson}},\ }\bibfield  {title} {\bibinfo {title} {{Experimental constraints
  on the properties of $P_c$ states}},\ }\href
  {https://doi.org/10.1140/epja/s10050-022-00723-9} {\bibfield  {journal}
  {\bibinfo  {journal} {Eur. Phys. J. A}\ }\textbf {\bibinfo {volume} {58}},\
  \bibinfo {pages} {68} (\bibinfo {year} {2022})},\ \Eprint
  {https://arxiv.org/abs/2112.11527} {arXiv:2112.11527 [hep-ph]} \BibitemShut
  {NoStop}%
\bibitem [{\citenamefont {Zhang}\ \emph {et~al.}(2023)\citenamefont {Zhang},
  \citenamefont {Liu}, \citenamefont {Hu}, \citenamefont {Wang},\ and\
  \citenamefont {Mei\ss{}ner}}]{Zhang:2023czx}%
  \BibitemOpen
  \bibfield  {author} {\bibinfo {author} {\bibfnamefont {Z.}~\bibnamefont
  {Zhang}}, \bibinfo {author} {\bibfnamefont {J.}~\bibnamefont {Liu}}, \bibinfo
  {author} {\bibfnamefont {J.}~\bibnamefont {Hu}}, \bibinfo {author}
  {\bibfnamefont {Q.}~\bibnamefont {Wang}},\ and\ \bibinfo {author}
  {\bibfnamefont {U.-G.}\ \bibnamefont {Mei\ss{}ner}},\ }\bibfield  {title}
  {\bibinfo {title} {{Revealing the nature of hidden charm pentaquarks with
  machine learning}},\ }\href {https://doi.org/10.1016/j.scib.2023.04.018}
  {\bibfield  {journal} {\bibinfo  {journal} {Sci. Bull.}\ }\textbf {\bibinfo
  {volume} {68}},\ \bibinfo {pages} {981} (\bibinfo {year} {2023})}\BibitemShut
  {NoStop}%
\bibitem [{\citenamefont {Liu}\ \emph {et~al.}(2024)\citenamefont {Liu},
  \citenamefont {Pan},\ and\ \citenamefont {Geng}}]{Liu:2024ugt}%
  \BibitemOpen
  \bibfield  {author} {\bibinfo {author} {\bibfnamefont {M.-Z.}\ \bibnamefont
  {Liu}}, \bibinfo {author} {\bibfnamefont {Y.-W.}\ \bibnamefont {Pan}},\ and\
  \bibinfo {author} {\bibfnamefont {L.-S.}\ \bibnamefont {Geng}},\ }\bibfield
  {title} {\bibinfo {title} {{Studying the heavy quark spin symmetry multiplet
  of hadronic molecules D{\textasciimacron}(*){\ensuremath{\Sigma}}c(*) in the
  three-body decays of
  D{\textasciimacron}(*){\ensuremath{\Lambda}}c{\ensuremath{\pi}}}},\ }\href
  {https://doi.org/10.1103/PhysRevD.110.114022} {\bibfield  {journal} {\bibinfo
   {journal} {Phys. Rev. D}\ }\textbf {\bibinfo {volume} {110}},\ \bibinfo
  {pages} {114022} (\bibinfo {year} {2024})},\ \Eprint
  {https://arxiv.org/abs/2407.17318} {arXiv:2407.17318 [hep-ph]} \BibitemShut
  {NoStop}%
\bibitem [{\citenamefont {Li}\ \emph {et~al.}(2025)\citenamefont {Li},
  \citenamefont {Chang}, \citenamefont {Tong}, \citenamefont {Tan},
  \citenamefont {Wang},\ and\ \citenamefont {Wang}}]{Li:2025ejt}%
  \BibitemOpen
  \bibfield  {author} {\bibinfo {author} {\bibfnamefont {Q.}~\bibnamefont
  {Li}}, \bibinfo {author} {\bibfnamefont {C.-H.}\ \bibnamefont {Chang}},
  \bibinfo {author} {\bibfnamefont {X.}~\bibnamefont {Tong}}, \bibinfo {author}
  {\bibfnamefont {X.-Z.}\ \bibnamefont {Tan}}, \bibinfo {author} {\bibfnamefont
  {T.}~\bibnamefont {Wang}},\ and\ \bibinfo {author} {\bibfnamefont {G.-L.}\
  \bibnamefont {Wang}},\ }\bibfield  {title} {\bibinfo {title} {{Strong decays
  of $P_{\psi}^N(4440)^+$ and $P_\psi^N(4457)^+$ within the Bethe-Salpeter
  framework}},\ }\href@noop {} {\  (\bibinfo {year} {2025})},\ \Eprint
  {https://arxiv.org/abs/2503.08440} {arXiv:2503.08440 [hep-ph]} \BibitemShut
  {NoStop}%
\bibitem [{\citenamefont {Maiani}\ \emph {et~al.}(2015)\citenamefont {Maiani},
  \citenamefont {Polosa},\ and\ \citenamefont {Riquer}}]{Maiani:2015vwa}%
  \BibitemOpen
  \bibfield  {author} {\bibinfo {author} {\bibfnamefont {L.}~\bibnamefont
  {Maiani}}, \bibinfo {author} {\bibfnamefont {A.~D.}\ \bibnamefont {Polosa}},\
  and\ \bibinfo {author} {\bibfnamefont {V.}~\bibnamefont {Riquer}},\
  }\bibfield  {title} {\bibinfo {title} {{The New Pentaquarks in the Diquark
  Model}},\ }\href {https://doi.org/10.1016/j.physletb.2015.08.008} {\bibfield
  {journal} {\bibinfo  {journal} {Phys. Lett. B}\ }\textbf {\bibinfo {volume}
  {749}},\ \bibinfo {pages} {289} (\bibinfo {year} {2015})},\ \Eprint
  {https://arxiv.org/abs/1507.04980} {arXiv:1507.04980 [hep-ph]} \BibitemShut
  {NoStop}%
\bibitem [{\citenamefont {Ali}\ and\ \citenamefont
  {Parkhomenko}(2019)}]{Ali:2019npk}%
  \BibitemOpen
  \bibfield  {author} {\bibinfo {author} {\bibfnamefont {A.}~\bibnamefont
  {Ali}}\ and\ \bibinfo {author} {\bibfnamefont {A.~Y.}\ \bibnamefont
  {Parkhomenko}},\ }\bibfield  {title} {\bibinfo {title} {{Interpretation of
  the narrow $J/\psi p$ Peaks in $\Lambda_b \to J/\psi p K^-$ decay in the
  compact diquark model}},\ }\href
  {https://doi.org/10.1016/j.physletb.2019.05.002} {\bibfield  {journal}
  {\bibinfo  {journal} {Phys. Lett. B}\ }\textbf {\bibinfo {volume} {793}},\
  \bibinfo {pages} {365} (\bibinfo {year} {2019})},\ \Eprint
  {https://arxiv.org/abs/1904.00446} {arXiv:1904.00446 [hep-ph]} \BibitemShut
  {NoStop}%
\bibitem [{\citenamefont {Ali}\ \emph {et~al.}(2019)\citenamefont {Ali},
  \citenamefont {Ahmed}, \citenamefont {Aslam}, \citenamefont {Parkhomenko},\
  and\ \citenamefont {Rehman}}]{Ali:2019clg}%
  \BibitemOpen
  \bibfield  {author} {\bibinfo {author} {\bibfnamefont {A.}~\bibnamefont
  {Ali}}, \bibinfo {author} {\bibfnamefont {I.}~\bibnamefont {Ahmed}}, \bibinfo
  {author} {\bibfnamefont {M.~J.}\ \bibnamefont {Aslam}}, \bibinfo {author}
  {\bibfnamefont {A.~Y.}\ \bibnamefont {Parkhomenko}},\ and\ \bibinfo {author}
  {\bibfnamefont {A.}~\bibnamefont {Rehman}},\ }\bibfield  {title} {\bibinfo
  {title} {{Mass spectrum of the hidden-charm pentaquarks in the compact
  diquark model}},\ }\href {https://doi.org/10.1007/JHEP10(2019)256} {\bibfield
   {journal} {\bibinfo  {journal} {JHEP}\ }\textbf {\bibinfo {volume} {10}},\
  \bibinfo {pages} {256}},\ \Eprint {https://arxiv.org/abs/1907.06507}
  {arXiv:1907.06507 [hep-ph]} \BibitemShut {NoStop}%
\bibitem [{\citenamefont {Zhu}\ \emph {et~al.}(2019)\citenamefont {Zhu},
  \citenamefont {Liu}, \citenamefont {Huang},\ and\ \citenamefont
  {Qiao}}]{Zhu:2019iwm}%
  \BibitemOpen
  \bibfield  {author} {\bibinfo {author} {\bibfnamefont {R.}~\bibnamefont
  {Zhu}}, \bibinfo {author} {\bibfnamefont {X.}~\bibnamefont {Liu}}, \bibinfo
  {author} {\bibfnamefont {H.}~\bibnamefont {Huang}},\ and\ \bibinfo {author}
  {\bibfnamefont {C.-F.}\ \bibnamefont {Qiao}},\ }\bibfield  {title} {\bibinfo
  {title} {{Analyzing doubly heavy tetra- and penta-quark states by variational
  method}},\ }\href {https://doi.org/10.1016/j.physletb.2019.134869} {\bibfield
   {journal} {\bibinfo  {journal} {Phys. Lett. B}\ }\textbf {\bibinfo {volume}
  {797}},\ \bibinfo {pages} {134869} (\bibinfo {year} {2019})}\BibitemShut
  {NoStop}%
\bibitem [{\citenamefont {Wang}(2020)}]{Wang:2019got}%
  \BibitemOpen
  \bibfield  {author} {\bibinfo {author} {\bibfnamefont {Z.-G.}\ \bibnamefont
  {Wang}},\ }\bibfield  {title} {\bibinfo {title} {{Analysis of the
  $P_c(4312)$, $P_c(4440)$, $P_c(4457)$ and related hidden-charm pentaquark
  states with QCD sum rules}},\ }\href
  {https://doi.org/10.1142/S0217751X20500037} {\bibfield  {journal} {\bibinfo
  {journal} {Int. J. Mod. Phys. A}\ }\textbf {\bibinfo {volume} {35}},\
  \bibinfo {pages} {2050003} (\bibinfo {year} {2020})}\BibitemShut {NoStop}%
\bibitem [{\citenamefont {Giron}\ \emph {et~al.}(2019)\citenamefont {Giron},
  \citenamefont {Lebed},\ and\ \citenamefont {Peterson}}]{Giron:2019bcs}%
  \BibitemOpen
  \bibfield  {author} {\bibinfo {author} {\bibfnamefont {J.~F.}\ \bibnamefont
  {Giron}}, \bibinfo {author} {\bibfnamefont {R.~F.}\ \bibnamefont {Lebed}},\
  and\ \bibinfo {author} {\bibfnamefont {C.~T.}\ \bibnamefont {Peterson}},\
  }\bibfield  {title} {\bibinfo {title} {{The Dynamical Diquark Model: First
  Numerical Results}},\ }\href {https://doi.org/10.1007/JHEP05(2019)061}
  {\bibfield  {journal} {\bibinfo  {journal} {JHEP}\ }\textbf {\bibinfo
  {volume} {05}},\ \bibinfo {pages} {061}}\BibitemShut {NoStop}%
\bibitem [{\citenamefont {Cheng}\ and\ \citenamefont
  {Liu}(2019)}]{Cheng:2019obk}%
  \BibitemOpen
  \bibfield  {author} {\bibinfo {author} {\bibfnamefont {J.-B.}\ \bibnamefont
  {Cheng}}\ and\ \bibinfo {author} {\bibfnamefont {Y.-R.}\ \bibnamefont
  {Liu}},\ }\bibfield  {title} {\bibinfo {title} {{$P_c(4457)^+$,
  $P_c(4440)^+$, and $P_c(4312)^+$: molecules or compact pentaquarks?}},\
  }\href {https://doi.org/10.1103/PhysRevD.100.054002} {\bibfield  {journal}
  {\bibinfo  {journal} {Phys. Rev. D}\ }\textbf {\bibinfo {volume} {100}},\
  \bibinfo {pages} {054002} (\bibinfo {year} {2019})}\BibitemShut {NoStop}%
\bibitem [{\citenamefont {Stancu}(2019)}]{Stancu:2019qga}%
  \BibitemOpen
  \bibfield  {author} {\bibinfo {author} {\bibfnamefont {F.}~\bibnamefont
  {Stancu}},\ }\bibfield  {title} {\bibinfo {title} {{Spectrum of the $uudc
  \bar{c}$ hidden charm pentaquark with an SU(4) flavor-spin hyperfine
  interaction}},\ }\href {https://doi.org/10.1140/epjc/s10052-019-7474-0}
  {\bibfield  {journal} {\bibinfo  {journal} {Eur. Phys. J. C}\ }\textbf
  {\bibinfo {volume} {79}},\ \bibinfo {pages} {957} (\bibinfo {year}
  {2019})}\BibitemShut {NoStop}%
\bibitem [{\citenamefont {Shi}\ \emph {et~al.}(2021)\citenamefont {Shi},
  \citenamefont {Huang},\ and\ \citenamefont {Wang}}]{Shi:2021wyt}%
  \BibitemOpen
  \bibfield  {author} {\bibinfo {author} {\bibfnamefont {P.-P.}\ \bibnamefont
  {Shi}}, \bibinfo {author} {\bibfnamefont {F.}~\bibnamefont {Huang}},\ and\
  \bibinfo {author} {\bibfnamefont {W.-L.}\ \bibnamefont {Wang}},\ }\bibfield
  {title} {\bibinfo {title} {{Hidden charm pentaquark states in a diquark
  model}},\ }\href {https://doi.org/10.1140/epja/s10050-021-00542-4} {\bibfield
   {journal} {\bibinfo  {journal} {Eur. Phys. J. A}\ }\textbf {\bibinfo
  {volume} {57}},\ \bibinfo {pages} {237} (\bibinfo {year} {2021})},\ \Eprint
  {https://arxiv.org/abs/2107.08680} {arXiv:2107.08680 [hep-ph]} \BibitemShut
  {NoStop}%
\bibitem [{\citenamefont {Giron}\ and\ \citenamefont
  {Lebed}(2021)}]{Giron:2021fnl}%
  \BibitemOpen
  \bibfield  {author} {\bibinfo {author} {\bibfnamefont {J.~F.}\ \bibnamefont
  {Giron}}\ and\ \bibinfo {author} {\bibfnamefont {R.~F.}\ \bibnamefont
  {Lebed}},\ }\bibfield  {title} {\bibinfo {title} {{Fine structure of
  pentaquark multiplets in the dynamical diquark model}},\ }\href
  {https://doi.org/10.1103/PhysRevD.104.114028} {\bibfield  {journal} {\bibinfo
   {journal} {Phys. Rev. D}\ }\textbf {\bibinfo {volume} {104}},\ \bibinfo
  {pages} {114028} (\bibinfo {year} {2021})},\ \Eprint
  {https://arxiv.org/abs/2110.05557} {arXiv:2110.05557 [hep-ph]} \BibitemShut
  {NoStop}%
\bibitem [{\citenamefont {Eides}\ \emph {et~al.}(2016)\citenamefont {Eides},
  \citenamefont {Petrov},\ and\ \citenamefont {Polyakov}}]{Eides:2015dtr}%
  \BibitemOpen
  \bibfield  {author} {\bibinfo {author} {\bibfnamefont {M.~I.}\ \bibnamefont
  {Eides}}, \bibinfo {author} {\bibfnamefont {V.~Y.}\ \bibnamefont {Petrov}},\
  and\ \bibinfo {author} {\bibfnamefont {M.~V.}\ \bibnamefont {Polyakov}},\
  }\bibfield  {title} {\bibinfo {title} {{Narrow Nucleon-$\psi(2S)$ Bound State
  and LHCb Pentaquarks}},\ }\href {https://doi.org/10.1103/PhysRevD.93.054039}
  {\bibfield  {journal} {\bibinfo  {journal} {Phys. Rev. D}\ }\textbf {\bibinfo
  {volume} {93}},\ \bibinfo {pages} {054039} (\bibinfo {year}
  {2016})}\BibitemShut {NoStop}%
\bibitem [{\citenamefont {Wu}\ \emph {et~al.}(2017)\citenamefont {Wu},
  \citenamefont {Liu}, \citenamefont {Chen}, \citenamefont {Liu},\ and\
  \citenamefont {Zhu}}]{Wu:2017weo}%
  \BibitemOpen
  \bibfield  {author} {\bibinfo {author} {\bibfnamefont {J.}~\bibnamefont
  {Wu}}, \bibinfo {author} {\bibfnamefont {Y.-R.}\ \bibnamefont {Liu}},
  \bibinfo {author} {\bibfnamefont {K.}~\bibnamefont {Chen}}, \bibinfo {author}
  {\bibfnamefont {X.}~\bibnamefont {Liu}},\ and\ \bibinfo {author}
  {\bibfnamefont {S.-L.}\ \bibnamefont {Zhu}},\ }\bibfield  {title} {\bibinfo
  {title} {{Hidden-charm pentaquarks and their hidden-bottom and $B_c$-like
  partner states}},\ }\href {https://doi.org/10.1103/PhysRevD.95.034002}
  {\bibfield  {journal} {\bibinfo  {journal} {Phys. Rev. D}\ }\textbf {\bibinfo
  {volume} {95}},\ \bibinfo {pages} {034002} (\bibinfo {year} {2017})},\
  \Eprint {https://arxiv.org/abs/1701.03873} {arXiv:1701.03873 [hep-ph]}
  \BibitemShut {NoStop}%
\bibitem [{\citenamefont {Eides}\ \emph {et~al.}(2020)\citenamefont {Eides},
  \citenamefont {Petrov},\ and\ \citenamefont {Polyakov}}]{Eides:2019tgv}%
  \BibitemOpen
  \bibfield  {author} {\bibinfo {author} {\bibfnamefont {M.~I.}\ \bibnamefont
  {Eides}}, \bibinfo {author} {\bibfnamefont {V.~Y.}\ \bibnamefont {Petrov}},\
  and\ \bibinfo {author} {\bibfnamefont {M.~V.}\ \bibnamefont {Polyakov}},\
  }\bibfield  {title} {\bibinfo {title} {{New LHCb pentaquarks as
  hadrocharmonium states}},\ }\href {https://doi.org/10.1142/S0217732320501515}
  {\bibfield  {journal} {\bibinfo  {journal} {Mod. Phys. Lett. A}\ }\textbf
  {\bibinfo {volume} {35}},\ \bibinfo {pages} {2050151} (\bibinfo {year}
  {2020})}\BibitemShut {NoStop}%
\bibitem [{\citenamefont {Ferretti}\ \emph {et~al.}(2019)\citenamefont
  {Ferretti}, \citenamefont {Santopinto}, \citenamefont {Naeem~Anwar},\ and\
  \citenamefont {Bedolla}}]{Ferretti:2018ojb}%
  \BibitemOpen
  \bibfield  {author} {\bibinfo {author} {\bibfnamefont {J.}~\bibnamefont
  {Ferretti}}, \bibinfo {author} {\bibfnamefont {E.}~\bibnamefont
  {Santopinto}}, \bibinfo {author} {\bibfnamefont {M.}~\bibnamefont
  {Naeem~Anwar}},\ and\ \bibinfo {author} {\bibfnamefont {M.~A.}\ \bibnamefont
  {Bedolla}},\ }\bibfield  {title} {\bibinfo {title} {{The baryo-quarkonium
  picture for hidden-charm and bottom pentaquarks and LHCb $P_{\rm c}(4380)$
  and $P_{\rm c}(4450)$ states}},\ }\href
  {https://doi.org/10.1016/j.physletb.2018.09.047} {\bibfield  {journal}
  {\bibinfo  {journal} {Phys. Lett. B}\ }\textbf {\bibinfo {volume} {789}},\
  \bibinfo {pages} {562} (\bibinfo {year} {2019})}\BibitemShut {NoStop}%
\bibitem [{\citenamefont {Garcilazo}\ and\ \citenamefont
  {Valcarce}(2022)}]{Garcilazo:2022kra}%
  \BibitemOpen
  \bibfield  {author} {\bibinfo {author} {\bibfnamefont {H.}~\bibnamefont
  {Garcilazo}}\ and\ \bibinfo {author} {\bibfnamefont {A.}~\bibnamefont
  {Valcarce}},\ }\bibfield  {title} {\bibinfo {title} {{Constituent quark-model
  hidden-flavor pentaquarks}},\ }\href
  {https://doi.org/10.1103/PhysRevD.105.114016} {\bibfield  {journal} {\bibinfo
   {journal} {Phys. Rev. D}\ }\textbf {\bibinfo {volume} {105}},\ \bibinfo
  {pages} {114016} (\bibinfo {year} {2022})},\ \Eprint
  {https://arxiv.org/abs/2207.02757} {arXiv:2207.02757 [hep-ph]} \BibitemShut
  {NoStop}%
\bibitem [{\citenamefont {Germani}\ \emph {et~al.}(2024)\citenamefont
  {Germani}, \citenamefont {Niliani},\ and\ \citenamefont
  {Polosa}}]{Germani:2024miu}%
  \BibitemOpen
  \bibfield  {author} {\bibinfo {author} {\bibfnamefont {D.}~\bibnamefont
  {Germani}}, \bibinfo {author} {\bibfnamefont {F.}~\bibnamefont {Niliani}},\
  and\ \bibinfo {author} {\bibfnamefont {A.~D.}\ \bibnamefont {Polosa}},\
  }\bibfield  {title} {\bibinfo {title} {{A model of pentaquarks}},\ }\href
  {https://doi.org/10.1140/epjc/s10052-024-13103-y} {\bibfield  {journal}
  {\bibinfo  {journal} {Eur. Phys. J. C}\ }\textbf {\bibinfo {volume} {84}},\
  \bibinfo {pages} {755} (\bibinfo {year} {2024})},\ \Eprint
  {https://arxiv.org/abs/2403.04068} {arXiv:2403.04068 [hep-ph]} \BibitemShut
  {NoStop}%
\bibitem [{\citenamefont {Nakamura}(2021)}]{Nakamura:2021qvy}%
  \BibitemOpen
  \bibfield  {author} {\bibinfo {author} {\bibfnamefont {S.~X.}\ \bibnamefont
  {Nakamura}},\ }\bibfield  {title} {\bibinfo {title} {{$P_c(4312)^+$,
  $P_c(4380)^+$, and $P_c(4457)^+$ as double triangle cusps}},\ }\href
  {https://doi.org/10.1103/PhysRevD.103.L111503} {\bibfield  {journal}
  {\bibinfo  {journal} {Phys. Rev. D}\ }\textbf {\bibinfo {volume} {103}},\
  \bibinfo {pages} {111503} (\bibinfo {year} {2021})},\ \Eprint
  {https://arxiv.org/abs/2103.06817} {arXiv:2103.06817 [hep-ph]} \BibitemShut
  {NoStop}%
\bibitem [{\citenamefont {Fern{\'a}ndez-Ram{\'\i}rez}\ \emph
  {et~al.}(2019)\citenamefont {Fern{\'a}ndez-Ram{\'\i}rez}, \citenamefont
  {Pilloni}, \citenamefont {Albaladejo}, \citenamefont {Jackura}, \citenamefont
  {Mathieu}, \citenamefont {Mikhasenko}, \citenamefont {Silva-Castro},\ and\
  \citenamefont {Szczepaniak}}]{Fernandez-Ramirez:2019koa}%
  \BibitemOpen
  \bibfield  {author} {\bibinfo {author} {\bibfnamefont {C.}~\bibnamefont
  {Fern{\'a}ndez-Ram{\'\i}rez}}, \bibinfo {author} {\bibfnamefont
  {A.}~\bibnamefont {Pilloni}}, \bibinfo {author} {\bibfnamefont
  {M.}~\bibnamefont {Albaladejo}}, \bibinfo {author} {\bibfnamefont
  {A.}~\bibnamefont {Jackura}}, \bibinfo {author} {\bibfnamefont
  {V.}~\bibnamefont {Mathieu}}, \bibinfo {author} {\bibfnamefont
  {M.}~\bibnamefont {Mikhasenko}}, \bibinfo {author} {\bibfnamefont {J.~A.}\
  \bibnamefont {Silva-Castro}},\ and\ \bibinfo {author} {\bibfnamefont {A.~P.}\
  \bibnamefont {Szczepaniak}} (\bibinfo {collaboration} {JPAC}),\ }\bibfield
  {title} {\bibinfo {title} {{Interpretation of the LHCb $P_c$(4312)$^+$
  Signal}},\ }\href {https://doi.org/10.1103/PhysRevLett.123.092001} {\bibfield
   {journal} {\bibinfo  {journal} {Phys. Rev. Lett.}\ }\textbf {\bibinfo
  {volume} {123}},\ \bibinfo {pages} {092001} (\bibinfo {year} {2019})},\
  \Eprint {https://arxiv.org/abs/1904.10021} {arXiv:1904.10021 [hep-ph]}
  \BibitemShut {NoStop}%
\bibitem [{\citenamefont {Pavon~Valderrama}(2019)}]{PavonValderrama:2019nbk}%
  \BibitemOpen
  \bibfield  {author} {\bibinfo {author} {\bibfnamefont {M.}~\bibnamefont
  {Pavon~Valderrama}},\ }\bibfield  {title} {\bibinfo {title} {{One pion
  exchange and the quantum numbers of the P$_c$(4440) and P$_c$(4457)
  pentaquarks}},\ }\href {https://doi.org/10.1103/PhysRevD.100.094028}
  {\bibfield  {journal} {\bibinfo  {journal} {Phys. Rev. D}\ }\textbf {\bibinfo
  {volume} {100}},\ \bibinfo {pages} {094028} (\bibinfo {year} {2019})},\
  \Eprint {https://arxiv.org/abs/1907.05294} {arXiv:1907.05294 [hep-ph]}
  \BibitemShut {NoStop}%
\bibitem [{\citenamefont {Liu}\ \emph {et~al.}(2023)\citenamefont {Liu},
  \citenamefont {Lu}, \citenamefont {Liu},\ and\ \citenamefont
  {Geng}}]{Liu:2023wfo}%
  \BibitemOpen
  \bibfield  {author} {\bibinfo {author} {\bibfnamefont {Z.-W.}\ \bibnamefont
  {Liu}}, \bibinfo {author} {\bibfnamefont {J.-X.}\ \bibnamefont {Lu}},
  \bibinfo {author} {\bibfnamefont {M.-Z.}\ \bibnamefont {Liu}},\ and\ \bibinfo
  {author} {\bibfnamefont {L.-S.}\ \bibnamefont {Geng}},\ }\bibfield  {title}
  {\bibinfo {title} {{Distinguishing the spins of Pc(4440) and Pc(4457) with
  femtoscopic correlation functions}},\ }\href
  {https://doi.org/10.1103/PhysRevD.108.L031503} {\bibfield  {journal}
  {\bibinfo  {journal} {Phys. Rev. D}\ }\textbf {\bibinfo {volume} {108}},\
  \bibinfo {pages} {L031503} (\bibinfo {year} {2023})},\ \Eprint
  {https://arxiv.org/abs/2305.19048} {arXiv:2305.19048 [hep-ph]} \BibitemShut
  {NoStop}%
\bibitem [{\citenamefont {Yang}\ \emph {et~al.}(2024)\citenamefont {Yang},
  \citenamefont {Song}, \citenamefont {Liang},\ and\ \citenamefont
  {Oset}}]{Yang:2024nss}%
  \BibitemOpen
  \bibfield  {author} {\bibinfo {author} {\bibfnamefont {Z.-Y.}\ \bibnamefont
  {Yang}}, \bibinfo {author} {\bibfnamefont {J.}~\bibnamefont {Song}}, \bibinfo
  {author} {\bibfnamefont {W.-H.}\ \bibnamefont {Liang}},\ and\ \bibinfo
  {author} {\bibfnamefont {E.}~\bibnamefont {Oset}},\ }\bibfield  {title}
  {\bibinfo {title} {{$P_c(4440)$ and $P_c(4457)$ decay into $\bar{D}\Sigma_c$
  and $\bar{D}\Lambda_c$ and the spin of the $P_c$ states}},\ }\href@noop {} {\
   (\bibinfo {year} {2024})},\ \Eprint {https://arxiv.org/abs/2412.15731}
  {arXiv:2412.15731 [hep-ph]} \BibitemShut {NoStop}%
\bibitem [{\citenamefont {Xu}\ \emph {et~al.}(2025)\citenamefont {Xu},
  \citenamefont {Meng}, \citenamefont {Zhu}, \citenamefont {Li},\ and\
  \citenamefont {Chen}}]{Xu:2025mhc}%
  \BibitemOpen
  \bibfield  {author} {\bibinfo {author} {\bibfnamefont {R.}~\bibnamefont
  {Xu}}, \bibinfo {author} {\bibfnamefont {L.}~\bibnamefont {Meng}}, \bibinfo
  {author} {\bibfnamefont {H.-X.}\ \bibnamefont {Zhu}}, \bibinfo {author}
  {\bibfnamefont {N.}~\bibnamefont {Li}},\ and\ \bibinfo {author}
  {\bibfnamefont {W.}~\bibnamefont {Chen}},\ }\bibfield  {title} {\bibinfo
  {title} {{Toward modeling the short-range interactions of hidden and open
  charm pentaquark molecular states}},\ }\href
  {https://doi.org/10.1103/PhysRevD.111.094015} {\bibfield  {journal} {\bibinfo
   {journal} {Phys. Rev. D}\ }\textbf {\bibinfo {volume} {111}},\ \bibinfo
  {pages} {094015} (\bibinfo {year} {2025})},\ \Eprint
  {https://arxiv.org/abs/2503.04264} {arXiv:2503.04264 [hep-ph]} \BibitemShut
  {NoStop}%
\bibitem [{\citenamefont {Xing}\ \emph {et~al.}(2022)\citenamefont {Xing},
  \citenamefont {Liang}, \citenamefont {Liu}, \citenamefont {Sun},\ and\
  \citenamefont {Yang}}]{Xing:2022ijm}%
  \BibitemOpen
  \bibfield  {author} {\bibinfo {author} {\bibfnamefont {H.}~\bibnamefont
  {Xing}}, \bibinfo {author} {\bibfnamefont {J.}~\bibnamefont {Liang}},
  \bibinfo {author} {\bibfnamefont {L.}~\bibnamefont {Liu}}, \bibinfo {author}
  {\bibfnamefont {P.}~\bibnamefont {Sun}},\ and\ \bibinfo {author}
  {\bibfnamefont {Y.-B.}\ \bibnamefont {Yang}},\ }\bibfield  {title} {\bibinfo
  {title} {{First observation of the hidden-charm pentaquarks on lattice}},\
  }\href@noop {} {\  (\bibinfo {year} {2022})},\ \Eprint
  {https://arxiv.org/abs/2210.08555} {arXiv:2210.08555 [hep-lat]} \BibitemShut
  {NoStop}%
\bibitem [{\citenamefont {Sakai}\ \emph {et~al.}(2019)\citenamefont {Sakai},
  \citenamefont {Jing},\ and\ \citenamefont {Guo}}]{Sakai:2019qph}%
  \BibitemOpen
  \bibfield  {author} {\bibinfo {author} {\bibfnamefont {S.}~\bibnamefont
  {Sakai}}, \bibinfo {author} {\bibfnamefont {H.-J.}\ \bibnamefont {Jing}},\
  and\ \bibinfo {author} {\bibfnamefont {F.-K.}\ \bibnamefont {Guo}},\
  }\bibfield  {title} {\bibinfo {title} {{Decays of $P_c$ into $J/\psi N$ and
  $\eta_cN$ with heavy quark spin symmetry}},\ }\href
  {https://doi.org/10.1103/PhysRevD.100.074007} {\bibfield  {journal} {\bibinfo
   {journal} {Phys. Rev. D}\ }\textbf {\bibinfo {volume} {100}},\ \bibinfo
  {pages} {074007} (\bibinfo {year} {2019})},\ \Eprint
  {https://arxiv.org/abs/1907.03414} {arXiv:1907.03414 [hep-ph]} \BibitemShut
  {NoStop}%
\bibitem [{\citenamefont {Xie}\ \emph {et~al.}(2022)\citenamefont {Xie},
  \citenamefont {Ling}, \citenamefont {Liu},\ and\ \citenamefont
  {Geng}}]{Xie:2022hhv}%
  \BibitemOpen
  \bibfield  {author} {\bibinfo {author} {\bibfnamefont {J.-M.}\ \bibnamefont
  {Xie}}, \bibinfo {author} {\bibfnamefont {X.-Z.}\ \bibnamefont {Ling}},
  \bibinfo {author} {\bibfnamefont {M.-Z.}\ \bibnamefont {Liu}},\ and\ \bibinfo
  {author} {\bibfnamefont {L.-S.}\ \bibnamefont {Geng}},\ }\bibfield  {title}
  {\bibinfo {title} {{Search for hidden-charm pentaquark states in three-body
  final states}},\ }\href {https://doi.org/10.1140/epjc/s10052-022-11026-0}
  {\bibfield  {journal} {\bibinfo  {journal} {Eur. Phys. J. C}\ }\textbf
  {\bibinfo {volume} {82}},\ \bibinfo {pages} {1061} (\bibinfo {year}
  {2022})},\ \Eprint {https://arxiv.org/abs/2204.12356} {arXiv:2204.12356
  [hep-ph]} \BibitemShut {NoStop}%
\bibitem [{\citenamefont {Aaij}\ \emph {et~al.}(2024)\citenamefont {Aaij} \emph
  {et~al.}}]{LHCb:2024pnt}%
  \BibitemOpen
  \bibfield  {author} {\bibinfo {author} {\bibfnamefont {R.}~\bibnamefont
  {Aaij}} \emph {et~al.} (\bibinfo {collaboration} {LHCb}),\ }\bibfield
  {title} {\bibinfo {title} {{Search for prompt production of pentaquarks in
  charm hadron final states}},\ }\href
  {https://doi.org/10.1103/PhysRevD.110.032001} {\bibfield  {journal} {\bibinfo
   {journal} {Phys. Rev. D}\ }\textbf {\bibinfo {volume} {110}},\ \bibinfo
  {pages} {032001} (\bibinfo {year} {2024})},\ \Eprint
  {https://arxiv.org/abs/2404.07131} {arXiv:2404.07131 [hep-ex]} \BibitemShut
  {NoStop}%
\bibitem [{\citenamefont {Berwein}\ \emph {et~al.}(2024)\citenamefont
  {Berwein}, \citenamefont {Brambilla}, \citenamefont {Mohapatra},\ and\
  \citenamefont {Vairo}}]{Berwein:2024ztx}%
  \BibitemOpen
  \bibfield  {author} {\bibinfo {author} {\bibfnamefont {M.}~\bibnamefont
  {Berwein}}, \bibinfo {author} {\bibfnamefont {N.}~\bibnamefont {Brambilla}},
  \bibinfo {author} {\bibfnamefont {A.}~\bibnamefont {Mohapatra}},\ and\
  \bibinfo {author} {\bibfnamefont {A.}~\bibnamefont {Vairo}},\ }\bibfield
  {title} {\bibinfo {title} {{Hybrids, tetraquarks, pentaquarks, doubly heavy
  baryons, and quarkonia in Born-Oppenheimer effective theory}},\ }\href
  {https://doi.org/10.1103/PhysRevD.110.094040} {\bibfield  {journal} {\bibinfo
   {journal} {Phys. Rev. D}\ }\textbf {\bibinfo {volume} {110}},\ \bibinfo
  {pages} {094040} (\bibinfo {year} {2024})},\ \Eprint
  {https://arxiv.org/abs/2408.04719} {arXiv:2408.04719 [hep-ph]} \BibitemShut
  {NoStop}%
\bibitem [{\citenamefont {Griffiths}\ \emph {et~al.}(1983)\citenamefont
  {Griffiths}, \citenamefont {Michael},\ and\ \citenamefont
  {Rakow}}]{Griffiths:1983ah}%
  \BibitemOpen
  \bibfield  {author} {\bibinfo {author} {\bibfnamefont {L.~A.}\ \bibnamefont
  {Griffiths}}, \bibinfo {author} {\bibfnamefont {C.}~\bibnamefont {Michael}},\
  and\ \bibinfo {author} {\bibfnamefont {P.~E.~L.}\ \bibnamefont {Rakow}},\
  }\bibfield  {title} {\bibinfo {title} {{Mesons With Excited Glue}},\ }\href
  {https://doi.org/10.1016/0370-2693(83)90680-9} {\bibfield  {journal}
  {\bibinfo  {journal} {Phys. Lett. B}\ }\textbf {\bibinfo {volume} {129}},\
  \bibinfo {pages} {351} (\bibinfo {year} {1983})}\BibitemShut {NoStop}%
\bibitem [{\citenamefont {Juge}\ \emph {et~al.}(1999)\citenamefont {Juge},
  \citenamefont {Kuti},\ and\ \citenamefont {Morningstar}}]{Juge:1999ie}%
  \BibitemOpen
  \bibfield  {author} {\bibinfo {author} {\bibfnamefont {K.~J.}\ \bibnamefont
  {Juge}}, \bibinfo {author} {\bibfnamefont {J.}~\bibnamefont {Kuti}},\ and\
  \bibinfo {author} {\bibfnamefont {C.~J.}\ \bibnamefont {Morningstar}},\
  }\bibfield  {title} {\bibinfo {title} {{Ab initio study of hybrid anti-b g b
  mesons}},\ }\href {https://doi.org/10.1103/PhysRevLett.82.4400} {\bibfield
  {journal} {\bibinfo  {journal} {Phys. Rev. Lett.}\ }\textbf {\bibinfo
  {volume} {82}},\ \bibinfo {pages} {4400} (\bibinfo {year} {1999})},\ \Eprint
  {https://arxiv.org/abs/hep-ph/9902336} {arXiv:hep-ph/9902336} \BibitemShut
  {NoStop}%
\bibitem [{\citenamefont {Brambilla}\ \emph {et~al.}(2000)\citenamefont
  {Brambilla}, \citenamefont {Pineda}, \citenamefont {Soto},\ and\
  \citenamefont {Vairo}}]{Brambilla:1999xf}%
  \BibitemOpen
  \bibfield  {author} {\bibinfo {author} {\bibfnamefont {N.}~\bibnamefont
  {Brambilla}}, \bibinfo {author} {\bibfnamefont {A.}~\bibnamefont {Pineda}},
  \bibinfo {author} {\bibfnamefont {J.}~\bibnamefont {Soto}},\ and\ \bibinfo
  {author} {\bibfnamefont {A.}~\bibnamefont {Vairo}},\ }\bibfield  {title}
  {\bibinfo {title} {{Potential NRQCD: An Effective theory for heavy
  quarkonium}},\ }\href {https://doi.org/10.1016/S0550-3213(99)00693-8}
  {\bibfield  {journal} {\bibinfo  {journal} {Nucl. Phys. B}\ }\textbf
  {\bibinfo {volume} {566}},\ \bibinfo {pages} {275} (\bibinfo {year}
  {2000})},\ \Eprint {https://arxiv.org/abs/hep-ph/9907240}
  {arXiv:hep-ph/9907240} \BibitemShut {NoStop}%
\bibitem [{\citenamefont {Juge}\ \emph {et~al.}(2003)\citenamefont {Juge},
  \citenamefont {Kuti},\ and\ \citenamefont {Morningstar}}]{Juge:2002br}%
  \BibitemOpen
  \bibfield  {author} {\bibinfo {author} {\bibfnamefont {K.~J.}\ \bibnamefont
  {Juge}}, \bibinfo {author} {\bibfnamefont {J.}~\bibnamefont {Kuti}},\ and\
  \bibinfo {author} {\bibfnamefont {C.}~\bibnamefont {Morningstar}},\
  }\bibfield  {title} {\bibinfo {title} {{Fine structure of the QCD string
  spectrum}},\ }\href {https://doi.org/10.1103/PhysRevLett.90.161601}
  {\bibfield  {journal} {\bibinfo  {journal} {Phys. Rev. Lett.}\ }\textbf
  {\bibinfo {volume} {90}},\ \bibinfo {pages} {161601} (\bibinfo {year}
  {2003})},\ \Eprint {https://arxiv.org/abs/hep-lat/0207004}
  {arXiv:hep-lat/0207004} \BibitemShut {NoStop}%
\bibitem [{\citenamefont {Braaten}(2013)}]{Braaten:2013boa}%
  \BibitemOpen
  \bibfield  {author} {\bibinfo {author} {\bibfnamefont {E.}~\bibnamefont
  {Braaten}},\ }\bibfield  {title} {\bibinfo {title} {{How the $Z_c$(3900)
  Reveals the Spectra of Quarkonium Hybrid and Tetraquark Mesons}},\ }\href
  {https://doi.org/10.1103/PhysRevLett.111.162003} {\bibfield  {journal}
  {\bibinfo  {journal} {Phys. Rev. Lett.}\ }\textbf {\bibinfo {volume} {111}},\
  \bibinfo {pages} {162003} (\bibinfo {year} {2013})},\ \Eprint
  {https://arxiv.org/abs/1305.6905} {arXiv:1305.6905 [hep-ph]} \BibitemShut
  {NoStop}%
\bibitem [{\citenamefont {Braaten}\ \emph
  {et~al.}(2014{\natexlab{a}})\citenamefont {Braaten}, \citenamefont
  {Langmack},\ and\ \citenamefont {Smith}}]{Braaten:2014qka}%
  \BibitemOpen
  \bibfield  {author} {\bibinfo {author} {\bibfnamefont {E.}~\bibnamefont
  {Braaten}}, \bibinfo {author} {\bibfnamefont {C.}~\bibnamefont {Langmack}},\
  and\ \bibinfo {author} {\bibfnamefont {D.~H.}\ \bibnamefont {Smith}},\
  }\bibfield  {title} {\bibinfo {title} {{Born-Oppenheimer Approximation for
  the XYZ Mesons}},\ }\href {https://doi.org/10.1103/PhysRevD.90.014044}
  {\bibfield  {journal} {\bibinfo  {journal} {Phys. Rev. D}\ }\textbf {\bibinfo
  {volume} {90}},\ \bibinfo {pages} {014044} (\bibinfo {year}
  {2014}{\natexlab{a}})},\ \Eprint {https://arxiv.org/abs/1402.0438}
  {arXiv:1402.0438 [hep-ph]} \BibitemShut {NoStop}%
\bibitem [{\citenamefont {Braaten}\ \emph
  {et~al.}(2014{\natexlab{b}})\citenamefont {Braaten}, \citenamefont
  {Langmack},\ and\ \citenamefont {Smith}}]{Braaten:2014ita}%
  \BibitemOpen
  \bibfield  {author} {\bibinfo {author} {\bibfnamefont {E.}~\bibnamefont
  {Braaten}}, \bibinfo {author} {\bibfnamefont {C.}~\bibnamefont {Langmack}},\
  and\ \bibinfo {author} {\bibfnamefont {D.~H.}\ \bibnamefont {Smith}},\
  }\bibfield  {title} {\bibinfo {title} {{Selection Rules for Hadronic
  Transitions of XYZ Mesons}},\ }\href
  {https://doi.org/10.1103/PhysRevLett.112.222001} {\bibfield  {journal}
  {\bibinfo  {journal} {Phys. Rev. Lett.}\ }\textbf {\bibinfo {volume} {112}},\
  \bibinfo {pages} {222001} (\bibinfo {year} {2014}{\natexlab{b}})},\ \Eprint
  {https://arxiv.org/abs/1401.7351} {arXiv:1401.7351 [hep-ph]} \BibitemShut
  {NoStop}%
\bibitem [{\citenamefont {Meyer}\ and\ \citenamefont
  {Swanson}(2015)}]{Meyer:2015eta}%
  \BibitemOpen
  \bibfield  {author} {\bibinfo {author} {\bibfnamefont {C.~A.}\ \bibnamefont
  {Meyer}}\ and\ \bibinfo {author} {\bibfnamefont {E.~S.}\ \bibnamefont
  {Swanson}},\ }\bibfield  {title} {\bibinfo {title} {{Hybrid Mesons}},\ }\href
  {https://doi.org/10.1016/j.ppnp.2015.03.001} {\bibfield  {journal} {\bibinfo
  {journal} {Prog. Part. Nucl. Phys.}\ }\textbf {\bibinfo {volume} {82}},\
  \bibinfo {pages} {21} (\bibinfo {year} {2015})},\ \Eprint
  {https://arxiv.org/abs/1502.07276} {arXiv:1502.07276 [hep-ph]} \BibitemShut
  {NoStop}%
\bibitem [{\citenamefont {Alasiri}\ \emph {et~al.}(2024)\citenamefont
  {Alasiri}, \citenamefont {Braaten},\ and\ \citenamefont
  {Mohapatra}}]{Alasiri:2024nue}%
  \BibitemOpen
  \bibfield  {author} {\bibinfo {author} {\bibfnamefont {F.}~\bibnamefont
  {Alasiri}}, \bibinfo {author} {\bibfnamefont {E.}~\bibnamefont {Braaten}},\
  and\ \bibinfo {author} {\bibfnamefont {A.}~\bibnamefont {Mohapatra}},\
  }\bibfield  {title} {\bibinfo {title} {{Born-Oppenheimer potentials for SU(3)
  gauge theory}},\ }\href {https://doi.org/10.1103/PhysRevD.110.054029}
  {\bibfield  {journal} {\bibinfo  {journal} {Phys. Rev. D}\ }\textbf {\bibinfo
  {volume} {110}},\ \bibinfo {pages} {054029} (\bibinfo {year} {2024})},\
  \Eprint {https://arxiv.org/abs/2406.05123} {arXiv:2406.05123 [hep-ph]}
  \BibitemShut {NoStop}%
\bibitem [{\citenamefont {Berwein}\ \emph {et~al.}(2015)\citenamefont
  {Berwein}, \citenamefont {Brambilla}, \citenamefont {Tarr\'us~Castell\`a},\
  and\ \citenamefont {Vairo}}]{Berwein:2015vca}%
  \BibitemOpen
  \bibfield  {author} {\bibinfo {author} {\bibfnamefont {M.}~\bibnamefont
  {Berwein}}, \bibinfo {author} {\bibfnamefont {N.}~\bibnamefont {Brambilla}},
  \bibinfo {author} {\bibfnamefont {J.}~\bibnamefont {Tarr\'us~Castell\`a}},\
  and\ \bibinfo {author} {\bibfnamefont {A.}~\bibnamefont {Vairo}},\ }\bibfield
   {title} {\bibinfo {title} {{Quarkonium Hybrids with Nonrelativistic
  Effective Field Theories}},\ }\href
  {https://doi.org/10.1103/PhysRevD.92.114019} {\bibfield  {journal} {\bibinfo
  {journal} {Phys. Rev. D}\ }\textbf {\bibinfo {volume} {92}},\ \bibinfo
  {pages} {114019} (\bibinfo {year} {2015})},\ \Eprint
  {https://arxiv.org/abs/1510.04299} {arXiv:1510.04299 [hep-ph]} \BibitemShut
  {NoStop}%
\bibitem [{\citenamefont {Brambilla}\ \emph {et~al.}(2018)\citenamefont
  {Brambilla}, \citenamefont {Krein}, \citenamefont {Tarr\'us~Castell\`a},\
  and\ \citenamefont {Vairo}}]{Brambilla:2017uyf}%
  \BibitemOpen
  \bibfield  {author} {\bibinfo {author} {\bibfnamefont {N.}~\bibnamefont
  {Brambilla}}, \bibinfo {author} {\bibfnamefont {G.~a.}\ \bibnamefont
  {Krein}}, \bibinfo {author} {\bibfnamefont {J.}~\bibnamefont
  {Tarr\'us~Castell\`a}},\ and\ \bibinfo {author} {\bibfnamefont
  {A.}~\bibnamefont {Vairo}},\ }\bibfield  {title} {\bibinfo {title}
  {{Born-Oppenheimer approximation in an effective field theory language}},\
  }\href {https://doi.org/10.1103/PhysRevD.97.016016} {\bibfield  {journal}
  {\bibinfo  {journal} {Phys. Rev. D}\ }\textbf {\bibinfo {volume} {97}},\
  \bibinfo {pages} {016016} (\bibinfo {year} {2018})},\ \Eprint
  {https://arxiv.org/abs/1707.09647} {arXiv:1707.09647 [hep-ph]} \BibitemShut
  {NoStop}%
\bibitem [{\citenamefont {Tarr\'us~Castell\`a}(2019)}]{TarrusCastella:2019rit}%
  \BibitemOpen
  \bibfield  {author} {\bibinfo {author} {\bibfnamefont {J.}~\bibnamefont
  {Tarr\'us~Castell\`a}},\ }\bibfield  {title} {\bibinfo {title} {{Heavy
  hybrids and tetraquarks in effective field theory}},\ }\href
  {https://doi.org/10.1051/epjconf/201920201005} {\bibfield  {journal}
  {\bibinfo  {journal} {EPJ Web Conf.}\ }\textbf {\bibinfo {volume} {202}},\
  \bibinfo {pages} {01005} (\bibinfo {year} {2019})},\ \Eprint
  {https://arxiv.org/abs/1901.09761} {arXiv:1901.09761 [hep-ph]} \BibitemShut
  {NoStop}%
\bibitem [{\citenamefont {Soto}\ and\ \citenamefont
  {Tarr\'us~Castell\`a}(2020{\natexlab{a}})}]{Soto:2020xpm}%
  \BibitemOpen
  \bibfield  {author} {\bibinfo {author} {\bibfnamefont {J.}~\bibnamefont
  {Soto}}\ and\ \bibinfo {author} {\bibfnamefont {J.}~\bibnamefont
  {Tarr\'us~Castell\`a}},\ }\bibfield  {title} {\bibinfo {title}
  {{Nonrelativistic effective field theory for heavy exotic hadrons}},\ }\href
  {https://doi.org/10.1103/PhysRevD.102.014012} {\bibfield  {journal} {\bibinfo
   {journal} {Phys. Rev. D}\ }\textbf {\bibinfo {volume} {102}},\ \bibinfo
  {pages} {014012} (\bibinfo {year} {2020}{\natexlab{a}})},\ \Eprint
  {https://arxiv.org/abs/2005.00552} {arXiv:2005.00552 [hep-ph]} \BibitemShut
  {NoStop}%
\bibitem [{\citenamefont {Mohapatra}(2025)}]{Mohapatra:2025iar}%
  \BibitemOpen
  \bibfield  {author} {\bibinfo {author} {\bibfnamefont {A.}~\bibnamefont
  {Mohapatra}},\ }\bibfield  {title} {\bibinfo {title} {{One Born Oppenheimer
  Effective Theory to rule all Exotics}},\ }in\ \href@noop {} {\emph {\bibinfo
  {booktitle} {{16th Conference on Quark Confinement and the Hadron
  Spectrum}}}}\ (\bibinfo {year} {2025})\ \Eprint
  {https://arxiv.org/abs/2503.22847} {arXiv:2503.22847 [hep-ph]} \BibitemShut
  {NoStop}%
\bibitem [{\citenamefont {Oncala}\ and\ \citenamefont
  {Soto}(2017)}]{Oncala:2017hop}%
  \BibitemOpen
  \bibfield  {author} {\bibinfo {author} {\bibfnamefont {R.}~\bibnamefont
  {Oncala}}\ and\ \bibinfo {author} {\bibfnamefont {J.}~\bibnamefont {Soto}},\
  }\bibfield  {title} {\bibinfo {title} {{Heavy Quarkonium Hybrids: Spectrum,
  Decay and Mixing}},\ }\href {https://doi.org/10.1103/PhysRevD.96.014004}
  {\bibfield  {journal} {\bibinfo  {journal} {Phys. Rev. D}\ }\textbf {\bibinfo
  {volume} {96}},\ \bibinfo {pages} {014004} (\bibinfo {year} {2017})},\
  \Eprint {https://arxiv.org/abs/1702.03900} {arXiv:1702.03900 [hep-ph]}
  \BibitemShut {NoStop}%
\bibitem [{\citenamefont {Brambilla}\ \emph {et~al.}(2019)\citenamefont
  {Brambilla}, \citenamefont {Lai}, \citenamefont {Segovia}, \citenamefont
  {Tarr\'us~Castell\`a},\ and\ \citenamefont {Vairo}}]{Brambilla:2018pyn}%
  \BibitemOpen
  \bibfield  {author} {\bibinfo {author} {\bibfnamefont {N.}~\bibnamefont
  {Brambilla}}, \bibinfo {author} {\bibfnamefont {W.~K.}\ \bibnamefont {Lai}},
  \bibinfo {author} {\bibfnamefont {J.}~\bibnamefont {Segovia}}, \bibinfo
  {author} {\bibfnamefont {J.}~\bibnamefont {Tarr\'us~Castell\`a}},\ and\
  \bibinfo {author} {\bibfnamefont {A.}~\bibnamefont {Vairo}},\ }\bibfield
  {title} {\bibinfo {title} {{Spin structure of heavy-quark hybrids}},\ }\href
  {https://doi.org/10.1103/PhysRevD.99.014017} {\bibfield  {journal} {\bibinfo
  {journal} {Phys. Rev. D}\ }\textbf {\bibinfo {volume} {99}},\ \bibinfo
  {pages} {014017} (\bibinfo {year} {2019})},\ \bibinfo {note} {[Erratum:
  Phys.Rev.D 101, 099902 (2020)]},\ \Eprint {https://arxiv.org/abs/1805.07713}
  {arXiv:1805.07713 [hep-ph]} \BibitemShut {NoStop}%
\bibitem [{\citenamefont {Brambilla}\ \emph
  {et~al.}(2020{\natexlab{b}})\citenamefont {Brambilla}, \citenamefont {Lai},
  \citenamefont {Segovia},\ and\ \citenamefont
  {Tarr\'us~Castell\`a}}]{Brambilla:2019jfi}%
  \BibitemOpen
  \bibfield  {author} {\bibinfo {author} {\bibfnamefont {N.}~\bibnamefont
  {Brambilla}}, \bibinfo {author} {\bibfnamefont {W.~K.}\ \bibnamefont {Lai}},
  \bibinfo {author} {\bibfnamefont {J.}~\bibnamefont {Segovia}},\ and\ \bibinfo
  {author} {\bibfnamefont {J.}~\bibnamefont {Tarr\'us~Castell\`a}},\ }\bibfield
   {title} {\bibinfo {title} {{QCD spin effects in the heavy hybrid potentials
  and spectra}},\ }\href {https://doi.org/10.1103/PhysRevD.101.054040}
  {\bibfield  {journal} {\bibinfo  {journal} {Phys. Rev. D}\ }\textbf {\bibinfo
  {volume} {101}},\ \bibinfo {pages} {054040} (\bibinfo {year}
  {2020}{\natexlab{b}})},\ \Eprint {https://arxiv.org/abs/1908.11699}
  {arXiv:1908.11699 [hep-ph]} \BibitemShut {NoStop}%
\bibitem [{\citenamefont {Soto}\ and\ \citenamefont
  {Valls}(2023)}]{Soto:2023lbh}%
  \BibitemOpen
  \bibfield  {author} {\bibinfo {author} {\bibfnamefont {J.}~\bibnamefont
  {Soto}}\ and\ \bibinfo {author} {\bibfnamefont {S.~T.}\ \bibnamefont
  {Valls}},\ }\bibfield  {title} {\bibinfo {title} {{Hyperfine splittings of
  heavy quarkonium hybrids}},\ }\href
  {https://doi.org/10.1103/PhysRevD.108.014025} {\bibfield  {journal} {\bibinfo
   {journal} {Phys. Rev. D}\ }\textbf {\bibinfo {volume} {108}},\ \bibinfo
  {pages} {014025} (\bibinfo {year} {2023})},\ \Eprint
  {https://arxiv.org/abs/2302.01765} {arXiv:2302.01765 [hep-ph]} \BibitemShut
  {NoStop}%
\bibitem [{\citenamefont {Schlosser}\ and\ \citenamefont
  {Wagner}(2025)}]{Schlosser:2025tca}%
  \BibitemOpen
  \bibfield  {author} {\bibinfo {author} {\bibfnamefont {C.}~\bibnamefont
  {Schlosser}}\ and\ \bibinfo {author} {\bibfnamefont {M.}~\bibnamefont
  {Wagner}},\ }\bibfield  {title} {\bibinfo {title} {{Hybrid spin-dependent and
  hybrid-quarkonium mixing potentials at order $(1 /m_Q)^1$ from SU(3) lattice
  gauge theory}},\ }\href@noop {} {\  (\bibinfo {year} {2025})},\ \Eprint
  {https://arxiv.org/abs/2501.08844} {arXiv:2501.08844 [hep-lat]} \BibitemShut
  {NoStop}%
\bibitem [{\citenamefont {Brambilla}\ \emph
  {et~al.}(2023{\natexlab{a}})\citenamefont {Brambilla}, \citenamefont {Lai},
  \citenamefont {Mohapatra},\ and\ \citenamefont {Vairo}}]{Brambilla:2022hhi}%
  \BibitemOpen
  \bibfield  {author} {\bibinfo {author} {\bibfnamefont {N.}~\bibnamefont
  {Brambilla}}, \bibinfo {author} {\bibfnamefont {W.~K.}\ \bibnamefont {Lai}},
  \bibinfo {author} {\bibfnamefont {A.}~\bibnamefont {Mohapatra}},\ and\
  \bibinfo {author} {\bibfnamefont {A.}~\bibnamefont {Vairo}},\ }\bibfield
  {title} {\bibinfo {title} {{Heavy hybrid decays to quarkonia}},\ }\href
  {https://doi.org/10.1103/PhysRevD.107.054034} {\bibfield  {journal} {\bibinfo
   {journal} {Phys. Rev. D}\ }\textbf {\bibinfo {volume} {107}},\ \bibinfo
  {pages} {054034} (\bibinfo {year} {2023}{\natexlab{a}})},\ \Eprint
  {https://arxiv.org/abs/2212.09187} {arXiv:2212.09187 [hep-ph]} \BibitemShut
  {NoStop}%
\bibitem [{\citenamefont {Tarr\'us~Castell\`a}\ and\ \citenamefont
  {Passemar}(2021)}]{TarrusCastella:2021pld}%
  \BibitemOpen
  \bibfield  {author} {\bibinfo {author} {\bibfnamefont {J.}~\bibnamefont
  {Tarr\'us~Castell\`a}}\ and\ \bibinfo {author} {\bibfnamefont
  {E.}~\bibnamefont {Passemar}},\ }\bibfield  {title} {\bibinfo {title}
  {{Exotic to standard bottomonium transitions}},\ }\href
  {https://doi.org/10.1103/PhysRevD.104.034019} {\bibfield  {journal} {\bibinfo
   {journal} {Phys. Rev. D}\ }\textbf {\bibinfo {volume} {104}},\ \bibinfo
  {pages} {034019} (\bibinfo {year} {2021})},\ \Eprint
  {https://arxiv.org/abs/2104.03975} {arXiv:2104.03975 [hep-ph]} \BibitemShut
  {NoStop}%
\bibitem [{\citenamefont {Bruschini}(2024{\natexlab{a}})}]{Bruschini:2023tmm}%
  \BibitemOpen
  \bibfield  {author} {\bibinfo {author} {\bibfnamefont {R.}~\bibnamefont
  {Bruschini}},\ }\bibfield  {title} {\bibinfo {title} {{Why quarkonium hybrid
  coupling to two S-wave heavy-light mesons is not suppressed}},\ }\href
  {https://doi.org/10.1103/PhysRevD.109.L031501} {\bibfield  {journal}
  {\bibinfo  {journal} {Phys. Rev. D}\ }\textbf {\bibinfo {volume} {109}},\
  \bibinfo {pages} {L031501} (\bibinfo {year} {2024}{\natexlab{a}})},\ \Eprint
  {https://arxiv.org/abs/2306.17120} {arXiv:2306.17120 [hep-ph]} \BibitemShut
  {NoStop}%
\bibitem [{\citenamefont {Tarr\'us~Castell\`a}(2024)}]{TarrusCastella:2024zps}%
  \BibitemOpen
  \bibfield  {author} {\bibinfo {author} {\bibfnamefont {J.}~\bibnamefont
  {Tarr\'us~Castell\`a}},\ }\bibfield  {title} {\bibinfo {title} {{Effect of
  continuum states on the double-heavy hadron spectra}},\ }\href
  {https://doi.org/10.1007/JHEP06(2024)107} {\bibfield  {journal} {\bibinfo
  {journal} {JHEP}\ }\textbf {\bibinfo {volume} {06}},\ \bibinfo {pages}
  {107}},\ \Eprint {https://arxiv.org/abs/2401.13393} {arXiv:2401.13393
  [hep-ph]} \BibitemShut {NoStop}%
\bibitem [{\citenamefont {Braaten}\ and\ \citenamefont
  {Bruschini}(2024)}]{Braaten:2024stn}%
  \BibitemOpen
  \bibfield  {author} {\bibinfo {author} {\bibfnamefont {E.}~\bibnamefont
  {Braaten}}\ and\ \bibinfo {author} {\bibfnamefont {R.}~\bibnamefont
  {Bruschini}},\ }\bibfield  {title} {\bibinfo {title} {{Model-independent
  predictions for decays of hidden-heavy hadrons into pairs of heavy
  hadrons}},\ }\href {https://doi.org/10.1103/PhysRevD.109.094051} {\bibfield
  {journal} {\bibinfo  {journal} {Phys. Rev. D}\ }\textbf {\bibinfo {volume}
  {109}},\ \bibinfo {pages} {094051} (\bibinfo {year} {2024})},\ \Eprint
  {https://arxiv.org/abs/2403.12868} {arXiv:2403.12868 [hep-ph]} \BibitemShut
  {NoStop}%
\bibitem [{\citenamefont {Brambilla}\ \emph
  {et~al.}(2005{\natexlab{a}})\citenamefont {Brambilla}, \citenamefont
  {Vairo},\ and\ \citenamefont {R{\"o}sch}}]{Brambilla:2005yk}%
  \BibitemOpen
  \bibfield  {author} {\bibinfo {author} {\bibfnamefont {N.}~\bibnamefont
  {Brambilla}}, \bibinfo {author} {\bibfnamefont {A.}~\bibnamefont {Vairo}},\
  and\ \bibinfo {author} {\bibfnamefont {T.}~\bibnamefont {R{\"o}sch}},\
  }\bibfield  {title} {\bibinfo {title} {{Effective field theory Lagrangians
  for baryons with two and three heavy quarks}},\ }\href
  {https://doi.org/10.1103/PhysRevD.72.034021} {\bibfield  {journal} {\bibinfo
  {journal} {Phys. Rev. D}\ }\textbf {\bibinfo {volume} {72}},\ \bibinfo
  {pages} {034021} (\bibinfo {year} {2005}{\natexlab{a}})},\ \Eprint
  {https://arxiv.org/abs/hep-ph/0506065} {arXiv:hep-ph/0506065} \BibitemShut
  {NoStop}%
\bibitem [{\citenamefont {Soto}\ and\ \citenamefont
  {Tarr\'us~Castell\`a}(2020{\natexlab{b}})}]{Soto:2020pfa}%
  \BibitemOpen
  \bibfield  {author} {\bibinfo {author} {\bibfnamefont {J.}~\bibnamefont
  {Soto}}\ and\ \bibinfo {author} {\bibfnamefont {J.}~\bibnamefont
  {Tarr\'us~Castell\`a}},\ }\bibfield  {title} {\bibinfo {title} {{Effective
  field theory for double heavy baryons at strong coupling}},\ }\href
  {https://doi.org/10.1103/PhysRevD.104.059901} {\bibfield  {journal} {\bibinfo
   {journal} {Phys. Rev. D}\ }\textbf {\bibinfo {volume} {102}},\ \bibinfo
  {pages} {014013} (\bibinfo {year} {2020}{\natexlab{b}})},\ \bibinfo {note}
  {[Erratum: Phys.Rev.D 104, 059901 (2021)]},\ \Eprint
  {https://arxiv.org/abs/2005.00551} {arXiv:2005.00551 [hep-ph]} \BibitemShut
  {NoStop}%
\bibitem [{\citenamefont {Soto}\ and\ \citenamefont
  {Tarr\'us~Castell\`a}(2021)}]{Soto:2021cgk}%
  \BibitemOpen
  \bibfield  {author} {\bibinfo {author} {\bibfnamefont {J.}~\bibnamefont
  {Soto}}\ and\ \bibinfo {author} {\bibfnamefont {J.}~\bibnamefont
  {Tarr\'us~Castell\`a}},\ }\bibfield  {title} {\bibinfo {title} {{Effective
  QCD string and doubly heavy baryons}},\ }\href
  {https://doi.org/10.1103/PhysRevD.104.074027} {\bibfield  {journal} {\bibinfo
   {journal} {Phys. Rev. D}\ }\textbf {\bibinfo {volume} {104}},\ \bibinfo
  {pages} {074027} (\bibinfo {year} {2021})},\ \Eprint
  {https://arxiv.org/abs/2108.00496} {arXiv:2108.00496 [hep-ph]} \BibitemShut
  {NoStop}%
\bibitem [{\citenamefont {Bruschini}(2024{\natexlab{b}})}]{Bruschini:2024fyj}%
  \BibitemOpen
  \bibfield  {author} {\bibinfo {author} {\bibfnamefont {R.}~\bibnamefont
  {Bruschini}},\ }\bibfield  {title} {\bibinfo {title} {{Model-independent
  predictions for decays of double-heavy hadrons into pairs of heavy
  hadrons}},\ }\href {https://doi.org/10.1103/PhysRevD.110.074033} {\bibfield
  {journal} {\bibinfo  {journal} {Phys. Rev. D}\ }\textbf {\bibinfo {volume}
  {110}},\ \bibinfo {pages} {074033} (\bibinfo {year} {2024}{\natexlab{b}})},\
  \Eprint {https://arxiv.org/abs/2408.05150} {arXiv:2408.05150 [hep-ph]}
  \BibitemShut {NoStop}%
\bibitem [{\citenamefont {Brambilla}\ \emph {et~al.}(2024)\citenamefont
  {Brambilla}, \citenamefont {Mohapatra}, \citenamefont {Scirpa},\ and\
  \citenamefont {Vairo}}]{Brambilla:2024thx}%
  \BibitemOpen
  \bibfield  {author} {\bibinfo {author} {\bibfnamefont {N.}~\bibnamefont
  {Brambilla}}, \bibinfo {author} {\bibfnamefont {A.}~\bibnamefont
  {Mohapatra}}, \bibinfo {author} {\bibfnamefont {T.}~\bibnamefont {Scirpa}},\
  and\ \bibinfo {author} {\bibfnamefont {A.}~\bibnamefont {Vairo}},\ }\bibfield
   {title} {\bibinfo {title} {{The nature of $\chi_{c1}\left(3872\right)$ and
  $T_{cc}^+\left(3875\right)$}},\ }\href@noop {} {\  (\bibinfo {year}
  {2024})},\ \Eprint {https://arxiv.org/abs/2411.14306} {arXiv:2411.14306
  [hep-ph]} \BibitemShut {NoStop}%
\bibitem [{\citenamefont {Braaten}\ and\ \citenamefont
  {Bruschini}(2025)}]{Braaten:2024tbm}%
  \BibitemOpen
  \bibfield  {author} {\bibinfo {author} {\bibfnamefont {E.}~\bibnamefont
  {Braaten}}\ and\ \bibinfo {author} {\bibfnamefont {R.}~\bibnamefont
  {Bruschini}},\ }\bibfield  {title} {\bibinfo {title} {{Exotic hidden-heavy
  hadrons and where to find them}},\ }\href
  {https://doi.org/10.1016/j.physletb.2025.139386} {\bibfield  {journal}
  {\bibinfo  {journal} {Phys. Lett. B}\ }\textbf {\bibinfo {volume} {863}},\
  \bibinfo {pages} {139386} (\bibinfo {year} {2025})},\ \Eprint
  {https://arxiv.org/abs/2409.08002} {arXiv:2409.08002 [hep-ph]} \BibitemShut
  {NoStop}%
\bibitem [{\citenamefont {Caswell}\ and\ \citenamefont
  {Lepage}(1986)}]{Caswell:1985ui}%
  \BibitemOpen
  \bibfield  {author} {\bibinfo {author} {\bibfnamefont {W.~E.}\ \bibnamefont
  {Caswell}}\ and\ \bibinfo {author} {\bibfnamefont {G.~P.}\ \bibnamefont
  {Lepage}},\ }\bibfield  {title} {\bibinfo {title} {{Effective Lagrangians for
  Bound State Problems in QED, QCD, and Other Field Theories}},\ }\href
  {https://doi.org/10.1016/0370-2693(86)91297-9} {\bibfield  {journal}
  {\bibinfo  {journal} {Phys. Lett. B}\ }\textbf {\bibinfo {volume} {167}},\
  \bibinfo {pages} {437} (\bibinfo {year} {1986})}\BibitemShut {NoStop}%
\bibitem [{\citenamefont {Bodwin}\ \emph {et~al.}(1995)\citenamefont {Bodwin},
  \citenamefont {Braaten},\ and\ \citenamefont {Lepage}}]{Bodwin:1994jh}%
  \BibitemOpen
  \bibfield  {author} {\bibinfo {author} {\bibfnamefont {G.~T.}\ \bibnamefont
  {Bodwin}}, \bibinfo {author} {\bibfnamefont {E.}~\bibnamefont {Braaten}},\
  and\ \bibinfo {author} {\bibfnamefont {G.~P.}\ \bibnamefont {Lepage}},\
  }\bibfield  {title} {\bibinfo {title} {{Rigorous QCD analysis of inclusive
  annihilation and production of heavy quarkonium}},\ }\href
  {https://doi.org/10.1103/PhysRevD.55.5853} {\bibfield  {journal} {\bibinfo
  {journal} {Phys. Rev. D}\ }\textbf {\bibinfo {volume} {51}},\ \bibinfo
  {pages} {1125} (\bibinfo {year} {1995})},\ \bibinfo {note} {[Erratum:
  Phys.Rev.D 55, 5853 (1997)]},\ \Eprint {https://arxiv.org/abs/hep-ph/9407339}
  {arXiv:hep-ph/9407339} \BibitemShut {NoStop}%
\bibitem [{\citenamefont {Pineda}\ and\ \citenamefont
  {Soto}(1998)}]{Pineda:1997bj}%
  \BibitemOpen
  \bibfield  {author} {\bibinfo {author} {\bibfnamefont {A.}~\bibnamefont
  {Pineda}}\ and\ \bibinfo {author} {\bibfnamefont {J.}~\bibnamefont {Soto}},\
  }\bibfield  {title} {\bibinfo {title} {{Effective field theory for ultrasoft
  momenta in NRQCD and NRQED}},\ }\href
  {https://doi.org/10.1016/S0920-5632(97)01102-X} {\bibfield  {journal}
  {\bibinfo  {journal} {Nucl. Phys. B Proc. Suppl.}\ }\textbf {\bibinfo
  {volume} {64}},\ \bibinfo {pages} {428} (\bibinfo {year} {1998})},\ \Eprint
  {https://arxiv.org/abs/hep-ph/9707481} {arXiv:hep-ph/9707481} \BibitemShut
  {NoStop}%
\bibitem [{\citenamefont {Brambilla}\ \emph {et~al.}(2001)\citenamefont
  {Brambilla}, \citenamefont {Pineda}, \citenamefont {Soto},\ and\
  \citenamefont {Vairo}}]{Brambilla:2000gk}%
  \BibitemOpen
  \bibfield  {author} {\bibinfo {author} {\bibfnamefont {N.}~\bibnamefont
  {Brambilla}}, \bibinfo {author} {\bibfnamefont {A.}~\bibnamefont {Pineda}},
  \bibinfo {author} {\bibfnamefont {J.}~\bibnamefont {Soto}},\ and\ \bibinfo
  {author} {\bibfnamefont {A.}~\bibnamefont {Vairo}},\ }\bibfield  {title}
  {\bibinfo {title} {{The QCD potential at $O(1/m)$}},\ }\href
  {https://doi.org/10.1103/PhysRevD.63.014023} {\bibfield  {journal} {\bibinfo
  {journal} {Phys. Rev. D}\ }\textbf {\bibinfo {volume} {63}},\ \bibinfo
  {pages} {014023} (\bibinfo {year} {2001})},\ \Eprint
  {https://arxiv.org/abs/hep-ph/0002250} {arXiv:hep-ph/0002250} \BibitemShut
  {NoStop}%
\bibitem [{\citenamefont {Pineda}\ and\ \citenamefont
  {Vairo}(2001)}]{Pineda:2000sz}%
  \BibitemOpen
  \bibfield  {author} {\bibinfo {author} {\bibfnamefont {A.}~\bibnamefont
  {Pineda}}\ and\ \bibinfo {author} {\bibfnamefont {A.}~\bibnamefont {Vairo}},\
  }\bibfield  {title} {\bibinfo {title} {{The QCD potential at O (1 / $m^{2)}$
  : Complete spin dependent and spin independent result}},\ }\href
  {https://doi.org/10.1103/PhysRevD.64.039902} {\bibfield  {journal} {\bibinfo
  {journal} {Phys. Rev. D}\ }\textbf {\bibinfo {volume} {63}},\ \bibinfo
  {pages} {054007} (\bibinfo {year} {2001})},\ \bibinfo {note} {[Erratum:
  Phys.Rev.D 64, 039902 (2001)]},\ \Eprint
  {https://arxiv.org/abs/hep-ph/0009145} {arXiv:hep-ph/0009145} \BibitemShut
  {NoStop}%
\bibitem [{\citenamefont {Brambilla}\ \emph
  {et~al.}(2005{\natexlab{b}})\citenamefont {Brambilla}, \citenamefont
  {Pineda}, \citenamefont {Soto},\ and\ \citenamefont
  {Vairo}}]{Brambilla:2004jw}%
  \BibitemOpen
  \bibfield  {author} {\bibinfo {author} {\bibfnamefont {N.}~\bibnamefont
  {Brambilla}}, \bibinfo {author} {\bibfnamefont {A.}~\bibnamefont {Pineda}},
  \bibinfo {author} {\bibfnamefont {J.}~\bibnamefont {Soto}},\ and\ \bibinfo
  {author} {\bibfnamefont {A.}~\bibnamefont {Vairo}},\ }\bibfield  {title}
  {\bibinfo {title} {{Effective Field Theories for Heavy Quarkonium}},\ }\href
  {https://doi.org/10.1103/RevModPhys.77.1423} {\bibfield  {journal} {\bibinfo
  {journal} {Rev. Mod. Phys.}\ }\textbf {\bibinfo {volume} {77}},\ \bibinfo
  {pages} {1423} (\bibinfo {year} {2005}{\natexlab{b}})},\ \Eprint
  {https://arxiv.org/abs/hep-ph/0410047} {arXiv:hep-ph/0410047} \BibitemShut
  {NoStop}%
\bibitem [{\citenamefont {Born}\ and\ \citenamefont
  {Oppenheimer}(1927)}]{Born-Oppenheimer}%
  \BibitemOpen
  \bibfield  {author} {\bibinfo {author} {\bibfnamefont {M.}~\bibnamefont
  {Born}}\ and\ \bibinfo {author} {\bibfnamefont {R.}~\bibnamefont
  {Oppenheimer}},\ }\bibfield  {title} {\bibinfo {title} {Zur quantentheorie
  der molekeln},\ }\href
  {https://doi.org/https://doi.org/10.1002/andp.19273892002} {\bibfield
  {journal} {\bibinfo  {journal} {Annalen der Physik}\ }\textbf {\bibinfo
  {volume} {389}},\ \bibinfo {pages} {457} (\bibinfo {year} {1927})},\ \Eprint
  {https://arxiv.org/abs/https://onlinelibrary.wiley.com/doi/pdf/10.1002/andp.19273892002}
  {https://onlinelibrary.wiley.com/doi/pdf/10.1002/andp.19273892002}
  \BibitemShut {NoStop}%
\bibitem [{\citenamefont {Landau}\ and\ \citenamefont
  {Lifshits}(1991)}]{Landau:1991wop}%
  \BibitemOpen
  \bibfield  {author} {\bibinfo {author} {\bibfnamefont {L.~D.}\ \bibnamefont
  {Landau}}\ and\ \bibinfo {author} {\bibfnamefont {E.~M.}\ \bibnamefont
  {Lifshits}},\ }\href@noop {} {\emph {\bibinfo {title} {{Quantum Mechanics}:
  {Non-Relativistic Theory}}}},\ \bibinfo {series} {Course of Theoretical
  Physics}, Vol.\ \bibinfo {volume} {v.3}\ (\bibinfo  {publisher}
  {Butterworth-Heinemann},\ \bibinfo {address} {Oxford},\ \bibinfo {year}
  {1991})\BibitemShut {NoStop}%
\bibitem [{\citenamefont {Alberti}\ \emph {et~al.}(2017)\citenamefont
  {Alberti}, \citenamefont {Bali}, \citenamefont {Collins}, \citenamefont
  {Knechtli}, \citenamefont {Moir},\ and\ \citenamefont
  {S\"oldner}}]{Alberti:2016dru}%
  \BibitemOpen
  \bibfield  {author} {\bibinfo {author} {\bibfnamefont {M.}~\bibnamefont
  {Alberti}}, \bibinfo {author} {\bibfnamefont {G.~S.}\ \bibnamefont {Bali}},
  \bibinfo {author} {\bibfnamefont {S.}~\bibnamefont {Collins}}, \bibinfo
  {author} {\bibfnamefont {F.}~\bibnamefont {Knechtli}}, \bibinfo {author}
  {\bibfnamefont {G.}~\bibnamefont {Moir}},\ and\ \bibinfo {author}
  {\bibfnamefont {W.}~\bibnamefont {S\"oldner}},\ }\bibfield  {title} {\bibinfo
  {title} {{Hadroquarkonium from lattice QCD}},\ }\href
  {https://doi.org/10.1103/PhysRevD.95.074501} {\bibfield  {journal} {\bibinfo
  {journal} {Phys. Rev. D}\ }\textbf {\bibinfo {volume} {95}},\ \bibinfo
  {pages} {074501} (\bibinfo {year} {2017})},\ \Eprint
  {https://arxiv.org/abs/1608.06537} {arXiv:1608.06537 [hep-lat]} \BibitemShut
  {NoStop}%
\bibitem [{\citenamefont {Prelovsek}\ \emph {et~al.}(2020)\citenamefont
  {Prelovsek}, \citenamefont {Bahtiyar},\ and\ \citenamefont
  {Petkovic}}]{Prelovsek:2019ywc}%
  \BibitemOpen
  \bibfield  {author} {\bibinfo {author} {\bibfnamefont {S.}~\bibnamefont
  {Prelovsek}}, \bibinfo {author} {\bibfnamefont {H.}~\bibnamefont
  {Bahtiyar}},\ and\ \bibinfo {author} {\bibfnamefont {J.}~\bibnamefont
  {Petkovic}},\ }\bibfield  {title} {\bibinfo {title} {{Zb tetraquark channel
  from lattice QCD and Born-Oppenheimer approximation}},\ }\href
  {https://doi.org/10.1016/j.physletb.2020.135467} {\bibfield  {journal}
  {\bibinfo  {journal} {Phys. Lett. B}\ }\textbf {\bibinfo {volume} {805}},\
  \bibinfo {pages} {135467} (\bibinfo {year} {2020})},\ \Eprint
  {https://arxiv.org/abs/1912.02656} {arXiv:1912.02656 [hep-lat]} \BibitemShut
  {NoStop}%
\bibitem [{\citenamefont {Sadl}\ and\ \citenamefont
  {Prelovsek}(2021)}]{Sadl:2021bme}%
  \BibitemOpen
  \bibfield  {author} {\bibinfo {author} {\bibfnamefont {M.}~\bibnamefont
  {Sadl}}\ and\ \bibinfo {author} {\bibfnamefont {S.}~\bibnamefont
  {Prelovsek}},\ }\bibfield  {title} {\bibinfo {title} {{Tetraquark systems
  $\bar bb \bar du$ in the static limit and lattice QCD}},\ }\href
  {https://doi.org/10.1103/PhysRevD.104.114503} {\bibfield  {journal} {\bibinfo
   {journal} {Phys. Rev. D}\ }\textbf {\bibinfo {volume} {104}},\ \bibinfo
  {pages} {114503} (\bibinfo {year} {2021})},\ \Eprint
  {https://arxiv.org/abs/2109.08560} {arXiv:2109.08560 [hep-lat]} \BibitemShut
  {NoStop}%
\bibitem [{\citenamefont {Foster}\ and\ \citenamefont
  {Michael}(1999)}]{Foster:1998wu}%
  \BibitemOpen
  \bibfield  {author} {\bibinfo {author} {\bibfnamefont {M.}~\bibnamefont
  {Foster}}\ and\ \bibinfo {author} {\bibfnamefont {C.}~\bibnamefont {Michael}}
  (\bibinfo {collaboration} {UKQCD}),\ }\bibfield  {title} {\bibinfo {title}
  {{Hadrons with a heavy color adjoint particle}},\ }\href
  {https://doi.org/10.1103/PhysRevD.59.094509} {\bibfield  {journal} {\bibinfo
  {journal} {Phys. Rev. D}\ }\textbf {\bibinfo {volume} {59}},\ \bibinfo
  {pages} {094509} (\bibinfo {year} {1999})},\ \Eprint
  {https://arxiv.org/abs/hep-lat/9811010} {arXiv:hep-lat/9811010} \BibitemShut
  {NoStop}%
\bibitem [{\citenamefont {Brambilla}\ \emph
  {et~al.}(2023{\natexlab{b}})\citenamefont {Brambilla}, \citenamefont
  {Delgado}, \citenamefont {Kronfeld}, \citenamefont {Leino}, \citenamefont
  {Petreczky}, \citenamefont {Steinbei\ss{}er}, \citenamefont {Vairo},\ and\
  \citenamefont {Weber}}]{Brambilla:2022het}%
  \BibitemOpen
  \bibfield  {author} {\bibinfo {author} {\bibfnamefont {N.}~\bibnamefont
  {Brambilla}}, \bibinfo {author} {\bibfnamefont {R.~L.}\ \bibnamefont
  {Delgado}}, \bibinfo {author} {\bibfnamefont {A.~S.}\ \bibnamefont
  {Kronfeld}}, \bibinfo {author} {\bibfnamefont {V.}~\bibnamefont {Leino}},
  \bibinfo {author} {\bibfnamefont {P.}~\bibnamefont {Petreczky}}, \bibinfo
  {author} {\bibfnamefont {S.}~\bibnamefont {Steinbei\ss{}er}}, \bibinfo
  {author} {\bibfnamefont {A.}~\bibnamefont {Vairo}},\ and\ \bibinfo {author}
  {\bibfnamefont {J.~H.}\ \bibnamefont {Weber}} (\bibinfo {collaboration}
  {TUMQCD}),\ }\bibfield  {title} {\bibinfo {title} {{Static energy in
  ($2+1+1$)-flavor lattice QCD: Scale setting and charm effects}},\ }\href
  {https://doi.org/10.1103/PhysRevD.107.074503} {\bibfield  {journal} {\bibinfo
   {journal} {Phys. Rev. D}\ }\textbf {\bibinfo {volume} {107}},\ \bibinfo
  {pages} {074503} (\bibinfo {year} {2023}{\natexlab{b}})},\ \Eprint
  {https://arxiv.org/abs/2206.03156} {arXiv:2206.03156 [hep-lat]} \BibitemShut
  {NoStop}%
\bibitem [{\citenamefont {Schlosser}\ and\ \citenamefont
  {Wagner}(2022)}]{Schlosser:2021wnr}%
  \BibitemOpen
  \bibfield  {author} {\bibinfo {author} {\bibfnamefont {C.}~\bibnamefont
  {Schlosser}}\ and\ \bibinfo {author} {\bibfnamefont {M.}~\bibnamefont
  {Wagner}},\ }\bibfield  {title} {\bibinfo {title} {{Hybrid static potentials
  in SU(3) lattice gauge theory at small quark-antiquark separations}},\ }\href
  {https://doi.org/10.1103/PhysRevD.105.054503} {\bibfield  {journal} {\bibinfo
   {journal} {Phys. Rev. D}\ }\textbf {\bibinfo {volume} {105}},\ \bibinfo
  {pages} {054503} (\bibinfo {year} {2022})},\ \Eprint
  {https://arxiv.org/abs/2111.00741} {arXiv:2111.00741 [hep-lat]} \BibitemShut
  {NoStop}%
\bibitem [{\citenamefont {Sharifian}\ \emph {et~al.}(2023)\citenamefont
  {Sharifian}, \citenamefont {Cardoso},\ and\ \citenamefont
  {Bicudo}}]{Sharifian:2023idc}%
  \BibitemOpen
  \bibfield  {author} {\bibinfo {author} {\bibfnamefont {A.}~\bibnamefont
  {Sharifian}}, \bibinfo {author} {\bibfnamefont {N.}~\bibnamefont {Cardoso}},\
  and\ \bibinfo {author} {\bibfnamefont {P.}~\bibnamefont {Bicudo}},\
  }\bibfield  {title} {\bibinfo {title} {{Eight very excited flux tube spectra
  and possible axions in SU(3) lattice gauge theory}},\ }\href
  {https://doi.org/10.1103/PhysRevD.107.114507} {\bibfield  {journal} {\bibinfo
   {journal} {Phys. Rev. D}\ }\textbf {\bibinfo {volume} {107}},\ \bibinfo
  {pages} {114507} (\bibinfo {year} {2023})},\ \Eprint
  {https://arxiv.org/abs/2303.15152} {arXiv:2303.15152 [hep-lat]} \BibitemShut
  {NoStop}%
\bibitem [{\citenamefont {Alasiri}\ \emph {et~al.}(2025)\citenamefont
  {Alasiri}, \citenamefont {Braaten},\ and\ \citenamefont
  {Bruschini}}]{Alasiri:2025roh}%
  \BibitemOpen
  \bibfield  {author} {\bibinfo {author} {\bibfnamefont {F.}~\bibnamefont
  {Alasiri}}, \bibinfo {author} {\bibfnamefont {E.}~\bibnamefont {Braaten}},\
  and\ \bibinfo {author} {\bibfnamefont {R.}~\bibnamefont {Bruschini}},\
  }\bibfield  {title} {\bibinfo {title} {{Hidden-Heavy Pentaquarks and Where to
  Find Them}},\ }\href@noop {} {\  (\bibinfo {year} {2025})},\ \Eprint
  {https://arxiv.org/abs/2507.06991} {arXiv:2507.06991 [hep-ph]} \BibitemShut
  {NoStop}%
\bibitem [{\citenamefont {Pineda}(2001)}]{Pineda:2001zq}%
  \BibitemOpen
  \bibfield  {author} {\bibinfo {author} {\bibfnamefont {A.}~\bibnamefont
  {Pineda}},\ }\bibfield  {title} {\bibinfo {title} {{Determination of the
  bottom quark mass from the $\Upsilon(1S)$ system}},\ }\href
  {https://doi.org/10.1088/1126-6708/2001/06/022} {\bibfield  {journal}
  {\bibinfo  {journal} {JHEP}\ }\textbf {\bibinfo {volume} {06}},\ \bibinfo
  {pages} {022}},\ \Eprint {https://arxiv.org/abs/hep-ph/0105008}
  {arXiv:hep-ph/0105008} \BibitemShut {NoStop}%
\bibitem [{\citenamefont {Bali}\ and\ \citenamefont
  {Pineda}(2004)}]{Bali:2003jq}%
  \BibitemOpen
  \bibfield  {author} {\bibinfo {author} {\bibfnamefont {G.~S.}\ \bibnamefont
  {Bali}}\ and\ \bibinfo {author} {\bibfnamefont {A.}~\bibnamefont {Pineda}},\
  }\bibfield  {title} {\bibinfo {title} {{QCD phenomenology of static sources
  and gluonic excitations at short distances}},\ }\href
  {https://doi.org/10.1103/PhysRevD.69.094001} {\bibfield  {journal} {\bibinfo
  {journal} {Phys. Rev. D}\ }\textbf {\bibinfo {volume} {69}},\ \bibinfo
  {pages} {094001} (\bibinfo {year} {2004})},\ \Eprint
  {https://arxiv.org/abs/hep-ph/0310130} {arXiv:hep-ph/0310130} \BibitemShut
  {NoStop}%
\bibitem [{\citenamefont {Yalikun}\ \emph {et~al.}(2021)\citenamefont
  {Yalikun}, \citenamefont {Lin}, \citenamefont {Guo}, \citenamefont {Kamiya},\
  and\ \citenamefont {Zou}}]{Yalikun:2021bfm}%
  \BibitemOpen
  \bibfield  {author} {\bibinfo {author} {\bibfnamefont {N.}~\bibnamefont
  {Yalikun}}, \bibinfo {author} {\bibfnamefont {Y.-H.}\ \bibnamefont {Lin}},
  \bibinfo {author} {\bibfnamefont {F.-K.}\ \bibnamefont {Guo}}, \bibinfo
  {author} {\bibfnamefont {Y.}~\bibnamefont {Kamiya}},\ and\ \bibinfo {author}
  {\bibfnamefont {B.-S.}\ \bibnamefont {Zou}},\ }\bibfield  {title} {\bibinfo
  {title} {{Coupled-channel effects of the
  {\ensuremath{\Sigma}}c(*)D{\textasciimacron}(*)-{\ensuremath{\Lambda}}c(2595)D{\textasciimacron}
  system and molecular nature of the Pc pentaquark states from one-boson
  exchange model}},\ }\href {https://doi.org/10.1103/PhysRevD.104.094039}
  {\bibfield  {journal} {\bibinfo  {journal} {Phys. Rev. D}\ }\textbf {\bibinfo
  {volume} {104}},\ \bibinfo {pages} {094039} (\bibinfo {year} {2021})},\
  \Eprint {https://arxiv.org/abs/2109.03504} {arXiv:2109.03504 [hep-ph]}
  \BibitemShut {NoStop}%
\bibitem [{\citenamefont {Segovia}\ \emph {et~al.}(2019)\citenamefont
  {Segovia}, \citenamefont {Steinbei{\ss}er},\ and\ \citenamefont
  {Vairo}}]{Segovia:2018qzb}%
  \BibitemOpen
  \bibfield  {author} {\bibinfo {author} {\bibfnamefont {J.}~\bibnamefont
  {Segovia}}, \bibinfo {author} {\bibfnamefont {S.}~\bibnamefont
  {Steinbei{\ss}er}},\ and\ \bibinfo {author} {\bibfnamefont {A.}~\bibnamefont
  {Vairo}},\ }\bibfield  {title} {\bibinfo {title} {{Electric dipole
  transitions of $1P$ bottomonia}},\ }\href
  {https://doi.org/10.1103/PhysRevD.99.074011} {\bibfield  {journal} {\bibinfo
  {journal} {Phys. Rev. D}\ }\textbf {\bibinfo {volume} {99}},\ \bibinfo
  {pages} {074011} (\bibinfo {year} {2019})},\ \Eprint
  {https://arxiv.org/abs/1811.07590} {arXiv:1811.07590 [hep-ph]} \BibitemShut
  {NoStop}%
\bibitem [{\citenamefont {Bali}\ \emph {et~al.}(2005)\citenamefont {Bali},
  \citenamefont {Neff}, \citenamefont {Duessel}, \citenamefont {Lippert},\ and\
  \citenamefont {Schilling}}]{Bali:2005fu}%
  \BibitemOpen
  \bibfield  {author} {\bibinfo {author} {\bibfnamefont {G.~S.}\ \bibnamefont
  {Bali}}, \bibinfo {author} {\bibfnamefont {H.}~\bibnamefont {Neff}}, \bibinfo
  {author} {\bibfnamefont {T.}~\bibnamefont {Duessel}}, \bibinfo {author}
  {\bibfnamefont {T.}~\bibnamefont {Lippert}},\ and\ \bibinfo {author}
  {\bibfnamefont {K.}~\bibnamefont {Schilling}} (\bibinfo {collaboration}
  {SESAM}),\ }\bibfield  {title} {\bibinfo {title} {{Observation of string
  breaking in QCD}},\ }\href {https://doi.org/10.1103/PhysRevD.71.114513}
  {\bibfield  {journal} {\bibinfo  {journal} {Phys. Rev. D}\ }\textbf {\bibinfo
  {volume} {71}},\ \bibinfo {pages} {114513} (\bibinfo {year} {2005})},\
  \Eprint {https://arxiv.org/abs/hep-lat/0505012} {arXiv:hep-lat/0505012}
  \BibitemShut {NoStop}%
\bibitem [{\citenamefont {Bulava}\ \emph {et~al.}(2024)\citenamefont {Bulava},
  \citenamefont {Knechtli}, \citenamefont {Koch}, \citenamefont {Morningstar},\
  and\ \citenamefont {Peardon}}]{Bulava:2024jpj}%
  \BibitemOpen
  \bibfield  {author} {\bibinfo {author} {\bibfnamefont {J.}~\bibnamefont
  {Bulava}}, \bibinfo {author} {\bibfnamefont {F.}~\bibnamefont {Knechtli}},
  \bibinfo {author} {\bibfnamefont {V.}~\bibnamefont {Koch}}, \bibinfo {author}
  {\bibfnamefont {C.}~\bibnamefont {Morningstar}},\ and\ \bibinfo {author}
  {\bibfnamefont {M.}~\bibnamefont {Peardon}},\ }\bibfield  {title} {\bibinfo
  {title} {{The quark-mass dependence of the potential energy between static
  colour sources in the QCD vacuum with light and strange quarks}},\ }\href
  {https://doi.org/10.1016/j.physletb.2024.138754} {\bibfield  {journal}
  {\bibinfo  {journal} {Phys. Lett. B}\ }\textbf {\bibinfo {volume} {854}},\
  \bibinfo {pages} {138754} (\bibinfo {year} {2024})},\ \Eprint
  {https://arxiv.org/abs/2403.00754} {arXiv:2403.00754 [hep-lat]} \BibitemShut
  {NoStop}%
\bibitem [{\citenamefont {Aaij}\ \emph {et~al.}(2022)\citenamefont {Aaij} \emph
  {et~al.}}]{LHCb:2021chn}%
  \BibitemOpen
  \bibfield  {author} {\bibinfo {author} {\bibfnamefont {R.}~\bibnamefont
  {Aaij}} \emph {et~al.} (\bibinfo {collaboration} {LHCb}),\ }\bibfield
  {title} {\bibinfo {title} {{Evidence for a new structure in the $J/\psi p$
  and $J/\psi \bar{p}$ systems in $B_s^0 \to J/\psi p \bar{p}$ decays}},\
  }\href {https://doi.org/10.1103/PhysRevLett.128.062001} {\bibfield  {journal}
  {\bibinfo  {journal} {Phys. Rev. Lett.}\ }\textbf {\bibinfo {volume} {128}},\
  \bibinfo {pages} {062001} (\bibinfo {year} {2022})},\ \Eprint
  {https://arxiv.org/abs/2108.04720} {arXiv:2108.04720 [hep-ex]} \BibitemShut
  {NoStop}%
\bibitem [{\citenamefont {Messchendorp}\ \emph {et~al.}(2025)\citenamefont
  {Messchendorp}, \citenamefont {Nerling},\ and\ \citenamefont
  {Ritman}}]{Messchendorp:2025vzi}%
  \BibitemOpen
  \bibfield  {author} {\bibinfo {author} {\bibfnamefont {J.}~\bibnamefont
  {Messchendorp}}, \bibinfo {author} {\bibfnamefont {F.}~\bibnamefont
  {Nerling}},\ and\ \bibinfo {author} {\bibfnamefont {J.}~\bibnamefont
  {Ritman}},\ }\bibfield  {title} {\bibinfo {title} {{A Cross-Community-Driven
  Hadron Physics Program at GSI/FAIR}},\ }\href
  {https://doi.org/10.1080/10619127.2025.2454877} {\bibfield  {journal}
  {\bibinfo  {journal} {Nucl. Phys. News}\ }\textbf {\bibinfo {volume} {35}},\
  \bibinfo {pages} {18} (\bibinfo {year} {2025})}\BibitemShut {NoStop}%
\end{thebibliography}%

\end{document}